\documentclass[11pt]{article}
\usepackage[letterpaper, portrait, margin=1in]{geometry}
\usepackage{algorithm}
\usepackage[noend]{algpseudocode}

\makeatletter
\AddToHook{env/algorithmic/before}{\def\@currentcounter{ALG@line}}
\makeatother

\usepackage{url}
\usepackage{amsmath,amssymb,amsthm}
\usepackage[hypertexnames=false,colorlinks=true,linkcolor=blue,citecolor=ForestGreen]{hyperref}
\usepackage[noabbrev,capitalise,nameinlink]{cleveref}
\usepackage{thmtools,thm-restate}
\usepackage{mathtools}
\usepackage{xspace}
\usepackage{verbatim}
\usepackage{mathrsfs}
\usepackage[usenames,dvipsnames,svgnames,table]{xcolor}
\usepackage{pgf}
\usepackage[dvipsnames]{xcolor}
\usepackage{todo}
\usepackage{tabularx}
\usepackage{enumitem}
\usepackage{derivative}
\usepackage{bm}
\usepackage{multirow}
\usepackage{diagbox}
\usepackage{nicematrix}
\usepackage{parskip}
\usepackage{adjustbox}
\usepackage{tikz}
\usepackage{tikz-3dplot}
\usepackage{transparent}
\usepackage{subcaption}

\usepackage{dsfont}

\newcommand*\ie{i.\kern.1em e., }
\newcommand*\eg{e.\kern.1em g., }
\newcommand*\cf{c.\kern.1em f.\ }
\newcommand*\almev{a.\kern.1em e.\ }
\newcommand*\iid{i.i.d.\ }

\makeatletter
\@ifundefined{newcounteralias}{}{\renewcommand\thmt@autorefsetup{\@xa\def\csname\thmt@envname autorefname\@xa\endcsname\@xa{\thmt@thmname}}}
\makeatother

\topsep=1em \theoremstyle{plain}
\newtheorem{theorem}{Theorem}[section]
\newtheorem{lemma}[theorem]{Lemma}

\newtheorem{proposition}[theorem]{Proposition}
\newtheorem{claim}[theorem]{Claim}
\newtheorem{corollary}[theorem]{Corollary}

\newtheorem{conjecture}[theorem]{Conjecture}

\newtheorem{definition}[theorem]{Definition}

\crefname{claim}{Claim}{Claims}
\crefname{fact}{Fact}{Facts}
\crefname{ALG@line}{Line}{Lines}

\theoremstyle{definition}

\newtheorem{remark}[theorem]{Remark}

\theoremstyle{plain}

\newcommand{\ignore}[1]{}

   \DeclareMathOperator{\poly}{poly}

\DeclareMathOperator{\sign}{sign}

\newcommand{\dist}{\mathsf{dist}}
\newcommand{\err}{\mathsf{err}}

\newcommand{\Accept}{\mathsf{Accept}}
\newcommand{\Reject}{\mathsf{Reject}}
\newcommand{\Abstain}{\mathsf{Tantrum}}

\newcommand{\Var}[1]{\mathrm{Var} \left[ #1 \right]}
\newcommand{\Varu}[2]{\underset{ #1 } {\mathrm{Var}} \left[ #2 \right]}
\newcommand{\Varuc}[3]{\underset{ #1 } {\mathrm{Var}} \left[ #2 \;\; \left| \;\; #3 \right. \right]}
\newcommand{\Cov}[1]{\mathrm{Cov} \left[ #1 \right]}
\newcommand{\Covu}[2]{\underset{ #1 } {\mathrm{Cov}} \left[ #2 \right]}

\newcommand{\Ex}[1]{\bE \left[ #1 \right]}
\newcommand{\Exu}[2]{\underset{#1} \bE \left[ #2 \right] }
\newcommand{\Exuc}[3]{\underset{#1} \bE \left[ #2 \;\; \left| \;\; #3
\right.\right] }

\renewcommand{\Pr}[1]{\bP \left[ #1 \right]} \newcommand{\Pru}[2]{\underset{ #1 }\bP \left[ #2 \right]}

\newcommand{\define}{\vcentcolon=}

\renewcommand{\epsilon}{\varepsilon}

\newcommand{\inn}[1]{\langle #1 \rangle}

\newcommand{\ind}[1]{\mathds{1} \left[ #1 \right] }

\newcommand{\zo}{\{0,1\}}
\newcommand{\pmset}{\{\pm 1\}}

\newcommand{\cA}{\ensuremath{\mathcal{A}}}

\newcommand{\cD}{\ensuremath{\mathcal{D}}}
\newcommand{\cE}{\ensuremath{\mathcal{E}}}

\newcommand{\cN}{\ensuremath{\mathcal{N}}}

\newcommand{\cS}{\ensuremath{\mathcal{S}}}

\newcommand{\cU}{\ensuremath{\mathcal{U}}}

\newcommand{\cX}{\ensuremath{\mathcal{X}}}

\newcommand{\bE}{\ensuremath{\mathbb{E}}}

\newcommand{\bN}{\ensuremath{\mathbb{N}}}

\newcommand{\bR}{\ensuremath{\mathbb{R}}}
\newcommand{\bS}{\ensuremath{\mathbb{S}}}

\newcommand{\bZ}{\ensuremath{\mathbb{Z}}}

\newcommand{\abs}[1]{{\left| #1 \right|}}
\newcommand{\vabs}[1]{{\left\| #1 \right\|}}
\newcommand{\abra}[1]{{\left\langle #1 \right\rangle}}
\newcommand{\pbra}[1]{{\left( #1 \right)}}
\newcommand{\cbra}[1]{{\left\{ #1 \right\}}}
\newcommand{\sbra}[1]{{\left[ #1 \right]}}

\newcommand{\wh}[1]{\widehat{#1}}
\newcommand{\wt}[1]{\widetilde{#1}}

\newcommand{\TV}{\dist_{\mathsf{TV}}}

\usepackage{authblk}

\newcommand{\diag}{\mathrm{diag}}
\def\eps{\epsilon}
\DeclareMathOperator{\Tr}{Tr}

\newcommand{\distdis}{\mathsf{DD}}

\newcommand{\trunc}{\mathrm{trunc}}
\newcommand{\ith}[2]{{#1}^{#2}}

\newcommand{\DYES}{\mathfrak D_{yes}}
\newcommand{\DNO}{\mathfrak D_{no}}
\newcommand{\yyes}{\bm y_{yes}}
\newcommand{\yno}{\bm y_{no}}
\newcommand{\Xyes}{\bm X_{yes}}
\newcommand{\Xno}{\bm X_{no}}
\newcommand{\vyes}{\bm v_{yes}}
\newcommand{\vno}{\bm v_{no}}
\newcommand{\pyes}{p_{yes}}
\newcommand{\pno}{p_{no}}
\newcommand{\qyes}{q_{yes}}
\newcommand{\qno}{q_{no}}

\algnewcommand{\AlgSection}[1]{\Statex
  \vspace{0.5\baselineskip}\hspace*{\dimexpr-\algorithmicindent/2\relax}\textit{#1}}

\title{Distribution-Robust Halfspace Testing}

\newcommand{\AffilColumbia}{*}
\newcommand{\AffilUBC}{$\dagger$}

\author[\AffilColumbia]{Xi Chen}
\author[\AffilColumbia]{Renato Ferreira Pinto Jr.}
\author[\AffilUBC]{Nathaniel Harms}
\author[\AffilColumbia]{Rocco A. Servedio}

\affil[\AffilColumbia]{Columbia University}
\affil[\AffilUBC]{University of British Columbia}

\date{}

\begin{document}

\maketitle

\begin{abstract}
Standard property testing algorithms are typically designed either for a
specific input distribution (\emph{distribution-specific} algorithms) or for an
arbitrary unknown distribution (\emph{distribution-free} algorithms). Halfspaces
are an important example where distribution-specific testers can be extremely
efficient, while distribution-free testers are expensive and essentially require
learning. We introduce an intermediate type of algorithm, a
\emph{distribution-robust} halfspace tester, which must provide the same
guarantees as a distribution-specific tester when the input distribution is the
one specified, but must also provide a trustworthy answer under arbitrary
distributions. This is achieved by allowing the tester to report that its input
distribution is not what it expected (inspired by the recent \emph{testable
learning} framework of Rubinfeld and Vasilyan).

We prove that distribution-robust testing of halfspaces under the standard
Gaussian requires only $O(n^{2/3})$ queries, compared to the known $\widetilde
\Omega(n)$ required for distribution-free testing. We also prove that
sample-based distribu\-tion-robust testing of halfspaces is harder than
sample-based testing under the standard Gaussian, requiring at least
$\Omega(n^{2/3})$ samples compared to the known $\widetilde \Theta(\sqrt n)$
bound  for the latter. In fact, we show that the latter $\widetilde
\Theta(\sqrt n)$ bound becomes $\Omega(n^{2/3})$ when the covariance matrix is
only mildly perturbed.
\end{abstract}

\textbf{AI Methodology:} The definitions, algorithms, lower bound construction,
and the entire text were created by humans. AI assistance was used for several
calculations (specifically \cref{sec:spectral,sec:test-statistic}, and
\cref{section:lb-poincare,section:lb-gradient-calculations}), and to help check
correctness of proofs. 

\thispagestyle{empty}
\setcounter{page}{0}
\newpage
{
\setcounter{tocdepth}{2}
\tableofcontents
}
\thispagestyle{empty}
\setcounter{page}{0}
\newpage
\setcounter{page}{1}

\allowdisplaybreaks

\section{Introduction}

One of the simplest and most practical methods of classifying data is with a
linear classifier, also known as a halfspace. Of course, this only works if the
data is actually linearly separable. A \emph{halfspace tester} is an algorithm
that checks whether this is the case. Unfortunately, as we will explain in more
detail below, known halfspace testers fall into two categories: they either make
no assumptions about the underlying probability distribution (in which case they
cannot be more efficient than learning), or make an assumption about the
underlying distribution that is effectively unverifiable. To get theoretical
guarantees on the output of a halfspace tester, one must therefore be certain
that the distributional assumptions are satisfied, or else give up and run the
na\"ive testing-by-learning algorithm, sacrificing efficiency. In this paper, we
study a new type of halfspace tester, called a \emph{distribution-robust
tester}, that is intended to resolve this issue.

Let us motivate the problem of halfspace testing from the perspective of learning theory, and then
we will describe the limitations of current testers, before introducing distribution-robust testing.
If we have data drawn from a probability distribution $\cD$ over $\bR^n$ and labelled by an unknown
function $f \colon \bR^n \to \pmset$, we could attempt to predict future labels $f(\bm x)$ on new
points $\bm x$ by training a linear classifier $g(x) \define \sign(\inn{w,x}-t)$, where we try to
choose $w \in \bR^n$ and $t \in \bR$ such that $g(\bm x) = f(\bm x)$ on new samples $\bm
x$. To formalize this goal, we aim to produce $g$ which minimizes the distance
\[
    \dist_\cD(f,g) \define \Pru{\bm x \sim \cD}{ f(\bm x) \neq g(\bm x) } .
\]
Algorithms for doing this have been studied extensively, but for this approach to succeed,
we~must assume that $f$ is close to a linear classifier.
The goal of a \emph{property testing} algorithm (in
particular, a \emph{halfspace tester}) is to test whether this assumption holds or not. As noted by
Goldreich, Goldwasser and Ron \cite{GGR98}, if we can do this more efficiently than simply running
the learning algorithm and checking if it worked (``trial-and-error''), then we can decide
whether our learning model will succeed without investing the resources to run the learner itself,
and move on to another model if necessary (this is known as \emph{model selection}). 

All known results for testing halfspaces (and, in fact, most known results on
testing properties of Boolean functions in general) fit into the following
framework:

\begin{definition}[Distribution-Specific Halfspace Testers]
Let $\mathfrak C$ be a class of probability distributions over $\bR^n$. Let algorithm $\cA$ be given
a proximity parameter $\epsilon > 0$, sample access to an input distribution $\cD$, and query access to a
function $f \colon \bR^n \to \pmset$. Algorithm $\cA$ is a \emph{$\mathfrak C$-specific halfspace tester}
if, for all  distributions $\cD \in \mathfrak C$, its output satisfies the following with
probability at least $3/4$:
\begin{description}
\item (Completeness) If $f$ is a halfspace, then $\cA$ outputs $\Accept$.
\item (Soundness) If $\dist_\cD(f,h) > \epsilon$ for all halfspaces $h$, then $\cA$ outputs
    $\Reject$.
\end{description}
An algorithm $\cA$ is \emph{sample-based} if it only queries samples that it
draws from $\cD$ (or equivalently if it receives labelled samples of the form
$(\bm x, f(\bm x))$ with $\bm x\sim \cD$ and makes no queries).
\end{definition}

Unfortunately, algorithms under this framework are quite brittle: If $\cD$ does
\emph{not} satisfy the promise $\mathfrak C$, then there are no guarantees on
the correctness of the output. Even worse, all known halfspace testers either
require as much data as is needed to \emph{learn} the halfspace, or they rely on
a promise $\mathfrak C$ that cannot be efficiently verified. Let us explain this
in more detail.

On one hand, if we have \emph{no} promise on the input distribution $\cD$ (\ie the class
$\mathfrak C$ is the class of all distributions\footnote{Throughout the paper, distributions $\mathcal D$ over $\mathbb R^n$ are Borel probability measures and functions $f\colon\mathbb R^n\to\{\pm1\}$ are Borel measurable;
this is needed only for quantities such as $\dist_{\mathcal D}(f,g)$ to be well defined.}
over $\bR^n$), then testing requires
$\widetilde \Omega(n)$ queries\footnote{In these comparisons we treat the proximity parameter $\eps$ as a constant and focus on the dependence on $n$.}; this setting is known as \emph{distribution-free} testing
\cite{BFH21,CP22,CFHPS26}. Since $\Theta(n)$  samples suffice to learn halfspaces regardless of  $\cD$, distribution-free testers provide almost no benefit over the ``trial-and-error''
strategy of running a learning algorithm and checking whether it worked.

On the other hand, when $\mathcal D$ is promised to be the standard
Gaussian or the uniform distribution over $\pmset^n$,
there are testers using only $O(1)$ queries \cite{MORS10}. If $\mathcal{D}$ is promised
to be the standard Gaussian \cite{BBBY12}, or more generally, any rotation-invariant distribution
\cite{Har19}, then there are sample-based testers
using $\widetilde \Theta(\sqrt n)$ samples. All of these bounds 
 are significantly better than the $\Theta(n)$ queries or samples  required to learn halfspaces under
any of these distributions. However, testing whether an unknown distribution is the standard
Gaussian cannot be done with any finite number of samples,
and testing whether an unknown distribution is the uniform distribution over $\pmset^n$ requires
$\Theta(2^{n/2})$ samples \cite{Paninski:08}, so these promises cannot be efficiently verified.

This motivates the definition of a \emph{distribution-robust tester}. The goal is to design testers
with the same guarantees as a distribution-specific tester, while avoiding both the brittleness of such
testers and the inefficiency of distribution-free ones. That is, we want:
\begin{enumerate}
\item If the promise $\cD \in \mathfrak C$ holds, the tester satisfies the same guarantees as
before;
\item If the promise does not hold, the tester is allowed to $\Abstain$ instead of answering $\Accept$ or $\Reject$,
but if it chooses to $\Accept$ or $\Reject$, then it must be correct, even if $\cD \notin
\mathfrak C$.
\end{enumerate}
A companion paper \cite{CFHS26general} defines distribution-robust testers in general, inspired by the recent \emph{testable learning} framework of Rubinfeld and
Vasilyan \cite{RV23} (on which more below), and proves a
variety of results, including separations between distribution-robust testing and standard property
testing models. In the current paper, we are focused only on halfspaces:

\begin{definition}[Distribution-Robust Halfspace Testers \cite{CFHS26general}]
\label{def:distribution-robust}
Let $\mathfrak C$ be a class of probability distributions over $\bR^n$. Let algorithm $\cA$ be given
a proximity parameter $\epsilon > 0$, sample access to an input distribution~$\cD$ (not necessarily in
$\mathfrak C$), and query access to a function $f \colon \bR^n \to \pmset$. It is a
\emph{$\mathfrak C$-robust~halfspace tester} if its output lies in
$\{\Accept, \Reject, \Abstain\}$ and the following hold with  probability at least $3/4$:
\begin{description}
\item (Completeness.) If $f$ is a halfspace, the tester outputs either $\Accept$ or $\Abstain$;
\item (Soundness.) If $\dist_\cD(f,h) > \epsilon$ for all halfspaces $h$, the tester outputs either
$\Reject$ or $\Abstain$;
\item (Assertiveness.) If $\cD \in \mathfrak C$, the tester does not output $\Abstain$.
\end{description}
\end{definition}

Another way to think of such a tester is that, if it outputs $\Accept$ or
$\Reject$, we can be confident that the answer is correct, even if the input
distribution $\cD$ does not satisfy the assumption $\cD \in \mathfrak C$. So its
$\Accept$ and $\Reject$ outputs are trustworthy even on distributions it was not
designed for.

\subsection{Results}
\label{section:intro-results}

The purpose of distribution-robust testers is to be significantly more efficient
than distribution-free testers, which for halfspaces require
$\widetilde\Omega(n)$ queries \cite{CP22}. \emph{A priori}, we had no intuition
for whether this can be achieved. Our first theorem shows that we do in fact
gain a significant advantage over distribution-free testing:

\begin{theorem}[Query Upper Bound]
\label{thm:intro-ub}
There is a $\cN(0, I_n)$-robust halfspace tester which uses only
$O(n^{2/3} \poly(1/\epsilon))$ non-adaptive queries.
\end{theorem}

Our second theorem shows that, for sample-based testers,  distribution-robust testing is harder than distribution-specific testing under $\cN(0,I_n)$
(for which $\widetilde \Theta(\sqrt n)$ samples suffice \cite{BBBY12,Har19}): 

\begin{theorem}[Sample Lower Bound]
\label{thm:intro-lb-robust}
For sufficiently small constant proximity parameter $\epsilon > 0$,
every sample-based $\cN(0, I_n)$-robust tester requires  $\Omega(n^{2/3})$ samples.
\end{theorem}

Recall that the companion paper \cite{CFHS26general} studies distribution-robust
testing in greater generalit\-y and, among other results, establishes
separations between distribution-robust testing and various other testing
models. One separation not shown there is between sample-based
distribution-specific testing and sample-based distribution-robust
testing. \cref{thm:intro-lb-robust} establishes this separation.

Beyond establishing this separation, the proof of \Cref{thm:intro-lb-robust} has
a consequence that surprised~us: the $\smash{\widetilde \Theta(\sqrt n)}$ upper
bound of \cite{BBBY12,Har19} is very fragile. Whereas we na\"ively expected
algori\-thms that work for $\cN(0, I_n)$ to generalize reasonably to $\cN(0,
\Sigma)$ when $\Sigma \approx I_n$, it turns out that this intuition fails even
when the minimum and maximum eigenvalues $\lambda_{\min}, \lambda_{\max}$ of
$\Sigma$ are within $[1,1 + \theta]$ for an arbitrarily small constant $\theta$:

\begin{theorem}[Sample Lower Bound for Gaussians]
\label{thm:intro-lb-gaussian}
For all constant $\theta \in (0,1)$, there is a sufficiently small constant
proximity parameter $\epsilon > 0$ such that every sample-based $\mathfrak
G$-specific halfspace tester requires  $\Omega(n^{2/3})$
samples, where $\mathfrak G$ is the following  class of distributions 
\begin{equation}\label{eq:haha9}\mathfrak G = \Big\{
\cN(0, \Sigma) \hspace{0.06cm}\big|\hspace{0.06cm} 1 \leq \lambda_{min}(\Sigma)
\leq \lambda_{max}(\Sigma) \leq 1+\theta  \Big\}.
\end{equation}
\end{theorem}

\subsection{Techniques}\label{sec:overview}

While our definition is inspired by testable learning, and there are now many
papers on testable learning of halfspaces (see \eg
\cite{RV23,DKKLZ:23testable,GKSV:23universal,DKLZ:24general}), there is no
significant overlap between our techniques and those for testable learning. This
is for two reasons. First, testable learning is intended to address time
complexity of agnostic learning, rather than sample or query complexity (the
sample complexity of agnostic learning of halfspaces is already well-understood,
but processing the samples is expensive). Second, we aim for sublinear  sample
complexity $o(n)$, compared to $\Omega(n)$ for learning.

\subsubsection{Distribution-Picky Testing}\label{sec:picky2}

Our upper and lower bounds both use more fine-grained models of property testing
that lie between distribution-specific and distribution-free testing. First, we
need the concept of \emph{distribution-picky testing}, which modifies the
definition of distribution-robust testing by removing the $\Abstain$ output
while requiring the algorithm to $\Accept$ only when $f$ is a halfspace
\emph{and} the input distribution $\cD$ is the algorithm's preferred
distribution $\cD = \cD^*$; that is, it is ``picky'' about which input
distributions to accept. Distribution-picky testers were defined and studied in
more generality in the companion paper \cite{CFHS26general}. In this paper we
need only a specialization of this definition for halfspace testing:

\begin{definition}[Distribution-Picky Halfspace Testers]
Let $\mathfrak C$ be a class of distributions over $\bR^n$ and let $\cD^* \in \mathfrak C$.  An
algorithm $\cA$ is a \emph{$(\cD^*, \mathfrak C)$-picky halfspace tester} if, given a proximity parameter $\epsilon > 0$,
sample access to a distribution $\cD \in \mathfrak C$, and query access to a  function $f \colon
\bR^n \to \pmset$, its output lies in $\{\Accept,\Reject\}$ and the following hold with probability at least $3/4$:
\begin{description}
\item (Completeness.) If $f$ is a halfspace and $\cD = \cD^*$, then $\cA$ outputs $\Accept$;
\item (Soundness.) If $\dist_\cD(f,h) > \epsilon$ for all halfspaces $h$, then $\cA$ outputs
$\Reject$.
\end{description}
\end{definition}
It is also convenient to amend the definition of distribution-robust testing to allow a 
promise class $\mathfrak A$ on the input distribution $\cD$.

\begin{definition}[Distribution-Robust Testing (Amended)]
Let $\mathfrak C \subseteq \mathfrak A$ be classes of distributions over $\bR^n$. A halfspace tester
is $(\mathfrak C, \mathfrak A)$-robust if it satisfies \cref{def:distribution-robust} for all input
distributions $\cD \in \mathfrak A$. (That is, \cref{def:distribution-robust} is obtained by taking
$\mathfrak A$ to be the set of all distributions over $\bR^n$.)
\end{definition}

It follows almost immediately from these definitions that we can state a hierarchy of
distribution-specific, -picky, and -robust testers; the sample- and query-complexity of these testers
satisfy:
\begin{equation}
\label{eq:testing-hierarchy}
    \text{$\cD^*$-specific} \leq \text{$(\cD^*, \mathfrak A)$-picky}
    \leq \text{$(\cD^*, \mathfrak A)$-robust}
    \leq \text{$\mathfrak A$-specific}
\end{equation}
The middle inequality is seen by just replacing the output $\Abstain$ in a distribution-robust
tester with $\Reject$, to obtain a distribution-picky tester. For testers that use queries, the
following converse (see \cref{prop:picky-strongness} for the formal statement) is shown in the
companion paper \cite{CFHS26general}; we provide a proof in \cref{sec:picky} for the sake of completeness
(and soundness and assertiveness):
\begin{equation}
\label{eq:robust-to-picky}
\text{$(\cD^*, \mathfrak A)$-robust} \lesssim \text{$(\cD^*, \mathfrak A)$-picky}.
\end{equation}
We have found this reduction to be extremely useful (in this paper and its
companion \cite{CFHS26general}) since it allows us to design distribution-robust
algorithms without worrying about whether to $\Reject$ or $\Abstain$ when we see
evidence in the input that it differs from instances we wish to $\Accept$. With
these two inequalities, to obtain
\Cref{thm:intro-ub,thm:intro-lb-robust,thm:intro-lb-gaussian} it now suffices to
prove:
\begin{enumerate}
\item Upper bounds for distribution-picky testing
where $\cD^* = \cN(0, I_n)$ and $\mathfrak A$ is the set of\\ all distributions; and
\item Lower bounds for distribution-picky testing
where $\cD^* = \cN(0, I_n)$ and $\mathfrak A$ is the set of\\ mean-$0$ Gaussians $\cN(0, \Sigma)$ with
all eigenvalues of $\Sigma$ in $[1,1+\theta]$ (i.e. $\mathfrak G$  in \Cref{eq:haha9}).
\end{enumerate}
We now discuss the techniques for these bounds in more detail.

\subsubsection{Upper Bound}

Our upper bound requires introducing several new algorithmic techniques. First,
thanks to the reduction \eqref{eq:robust-to-picky}, it suffices to provide a
distribution-picky halfspace tester.
Then, our picky tester uses a different technique from all prior halfspace testers.
All earlier sublinear halfspace testers use the same strategy: on function $f$ and distribution
$\cD$, they estimate the quantities $\smash{\Exu{\bm x \sim \cD}{f(\bm x)}}$ and
\begin{equation}
\label{eq:intro-center-of-mass}
\Exu{\bm x, \bm y \sim \cD}{ f(\bm x) f(\bm y) \bm x^\top \bm y } .
\end{equation}
When $\cD$ is the standard Gaussian, for any fixed value of $\Ex{f(\bm x)}$, halfspaces are the functions that 
maximize \eqref{eq:intro-center-of-mass}. Furthermore, when $f$ is $\epsilon$-far from being a
halfspace, this quantity will be lower than this maximum value by at least $\Omega(\epsilon^2)$ (see
\cite{MORS10, BBBY12, Eld15, Har19}). The quantity \eqref{eq:intro-center-of-mass} can be estimated
using $\widetilde O(\sqrt n)$ samples \cite{BBBY12,Har19}, or $O(1)$ queries \cite{MORS10}
(exploiting the fact that \eqref{eq:intro-center-of-mass} is the sum of squared degree-$1$ Hermite
coefficients). For $\pmset^n$,  \cite{MORS10} uses a more sophisticated algorithm to
translate this idea from the Gaussian into the discrete domain.

This strategy, however, does not work for distribution-picky testing, because the unknown $\cD$ may not be
rotation-invariant. Rotation invariance guarantees that halfspaces maximize
\eqref{eq:intro-center-of-mass} among all functions with the same mean $\Ex{
f(\bm x) }$ \cite{Har19}. Without it, halfspaces still
maximize \eqref{eq:intro-center-of-mass}, but only among all functions with the
same $\Ex{ f(\bm x) \bm x }$ (rather than $\Ex{ f(\bm x) }$), and unlike $\Ex{
f(\bm x)}$, it is not possible to approximate $\Ex{ f(\bm x) \bm x }$ using
$o(n)$ samples (since this would allow us to learn halfspaces under the
Gaussian).

\begin{figure}
\begin{center}
\includegraphics[scale=0.6]{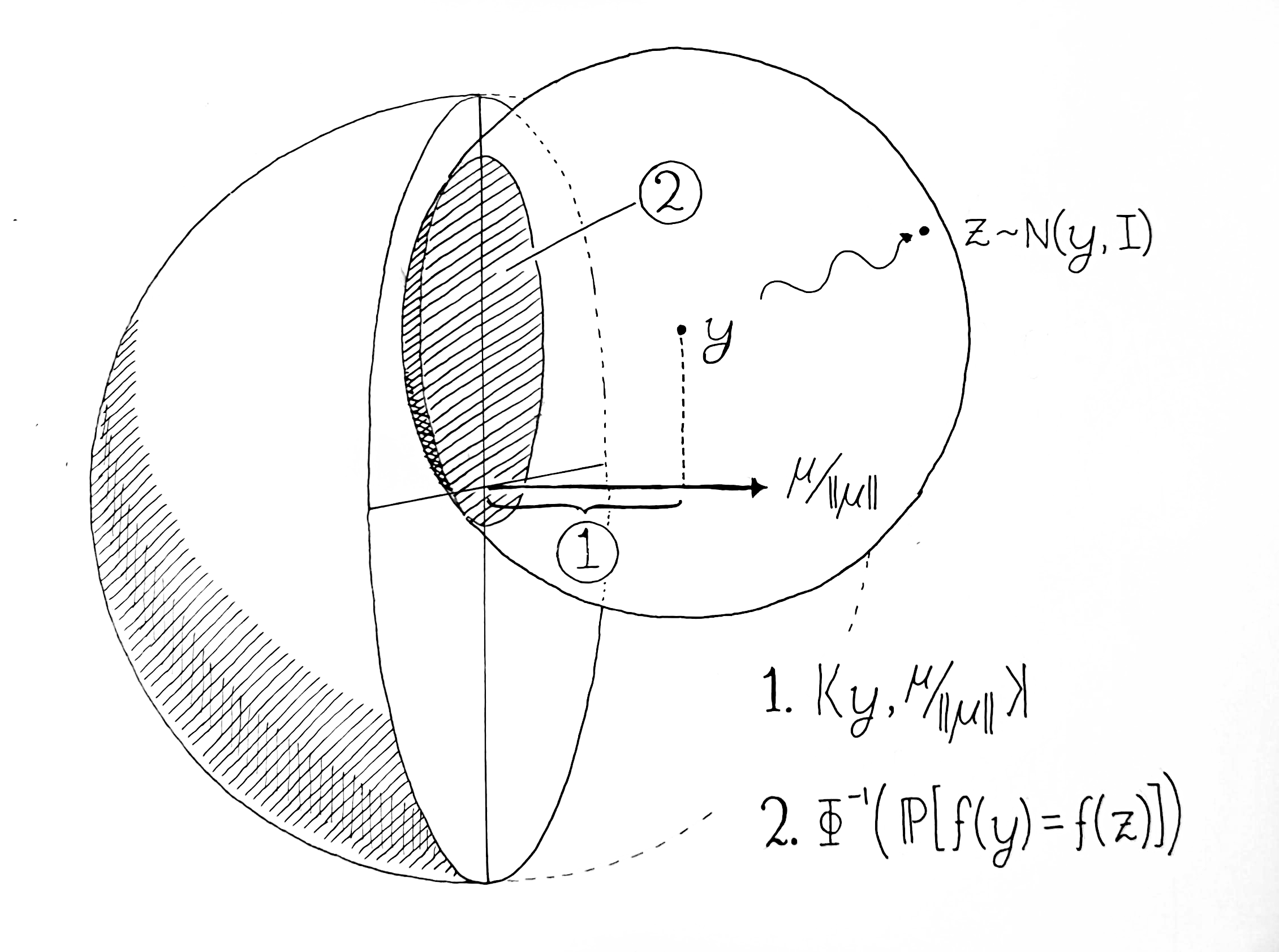}
\end{center}
\caption{Two ways to calculate the distance from $y$ to the hyperplane.}
\label{fig:intro-ub}
\end{figure}

Instead, our distribution-picky tester estimates the ``distance to the ideal hyperplane'' in \emph{two} different ways (see \cref{fig:intro-ub}),
and compares them using a ``distance disagreement'' measure. Let us explain what this means.
 For simplicity in
this discussion, assume that the function $f$ is balanced under the standard Gaussian: 
\[
    \Exu{\bm x\sim \cN(0,I_n)}{ f(\bm x)} = 0 .
\]

\textbf{First distance to hyperplane:} If $f$ is a halfspace (which, being
balanced, has its separating hyperplane through the origin), then
\[
    f(x) = \sign\inn{x, \mu/\|\mu\|},\quad \text{ where } \mu = \Exu{\bm x\sim \cN(0,I_n)}{ f(\bm x) \bm x } .
\]
In this case, the quantity $|\inn{x, \mu/\|\mu\|}|=f(x)\inn{x, \mu/\|\mu\|}$ gives the distance from $x$ to the hyperplane defining
$f$ (item \textcircled{1} in \cref{fig:intro-ub}).

\textbf{Second distance to hyperplane:} If $f$ is a halfspace, we can also
calculate the distance of $x$ to the hyperplane of $f$ using the function
\[
    \alpha(x) \define \Phi^{-1}\left( \Pru{\bm z \sim \cN(0, I_n)}{ f(x + \bm z) = f(x) } \right) ,
\]
where $\Phi^{-1}$ is the inverse cumulative distribution function (CDF) of $\cN(0, 1)$. In other words,
by knowing the mean of $f$ under the ``local'' Gaussian distribution $\cN(x, I_n)$ centered at $x$,
we can determine how far away $x$ is from the hyperplane defining $f$
(item \textcircled{2} in \cref{fig:intro-ub}).

\textbf{Distance disagreement:} If $f$ is indeed a halfspace, 
then these
distance calculations match:
\begin{equation}
\label{eq:intro-tech-ub}
    f(y)\inn{y, \mu/\|\mu\|} = \alpha(y),\quad\text{for all $y\in \mathbb{R}^n$.}
\end{equation}
However, if $f$ is not a halfspace, this equality need not hold, and indeed our
tester will estimate how far these two quantities are from agreeing, on average
over $\bm y\sim \cD$. Miraculously, we can prove the following: for any function
$f$ and distribution $\cD$, if $f$ is $\epsilon$-far from being a halfspace
under $\cD$, and $\alpha(\bm y)$ is not likely to be small under $\bm y \sim
\cD$ (which holds for the case when $f$ is a halfspace and $\cD$ is the standard
Gaussian, and is moreover easy to test), then we have
\[
    \Exu{\bm y \sim \cD}{ \big( f(\bm y) \inn{\bm y, \mu/\|\mu\|} - \alpha(\bm y) \big)^2 }
    = \Omega(\epsilon^3) ;
\]
(see \cref{lemma:distance-disagreement-no}).
We call the quantity on the left-hand side  the \emph{distance disagreement
measure}, written $\distdis(f, \cD$).  Our actual definition (\cref{def:DD}) is
more general: it replaces the true $\alpha(\cdot)$ with an estimator
$\beta(\cdot)$, and also introduces a threshold parameter $\hat{t}$ (as an
estimate for the true threshold $t$ of $f$ when it is a halfspace) to handle
general cases (when $f$ is not balanced).

\textbf{Estimating the distance disagreement:}
When $\cD = \cN(0,I_n)$, standard Gaussian calculations show that $\distdis(f,
\cD)$ is not difficult to estimate. However, when $\cD$ is an arbitrary distribution, the
possibility of inputs such as heavy-tailed distributions makes it challenging to
control the variance of the na\"ive estimator for $\distdis(f,\cD)$. The main idea behind the rest of
our proof is to test ``just enough'' properties of the distribution $\cD$ so
that, even if we cannot certify that it is (close to) Gaussian, we can certify
that $\cD$ is ``sufficiently nice'' to enable the algorithm to compute a good
enough proxy for $\distdis(f, \cD)$ that we can use to accept or reject based on
a threshold.

To this end, we
define a ``whittling down'' procedure (\cref{sec:whittling}) which constrains the distribution
$\cD$ to a distribution $\cD_*$ supported inside the intersection of
\begin{enumerate}[topsep=0pt,itemsep=0pt]
    \item A $\Theta(\sqrt{n})$-radius ball centered at the origin, as is natural for the Gaussian; and
    \item A $\Theta(n^{1/4}/\epsilon)$-width \emph{band} in the direction of an approximation of $\mu/\vabs{\mu}$.
\end{enumerate}
If $\cD$ is far from $\cD_*$, \ie too many points fall outside this intersection, we can
safely reject (since this cannot happen for the standard Gaussian); otherwise, distribution $\cD_*$ is well-behaved enough that we can bound
the spectral quantities controlling the variance of our estimator, such as
the Frobenius norm of its covariance matrix (\cref{lemma:frobenius-tester}).
With these guarantees in hand, we show in \cref{sec:test-statistic} how to control the
variance of a natural estimator of the distance disagreement measure, which suffices to
accomplish our task.

While the full analysis of the variance of our estimator is somewhat involved,
let us give some very informal intuition for how the $n^{2/3}$ dependence in our
query complexity comes to appear. In \cref{alg:tester}, we estimate the distance
disagreement in three parts, essentially corresponding to the square terms
$\abra{\bm{y}, \mu/\vabs{\mu}}^2$, $\alpha(\bm{y})^2$ and the cross product
term. The first term is the most challenging, since the algorithm
can only ascertain a few basic facts about the distribution $\cD$ of $\bm{y}$,
and it does not know (even a good approximation of) $\mu$.
We construct two (very coarse) estimates $\wh{\bm{w}}, \wh{\bm{z}}$ of the vector
$\mu/\vabs{\mu}$ using samples $\bm{x} f(\bm{x})$ for $\bm{x} \sim
\cN(0, I_n)$. Given a budget of $O(m)$ queries, this gives us
vectors of norm roughly $\sqrt{n/m}$. We then use $O(m)$ samples
$\ith{\bm{y}}{i} \sim \cD$ to construct the estimator
\[
    \frac{1}{m} \sum_{i=1}^m
        \abra{\ith{\bm{y}}{i}, \wh{\bm{w}}} \abra{\ith{\bm{y}}{i}, \wh{\bm{z}}}
\]
(where we omit terms related to the estimate $\hat{t}$ of the threshold $t$).
Now, even if $\ith{\bm{y}}{i}$'s are Gaussian (the ideal case), each
inner product above has norm roughly $\sqrt{n/m}$, and hence the estimator
is an average over $m$ samples of quantities of order $n/m$. This can be
expected to have variance roughly $$\frac{1}{m}
\cdot \pbra{\frac{n}{m}}^2 = \frac{n^2}{m^3},$$
which explains why we pick $m = \Theta(n^{2/3})$. (Of course, these $m$ samples
are not independent, so we need to invoke the law of total variance to handle
all components of the variance of our true estimator --- along with the threshold
estimate $\hat{t}$ and the other terms of $\distdis(f, \cD)$, which we do in
\cref{sec:test-statistic}.)

\subsubsection{Lower Bound}

The following theorem suffices to prove both lower bounds
\cref{thm:intro-lb-robust,thm:intro-lb-gaussian}.

\begin{restatable}{theorem}{thmintrolbmain}
\label{thm:intro-lb-main}
For all constant $\theta \in (0,1)$, there exists a constant proximity parameter
$\epsilon > 0$ such that any sample-based $(\cN(0, I_n), \mathfrak A)$-picky
halfspace tester, with the promise $\mathfrak A = \{ \cN(0, \Sigma) \colon 1
\leq \lambda_{min} \leq \lambda_{max} \leq 1+\theta \}$, requires at least
$\Omega(n^{2/3})$ samples.
\end{restatable}

To prove a lower bound on distribution-picky testing, we will need to define the following
``meta-distributions'' over distribution-function pairs $(\cD, f)$:
\begin{enumerate}
    \item A meta-distribution $\DYES$ over pairs $(\cD, f)$ where $\cD = \cN(0, I_n)$
        and $f$ is a halfspace (which should make the picky tester $\Accept$; and
    \item A meta-distribution $\DNO$ over pairs $(\cD, g)$ where
            $\dist_\cD(g,h) > \epsilon$ for all halfspaces $h$,
        (which should make the picky tester $\Reject$),
\end{enumerate}
such that $m = o(n^{2/3})$ samples $(\bm x, f(\bm x))$ from $(\cD, f) \sim \DYES$
are indistinguishable from $m$
samples $(\bm x, g(\bm x))$ from $(\cD, g) \sim \DNO$.

Our $\DYES$ meta-distribution is simple: draw a random normal vector $\bm v \sim
\cN(0, \tfrac{1}{n} I_n)$ and choose the halfspace $f(x) = \sign(\bm v^\top x)$
whose separating hyperplane goes through the origin. Our starting point for the
$\DNO$ distribution is to take $\cD$ to be a ``\emph{spiked} Gaussian''.

\textbf{Spiked Gaussian.} To obtain $\cD$, we first draw $\bm v \sim \cN(0, \tfrac{1}{n} I_n)$
and then take
\[
    \cD = \cN(0, I_n + \theta \bm v \bm v^\top) ,
\]
where $\theta \in (0,1)$ is an arbitrary constant; these distributions are known as spiked Gaussians.
It is known that $\Theta(n)$ samples are necessary to distinguish them from $\cN(0,I_n)$ (see \eg~\cite{CMW15}).

\textbf{Hard functions.} We must now choose the hard functions $g$ that will be
hard to distinguish from halfspaces. We start with a halfspace $g(x) =
\sign(w^\top x)$ and then perturb it to $g'$ to make it far from being a
halfspace. To guide this choice, recall how the $\widetilde O(\sqrt n)$-sample
halfspace tester for $\cN(0, I_n)$ works (\cite{BBBY12}, described above): it
approximates the quantities $\Ex{ f(\bm x)}$ and
\begin{equation}
\label{eq:intro-lb-com}
    \Ex{ f(\bm x) f(\bm y) \bm x^\top \bm y } ,
\end{equation}
which we can think of as the distance between the average 1-valued point and average 0-valued point;
this is maximized by halfspaces for a given value of $\Ex{ \bm x f(\bm x) }$. If we start with a
halfspace $g$ whose normal vector $w$ is orthogonal to the spike $\bm v$, then
adding noise to create $g'$ will change \eqref{eq:intro-lb-com} and be detected.
So we choose $g$ with normal vector $\bm v$ equal to the spike vector, so that:
\begin{enumerate}
    \item The added spike in the $\bm v$ direction will increase \eqref{eq:intro-lb-com}
        above the maximum achieved by any halfspace under $\cN(0, I_n)$; meanwhile
    \item Adding noise to $g$ to get $g'$ will decrease \eqref{eq:intro-lb-com} back down to
        match the value achieved by the halfspaces under $\cN(0, I_n)$.
\end{enumerate}
Essentially, the added (and undetectable) spike $\bm v$ lets us perturb a non-halfspace in the $\bm v$
direction so that it matches the statistics used by the Gaussian halfspace tester.
More specifically, we take the function
\[
    g(x) = \sign\left( \frac{\rho}{\sqrt{1+\theta}} \bm v^\top x + \sqrt{1-\rho^2} \bm z \right) ,
\]
where $\bm z \sim \cN(0, 1)$ is drawn \emph{independently} for each sample point $x$. This has the effect
of creating significant noise near the separating hyperplane, which drops off the farther $x$ gets.

\textbf{Indistinguishability.} To prove that the hard functions under the spiked Gaussian are
indistinguishable from the Gaussian halfspaces using $m = o(n^{2/3})$ samples, we start by defining two matrices $\Xyes, \Xno \in \bR^{m \times n}$, where:
\begin{itemize}
    \item Each row of $\Xyes$ is drawn independently from $\cN(0, I_n)$; and
    \item Each row of $\Xno$ is drawn independently from $\cN(0, I_n + \theta \bm v \bm v^\top )$ .
\end{itemize}
Thus, the $m$ sample points that the algorithm receives are recorded in these
matrices. To determine their $\pmset$ labels, we consider the following three
random variables in $\bR^m$:
\[
    \yyes \define \Xyes \bm v ,\quad \yno \define \Xno \bm v \quad \text{ and } \quad
            \yno^\rho \define\frac{\rho}{\sqrt{1+\theta}} \yno + \sqrt{1-\rho^2} \bm z ,
\]
where $\bm z \sim \cN(0, I_n)$. These vectors determine the labels: in the YES case,
the labels are $\sign(\yyes)$ (so the algorithm observes $\sign(\yyes), \Xyes$), and in the NO case,
the labels are $\sign(\yno^\rho)$.

Since $\Xyes$ and $\bm v$ are independent, the variable $\yyes = \Xyes \bm v$ is distributed as a
Gaussian $\cN(0, \bm A)$ with covariance depending on $\Xyes$. However, since $\Xno$ and $\bm v$ are
\emph{not} independent, the same does not hold for $\yno$: the probability density function (PDF) $\pno$
of $\yno$ is
\[
    \pno(y) \propto \pyes(y) \cdot
        \frac{1}{(1+\theta \|\bm v\|^2)^{m/2}} \exp\left( \frac{\theta}{2(1+\theta\|\bm v\|^2)} \|X\bm v\|^2 \right) ,
\]
where $\pyes$ is the PDF of $\yyes$. However, $\|\bm v\|^2 = 1 \pm o(1)$ with high probability, and
the simpler density
\[
    \widetilde{\pno}(y) \propto \pyes(y) \cdot \exp\left( \frac{\theta}{2(1+\theta)} \|X\bm v\|^2 \right) ,
\]
\emph{does} describe a Gaussian. We define a variable $\widetilde{\yno}$ with this simpler density, along with
\[
    \widetilde{\yno^\rho} = \frac{\rho}{\sqrt{1+\theta}} \widetilde{\yno} + \sqrt{1-\rho^2} \bm z ,
\]
which is also Gaussian because it is the sum of two Gaussians. Plugging in known bounds on the TV
distance between two Gaussians, we obtain
\[
    \TV\big(\yyes, \widetilde{\yno^\rho}\big) = O\left(\frac{m^{3/2}}{n}\right) .
\]
It remains to show that $\widetilde{\yno}^\rho \approx \yno^\rho$, so that the same bound holds for
the true distribution $\yno^\rho$ that determines the labels, and for this it suffices to show
that $\TV(\yno, \widetilde{\yno}) = o(1)$. To accomplish this, we use the following consequence of the
Gaussian Poincar\'e inequality: if $q$ is the PDF of $\cN(0, \Sigma)$ and $p$ is the differentiable
PDF of another distribution, then
\[
    \TV(p,q)^2 \leq \frac{1}{4} \|\Sigma\|_{op}
        \Exu{\bm y \sim p}{ \left\| \nabla \log(p(\bm y)/q(\bm y)) \right\|^2 } .
\]
Applying this lemma with $\pno$ and $\widetilde{\pno}$, we bound the gradient by
\[
    \left\| \nabla \log \frac{\pno(y)}{\widetilde{\pno}(y)} \right\|^2
    = O\left( \frac{\|y\|^2(m + \|y\|^2)^2}{n^2} \right) .
\]
With $\Ex{\|\yno\|^6} = O(m^3)$ (as is the case for $\cN(0, I_m)$), this concludes the proof.

\subsection{Future Work and Open Problems}

Despite recent progress, most problems in theoretical computer science remain
open. Our paper and its companion \cite{CFHS26general} introduce
distribution-robust testing, which open up yet more problems.

Let us take a moment to evangelize about ``testable algorithms'' more generally.
Rubinfeld and Vasilyan \cite{RV23} introduced \emph{testable learning} to deal
with the time complexity of agnostic~learning: time-efficient agnostic~learners
were known under certain distributional assumptions, but these assumptions were
not efficiently testable. Since then, there has been rapid progress on testable
learning, including testing for distributional assumptions \cite{RV23},
distribution shift \cite{KSV24}, and noise assumptions \cite{GKSV25}. We believe
there is immense value in this idea as a more general framework for algorithm
design: a ``testable algorithm'' is one which has some set of assumptions on its
input, and behaves as follows:
\begin{enumerate}
\item On every input, either $\Abstain$ or output a correct answer.
\item If the assumptions are satisfied, do not $\Abstain$.
\end{enumerate}
In some sense, these algorithms are ``self-certifying:'' they  check whether
their own assumptions hold. This is an excellent property for any algorithm to
have. The current paper and its companion \cite{CFHS26general} introduce this
framework to property testing (``testable testing''), and we are aware of only
two other papers applying it to non-learning algorithms: see \cite{FH24} for
distribution testing algorithms that test if their assumptions on label quality
hold, and \cite{ERV26} for testable algorithms estimating the number of
triangles in a graph.

Regarding property testing, our paper shows that distribution-robust testers can
indeed succeed in ``self-certifying,'' while being significantly more
efficient than distribution-free testing, for important and well-studied
properties like halfspaces. These algorithms also require interesting new
algorithmic techniques. We are curious to see how these ideas can be applied to
other important property testing problems.

Turning to concrete bounds, our paper suggests several obvious opportunities for
quantitative improvements, since our upper bound has no matching lower bound,
and our lower bound has no matching upper bound. When given a nearly-final draft
of our manuscript and asked to improve the results, \texttt{GPT 6 Astra} claims
the following, which we have not verified. Thus we state them as conjectures:
\begin{conjecture}\;
\begin{enumerate}
    \item Sample-based distribution-robust halfspace testing requires $\Omega(n)$ samples,
        which would improve our $\Omega(n^{2/3})$ lower bound.
    \item Adaptive distribution-robust halfspace testing with queries requires $\Theta(\log n)$ queries.
    \item Non-adaptive distribution-robust halfspace testing with queries requires between $\Omega((n/\log n)^{1/3})$
        and $O(\sqrt n)$ queries, which would improve our $O(n^{2/3})$ upper bound.
\end{enumerate}
\end{conjecture}

In light of our results and the companion paper \cite{CFHS26general}, we view the
possibility of these conjectured results as further encouragement for the study
of the distribution-robust testing model: it is amenable to strong positive
results, 
and it also enjoys rich separations from other models.
Therefore, we believe there remain many exciting avenues
of exploration for distribution-robust testing algorithms, and testable
algorithms more generally.

\section{Upper Bound: Distance Disagreement and Whittling}
\label{sec:upper-bound}

Our goal in this and the next section is to prove the following theorem:

\begin{theorem}\label{mainthm:upper}
    There is a distribution-picky tester (\cref{alg:tester}) for the class of halfspaces with respect to the
    standard Gaussian distribution over $\bR^n$ with sample and query complexity
    $O(n^{2/3}/\epsilon^{38})$.
\end{theorem}

{Note that we did not try to optimize the polynomial dependence on
$1/\epsilon$, but rather favored simplicity of exposition since the
$n^{2/3}$ dependence is the main message of this result.}

\def\calD{\mathcal{D}}

\Cref{thm:intro-ub} follows from our discussion in \Cref{sec:picky2} about distribution-robust vs distribution-picky testers.
Recall for the latter, the goal is to distinguish the YES case, where $f $ is a halfspace and $\calD$ is the standard Gaussian, from the NO case, where $\dist_\calD(f,h)>\eps$ for all halfspaces $h$.

We present \cref{alg:tester} and prove its correctness in the next section.
In this section, we focus on the following two ingredients that will play important roles in  \cref{alg:tester}:
\begin{flushleft}\begin{enumerate}
    \item The first ingredient is a key quantity $\distdis(f, \cD)$, called the \emph{distance disagreement measure} (see \Cref{def:DD}). This essentially separates the YES and NO cases by comparing the following two different ways of
        measuring the distance from a point $\bm{y}\sim \calD$  to the boundary of a target halfspace: 1)~by estimating  the normal vector of the halfspace and measuring the
        projection of $\bm{y}$ along that direction; 
        and 2)~by estimating the ``bias'' of the input function
        $f$ in the neighborhood of $\bm{y}$ and applying the inverse Gaussian CDF $\Phi^{-1}$.\footnote{Note that the distance disagreement measure defined in \Cref{def:DD}, 
        $\smash{\distdis^{\hat{t}}_\beta(f,\calD)}$, takes two additional objects: a threshold parameter $\hat{t}$ for $f$ and a local distance estimate $\beta$; we delay the precise definition to that point.}

        We introduce the distance disagreement measure in \Cref{sec:DD} and prove two lemmas (\Cref{lemma:distdis-YES} for the YES case and \Cref{lemma:distance-disagreement-no} for the NO case), which show that a good estimation of $\distdis(f,\calD)$ can distinguish the two cases.
        Unfortunately, estimating $\mathsf{DD}(f,\mathcal D)$ directly {on an arbitrary $\calD$ in the NO case} has a large variance that we cannot afford.
    \item To control the variance, our second ingredient is a ``whittled-down'' version $\calD_*$ of the input distribution $\calD$, by eliminating from $\cD$ points that are ``detectably
        non-Gaussian,'' in the sense that we can safely reject if such a point is ever observed. 
    After defining the whittled distribution $\calD_*$ in \Cref{def:whittled-distribution},\footnote{Again, the formal definition of the whittled distribution, $\smash{\calD_*^{\hat{t}}}$, uses a threshold parameter $\hat{t}$ for $f$.}
    we give a tester in \Cref{sec:whittling} to check whether $\calD$ is close to $\calD_*$
     (\Cref{lemma:testing-gaussian-versus-far-from-whittled}).
     
     In the YES case when $\calD$ is the standard Gaussian, the two are very close in variation distance so the tester accepts with high probability; when the two are far in variation distance, the tester rejects with high probability. So running this tester in \cref{alg:tester} ensures that $\calD$ and its whittled distribution $\calD_*$ are very close in variation distance. Indeed, they are so close that the rest of \cref{alg:tester} cannot tell the difference and can be viewed equivalently as running on sample access to $\calD_*$.\footnote{Of course, this is under the caveat that the number of samples \cref{alg:tester} draws in the rest of its execution is small compared to the inverse of variation distance between $\calD$ and $\calD_*$, which is the case.}

    Now even on the nicer whittled distribution $\calD_*$, estimating $\distdis(f,\calD_*)$ may still lead to large variance. To achieve the 
    upper bound in \Cref{mainthm:upper}, we give two additional testers in \Cref{sec:spectral} (\Cref{lemma:frobenius-tester} and \Cref{lemma:2-norm-tester}) to check, informally speaking, whether
    the whittled distribution $\calD_*$ satisfies certain spectral conditions.
Again, the YES case with the standard Gaussian passes both tests with high probability, while passing them ensures that two additional spectral conditions hold for $\calD_*$.
\end{enumerate}\end{flushleft}
Looking ahead,
        \Cref{alg:tester} presented in \Cref{sec:maintester} performs all three tests, rejecting if any of these tests fail. 
Upon passing all tests, we give an estimator for the distance disagreement measure and show that its variance is under control, which can be used to distinguish the YES and NO cases.

\subsection{Preliminaries}
\label{sec:upper-prelim}

\noindent\textbf{Notation.}
Given a real number $T > 0$, we write $\trunc_{\pm T} : \bR \to \bR$ for the truncation function
$$\trunc_{\pm T}(a) = \sign(a) \cdot \min\{T, |a|\}.$$ We write $\log$ for the
natural logarithm, and $\phi$ and $\Phi$ for the probability density function
and cumulative distribution function of the standard Gaussian distribution, respectively.\medskip

\noindent\textbf{Constants and asymptotics.}
Throughout this section,
$C$ denotes a suitable absolute constant
chosen sufficiently large as
required by various steps in the proof, and its value may change from line to line. We assume
throughout that $n \ge n_0$ and $\epsilon \le \epsilon_0$ for suitable absolute constants $n_0,
\epsilon_0$.

We also assume that $\epsilon \ge 1/n$, since otherwise our announced query complexity 
is
$\Theta({{n^{2/3}}/{\epsilon^{38}}}) \gg (n/\epsilon)^{10}$, in which case na\"ive testing by
proper PAC learning \cite{GGR98} suffices to perform distribution-free and hence distribution-robust
testing of the class of halfspaces.

\noindent\textbf{Thresholds.} We fix the following threshold quantities,
which will be useful throughout
the proof. These choices determine the final exponent of our dependence on $\epsilon$
in \cref{mainthm:upper}.
First, let
\begin{align*}
    T_\epsilon &\define 2\sqrt{\log(1/\epsilon)} \quad \text{and} \quad
    R_\epsilon \define 5\sqrt{\log(1/\epsilon)} \,,
\end{align*}
which are defined so that the two intervals $[-T_\epsilon, T_\epsilon]$ and $[-R_\epsilon, R_\epsilon]$ each
contain all but $O(\epsilon)$ mass of the standard Gaussian distribution. Second, the quantity
\[
    B = B_{n,\epsilon} \define \frac{n^{1/4}}{\epsilon} \,,
\]
which is defined so that, roughly speaking, we will be able to certify with high probability that
the input distribution is concentrated in a band of width $B$ in a certain special direction.

\subsection{Distance Disagreement Lemmas}\label{sec:DD}

Suppose $v \in \bR^n$ is a unit vector and $f$ is the halfspace $f(x) = \sign(\abra{x, v} - t)$ for
some threshold $t\in \bR$. Then given a point $y \in \bR^n$, there are at least two ways to think
about the distance from $y$ to the halfspace boundary $\cbra{x \in \bR^n : \abra{x, v} - t = 0}$:
using the inner product directly via $\abs{\abra{y, v} - t} = f(y) (\abra{y, v} - t)$, or via the
``bias'' of a local random walk starting from $y$, as measured by $\Pr{f(y+\bm{g}) = f(y)}$ for
a standard Gaussian $\bm{g}$, by inverting the Gaussian CDF $\Phi$.

The key idea is that not only do these quantities agree exactly when $f$ is a halfspace, but they
also disagree by a detectable margin when $f$ is \emph{far} from a halfspace and the pair $(f, \cD)$
passes a certain test. Next, we implement these ideas concretely.

\subsubsection{Normal Vector Estimator}

In this subsection we introduce a ``normal vector estimator'' and analyze its
behavior when $f$ is a halfspace, as needed for the YES case. The estimator is
defined for arbitrary Boolean functions and will also be used in the NO case. In
more detail, we define, for any given Boolean function $f$ and threshold
parameter $\hat{t}\in \mathbb{R}$,  a vector-valued estimator
$\wt{\bm{v}}(\hat{t})$ such that, when $f$ is a halfspace  $f(x)=\sign(\langle
x,v\rangle-t)$ and $\hat{t}$ is a good approximation of $t$, the expectation of
$\wt{\bm{v}}(\hat{t})$ approximates the unit normal vector $v$ of $f$. We note
that, while learning $v$ is infeasible with a sublinear number of queries, our
goal here is weaker: we only seek an estimator whose expectation is close to $v$
(i.e., which has small bias).

Define the function $\tau : \bR \to (0,1)$ by
\[
    \tau(t) \define \Exu{\bm{g} \sim \cN(0,1)}{\bm{g} \cdot \sign(\bm{g}-t)} \,.
\]
We note that $\tau(t)$ has the closed form $\tau(t) = 2\phi(t) = \sqrt{2/\pi}\;e^{-t^2/2}$.
To see that this is the case,
$$\tau(t) = \Exu{\bm{g}\sim\cN(0,1)}{\bm{g}} - 2\,\Exu{\bm{g}\sim\cN(0,1)}{\bm{g}\cdot \ind{\bm{g}<t}} = -2\int_{-\infty}^{t} x\,\phi(x)\,dx=2\phi(t),$$
using that $\phi'(x)=-x\phi(x)$.

The following estimate, which says that $\tau$ is not too small in the relevant range of $\hat{t}$,
is useful.

\begin{restatable}{lemma}{lemmataulowerbound}
    \label{lemma:tau-lower-bound}
    If $\hat{t} \in \bR$ satisfies $|\hat{t}| \le T_\epsilon$, then $
        \tau(\hat{t}) \ge \sqrt{2/\pi} \cdot \epsilon^2. $
\end{restatable}
\begin{proof}
With the closed form we have 
$\tau(\hat{t})\ge \sqrt{2/\pi}\cdot  e^{-T_\eps^2/2}
=\sqrt{2/\pi}\cdot e^{-2\log(1/\eps)}=\sqrt{2/\pi}\cdot \eps^2.$
\end{proof}

\begin{definition}[Normal vector estimator]
    \label{def:normal-vector-estimator}
    Let $\hat{t} \in \bR$. The \emph{normal vector estimator} for function $f$ with threshold estimate $\hat{t}$ is the vector-valued 
    random variable
    \[
        \wt{\bm{v}}(\hat{t}) \define \frac{\bm{x} f(\bm{x})}{\tau(\hat{t})} \,,
        \quad \text{where} \quad \bm{x} \sim \cN(0, I_n) \,.
    \]
\end{definition}

We show that if $\hat{t}$ is close to the true threshold $t$ of a halfspace $f$, then
$\wt{\bm{v}}(\hat{t})$ has small bias.

\begin{definition}
    Given a halfspace $f(x) = \sign(\abra{x, v} - t)$, where $v$ is a unit vector and $t \in
    \bR$, and parameter $\xi \ge 0$, we say that $\hat{t} \in \bR$ is a \emph{$\xi$-good threshold
    estimate} for $f$ if $|{t - \hat{t}}| \le \xi$.
\end{definition}

The following small technical lemma translates the guarantee of a good threshold estimate into an
approximation guarantee for the function $\tau$.

\begin{restatable}{lemma}{lemmataudiff}
    \label{lemma:tau-diff}
    Suppose that $f$ is a halfspace and $\hat{t}$ is a $\xi$-good threshold estimate for $f$. Then
    \[
        \abs{\tau(t) - \tau(\hat{t})} \le \sqrt{2/\pi}\cdot  \xi \big(|{\hat{t}}| + \xi\big) \,.
    \]
\end{restatable}
\begin{proof}
We have $\tau'(s)=-2s\phi(s)$. By the mean value theorem
there is $s$ between $t$ and $\hat t$ such that
\[
  \abs{\tau(t)-\tau(\hat t)} = |\tau'(s)|\,|t-\hat t| = 2|s|\,\phi(s)\,|t-\hat t|
  \le 2\big(|\hat t|+\xi\big)\cdot\frac{1}{\sqrt{2\pi}}\cdot \xi,
\]
using $|s|\le|\hat t|+|t-\hat t|\le|\hat t|+\xi$, $\phi\le 1/\sqrt{2\pi}$ and $|t-\hat t|\le\xi$.
\end{proof}

The following lemma quantifies the quality of $\Ex{\wt{\bm{v}}(\hat{t})}$ when $\hat{t}$ is
a good threshold estimate.

\begin{lemma}[Normal vector estimator in the YES case]
    \label{lemma:normal-vector-estimator-yes}
    Suppose that $f(x) = \sign(\abra{x, v} - t)$ where $v$ is a unit vector and $t
    \in \bR$. Let  $\hat{t} \in \bR$ be a $\xi$-good threshold estimate for
    $f$. Then
    \[
        \Ex{\wt{\bm{v}}(\hat{t})} = (1+a) v, \quad \text{for some }
        a \in \bR \text{ satisfying } \abs{a} \le \frac{\sqrt{2/\pi}\cdot \xi (|{\hat{t}}|+\xi)}{\tau(\hat{t})}
        \,.
    \]
    (So $\Ex{\wt{\bm{v}}(\hat{t})}$ always points in the same direction as $v$, and in particular, $\wt{\bm{v}}(\hat{t})$ is unbiased if $\hat{t}=t$.)
\end{lemma}
\begin{proof}
    Suppose without loss of generality that $v = e_1$, i.e. $f(x) = \sign(x_1 - t)$. Then, by
    symmetry,
    \[
        \Exu{\bm{x} \sim \cN(0, I_n)}{\bm{x} f(\bm{x})}
        = \Exu{\bm{x} \sim \cN(0, I_n)}{\bm{x}_1 e_1 \sign(\bm{x}_1 - t)}
        = \Exu{\bm{g} \sim \cN(0,1)}{\bm{g} \sign(\bm{g}-t)} e_1
        = \tau(t) e_1 \,.
    \]
    It follows that
    \[
        \Ex{\wt{\bm{v}}(\hat{t})}
        = \frac{\Ex{\bm{x} f(\bm{x})}}{\tau(\hat{t})}
        = \frac{\tau(t)}{\tau(\hat{t})} e_1
        = (1+a) e_1 \,,
    \]
    where 
    \[
        a \define \frac{\tau(t) - \tau(\hat{t})}{\tau(\hat{t})}
        \quad\text{satisfies}\quad
        \abs{a} \le \frac{\sqrt{2/\pi}\cdot \xi (|{\hat{t}}|+\xi)}{\tau(\hat{t})}
    \]
    by \cref{lemma:tau-diff}.
\end{proof}

Therefore, when $f$ is a halfspace, the quantity $$\big| \abra{y, \Ex{\wt{\bm{v}}(t)}} - t\big| = f(y)
\big({\abra{y, \Ex{\wt{\bm{v}}(t)}} - t}\big)$$ is the distance from $y$ to the boundary of $f$,
and by \Cref{lemma:normal-vector-estimator-yes}, the error from using an estimate $\hat{t}$ instead of $t$ can be controlled.
This is our first perspective on the distance to the boundary.

Before giving the second perspective, we record a corollary which shows that, for a sufficiently good threshold estimate $\hat{t}$ in the relevant range, the mean of our estimator has norm bounded by an absolute constant (which is a desirable property for an estimator of a unit vector).

\begin{corollary}
    \label{cor:yes-ex-v-small}
    Let $f$ be a halfspace and $\hat{t}$ be an $\epsilon^4$-good threshold estimate for $f$
  with $|{\hat{t}}| \le T_\epsilon$.~Then $$\vabs{\Ex{\wt{\bm{v}}(\hat{t})}}_2^2 \le 3.$$
\end{corollary}
\begin{proof}
By \Cref{lemma:normal-vector-estimator-yes}, $\Ex{\tilde {\bm{v}}(\hat
t)}=(1+a)v$ with $\|v\|_2=1$ and 
\[
|a|\le\frac{\sqrt{2/\pi}\cdot \varepsilon^4(|\hat t|+\varepsilon^4)}{\tau(\hat t)}
      \le\frac{\sqrt{2/\pi}\cdot \varepsilon^4\cdot 2T_\varepsilon}{\sqrt{2/\pi}\,\varepsilon^2}
      \le 2\varepsilon^2 T_\varepsilon,
\]
where we used $\eps^4\le T_\eps$ for $\eps\le \eps_0$ when $\eps_0$ is a
sufficiently small constant. As a result,
$$\vabs{\Ex{\wt{\bm{v}}(\hat{t})}}_2^2  =(1+a)^2\le 2+2a^2\le 2+8\eps^4 T_\eps^2
\le 3,$$ for $\eps\le\eps_0$ when $\eps_0$ is sufficiently small.
\end{proof}

\subsubsection{Local Uniformity and Local Distance Measure}

We now formalize the second perspective, as follows.

\begin{definition}
    For each point $y\in \mathbb{R}^n$, define the \emph{local uniformity} of $y$ with respect to $f$ as
    \[
        \ell_f(y) \define \Pru{\bm{g} \sim \cN(0, I_n)}{f(y+\bm{g}) = f(y)} \,.
    \]
    Then, define the \emph{local distance measure} of $y$ (with respect to $f$) as
    \[
        \alpha_f(y) \define \Phi^{-1}(\ell_f(y)) \,.
    \]
\end{definition}

For halfspaces, this quantity is nonnegative and equals the distance to the
boundary, as shown below; for general Boolean functions, it may be negative.
Intuitively, notice that a point $y$ at the boundary of a halfspace $f$
satisfies $\ell_f(y) = 1/2$ and hence $\alpha_f(y) = 0$, while taking $y$
arbitrarily far from the boundary gives $\ell_f(y) \to 1$ and so $\alpha_f(y)
\to +\infty$. More concretely, for a halfspace $f(x) = \sign(\abra{x, v} - t)$
with a unit vector $v$, we have $$f(y) \big(\abra{y, v} - t\big) =
\big|{\abra{y, v} - t} \big|= \alpha_f(y),$$ \ie the two perspectives on
distance to the boundary agree in the YES case.

\subsubsection{Distance Disagreement Measure}

We are ready to define the \emph{distance disagreement measure}. 
Ideally one may define it as
\begin{equation}\label{eq:naive}
        \Exu{\bm{y} \sim \cD}{\big({f(\bm{y}) (\abra{\bm{y}, \Ex{\wt{\bm{v}}(t)}} - t)
            - \alpha_f(\bm{y})}\big)^2}.
\end{equation}
A challenge in \Cref{eq:naive} is that $\alpha_f$ is na\"ively unbounded and hence hard to estimate at arbitrary points. 
To mitigate this issue, we will replace $\alpha_f$ with an estimator  ($\wh{\alpha}_{f,N}$, which we will discuss shortly in \Cref{sec:localestimate}). 
To this end, we parameterize the definition of the distance disagreement measure using a generic bounded measurable function $\beta$ in the place of $\alpha_f$:

\begin{definition}[Distance disagreement measure $\distdis^{\hat{t}}_{\beta}$]\label{def:DD}
    Given a bounded measurable function $\beta:\bR^n$  $\to \bR$, a threshold parameter $\smash{\hat{t} \in \bR}$,
    a Boolean function $f : \bR^n \to \pmset$ and a distribution $\cD$~over $\bR^n$, the \emph{distance
    disagreement measure} of  $(f, \cD)$ with threshold $\smash{\hat{t}}$ under function
    $\beta$ is
    \[
        \distdis^{\hat{t}}_\beta(f, \cD) \define
        \Exu{\bm{y} \sim \cD}{\big({f(\bm{y}) (\abra{\bm{y}, \Ex{\wt{\bm{v}}(\hat{t})}} - \hat{t})
            - \beta(\bm{y})\big)}^2}
        \,,
    \]
    where the random variable $\wt{\bm{v}}(\hat{t})$ is the normal vector estimator from
    \cref{def:normal-vector-estimator}.
\end{definition}

Our first distance disagreement lemma formalizes the guarantees 
in the YES case:
as long as $\hat{t}$~is~a good threshold estimate of $f$ and $\beta$ is a good approximation of $\alpha_f$ in terms of \Cref{eq:haha1} below, the distance disagreement measure
$\smash{\distdis^{\hat{t}}_\beta(f,\calD)}$ is small. {Then, our second distance disagreement lemma (\cref{lemma:distance-disagreement-no}) will give the corresponding result for the NO case.}

\begin{lemma}[Distance disagreement in the YES case]
    \label{lemma:distdis-YES}
    Suppose that $f$ is a halfspace and $\cD$ is~the standard Gaussian. Let $\xi \ge 0$ and suppose that $\smash{\hat{t}}$ is a
    $\xi$-good threshold estimate for $f$. Let $\Delta \ge 0$ and suppose that  $\beta
    : \bR^n \to \bR$ is a bounded measurable function that satisfies
\begin{equation}
\label{eq:haha1}
\Exu{\bm{x} \sim \cN(0, I_n)}{\pbra{\alpha_f(\bm{x}) -
    \beta(\bm{x})}^2} \le \Delta.
\end{equation}
Then we have
    \[
        \distdis^{\hat{t}}_\beta(f, \cD)
        \le 3\Delta + 3 \xi^2 \pbra{1 + \frac{|{\hat{t}}|+\xi}{\tau(\hat{t})}}^2 .
    \]
\end{lemma}
\begin{proof}
    Let $f(x) = \sign(\abra{x,v} - t)$ with unit  normal vector $v$. By
    \cref{lemma:normal-vector-estimator-yes}, we have
    \[
        \Ex{\wt{\bm{v}}(\hat{t})} = (1+a) v,
        \quad \text{where}\ 
        \abs{a} \le \frac{\sqrt{2/\pi}\cdot  \xi (|{\hat{t}}|+\xi)}{\tau(\hat{t})} \,.
    \]
    Also, as discussed above, in the YES case we have $\alpha_f(y) = f(y) (\abra{y,v} - t)$ for each
    point $y$. Then
    \begin{align*}
        \distdis^{\hat{t}}_\beta(f, \cD)
        &= \Exu{\bm{y} \sim \cD}{
            \pbra{f(\bm{y}) (\abra{y,\Ex{\wt{\bm{v}}(\hat{t})}} - \hat{t}) - \beta(\bm{y})}^2
        } \\
        &= \Exu{\bm{y} \sim \cD}{\pbra{
            \alpha_f(\bm{y}) - \beta(\bm{y}) + f(\bm{y}) (a\abra{\bm{y},v} + t - \hat{t})
        }^2} \\
        &\le 3 \Exu{\bm{y} \sim \cD}{
            \pbra{\alpha_f(\bm{y}) - \beta(\bm{y})}^2 + a^2 \abra{\bm{y},v}^2 + (t-\hat{t})^2
        } \\
        &\le 3\Delta + 3a^2 \Exu{\bm{g} \sim \cN(0,1)}{\bm{g}^2} + 3(t-\hat{t})^2 \\
        &\le 3\Delta + 3a^2 + 3\xi^2 \,.
    \end{align*}
    Substituting the bound on $a$ completes the proof.
\end{proof}

\begin{lemma}[Distance disagreement in the NO case]
    \label{lemma:distance-disagreement-no}
    Suppose $\dist_\cD(f, h) > \epsilon$ for all halfspaces $h$. Let $\beta:\mathbb{R}^n\rightarrow \mathbb{R}$ be any bounded measurable function that satisfies 
\begin{equation}
        \label{eq:alpha-not-too-small}
        \Pru{\bm{y} \sim \cD}{\beta(\bm{y}) < \frac{\epsilon}{4}}
        \le \frac{\epsilon}{2} \,.
    \end{equation}
    Then for any $\hat{t} \in \bR$, we have
    \[
        \distdis^{\hat{t}}_{\beta}(f, \cD) > \frac{\epsilon^3}{32} \,.
    \]
\end{lemma}
\begin{proof}
Define a halfspace $h$ by $h(x) \define \sign(\abra{x, \Ex{\wt{\bm{v}}(\smash{\hat{t}})}} -\hat{t}).$
Using $\dist_\calD(f,h)>\eps$, we have \begin{align*}
        \distdis^{\hat{t}}_{\beta}(f, \cD)
        &= \Exu{\bm{y} \sim \cD}{
            \big({f(\bm{y}) (\abra{\bm{y}, \Ex{\wt{\bm{v}}(\smash{\hat{t}})}} - \hat{t}) - \beta(\bm{y})\big)}^2
        } \\
        &\ge \Pru{\bm{y} \sim \cD}{
            f(\bm{y}) \ne h(\bm{y}) \text{ and } \beta(\bm{y}) \ge \frac{\epsilon}{4}
        }\cdot \left(\frac{\eps}{4}\right)^2 
        > \frac{\epsilon}{2} \cdot \pbra{\frac{\epsilon}{4}}^2
        = \frac{\epsilon^3}{32} \,. \qedhere
    \end{align*}
\end{proof}

Note that \cref{lemma:distance-disagreement-no} only requires $\beta$ to satisfy
\Cref{eq:alpha-not-too-small}. This turns out to be a fairly mild assumption to
make on $\beta$. (Looking ahead, we will instantiate $\beta$ with the local
distance estimator $\widehat{\alpha}_{f,N}$ introduced in
\Cref{sec:localestimate}. For this choice, we will give a simple test that
accepts with high probability in the YES case and rejects with high probability
when \Cref{eq:alpha-not-too-small} fails. Our algorithm can therefore
confidently reject whenever this test fails.)

\subsubsection{Estimating the Local Distance Measure}\label{sec:localestimate}

Before moving on, we describe how $\beta$ is actually implemented later in the algorithm. 

Under an arbitrary distribution $\cD$, in principle we have no control over which points $y$ may
require an estimate of $\alpha_f$, which is an unbounded function. If we estimate $\alpha_f$ by
sampling points $\bm{g}$ to approximate $\Pr{f(y+\bm{g}) = f(y)}$ and then evaluating $\Phi^{-1}$
(which has unbounded sensitivity near $0$ and $1$), then na\"ively we may incur arbitrarily large
additive error. For our purposes, it suffices to estimate $\alpha_f$ in the ``relevant region'' to
$\poly(\epsilon)$ accuracy, which motivates the following definition.

\begin{definition}
    \label{def:alpha-bar}
    Define the \emph{truncated distance measure} of point $y$ with respect to $f$ as
    \[
        \overline{\alpha}_f(y)
        \define \trunc_{\pm R_\epsilon}(\alpha_f(y))
        = \trunc_{\pm R_\epsilon}\pbra{\Phi^{-1}(\ell_f(y))}
        \,.
    \]
    Here and going forward, in the composition $\trunc_{\pm R_\epsilon} \circ \Phi^{-1}$ we extend
    $\Phi^{-1}$ to take values $-\infty$ and $+\infty$ at $0$ and $1$, respectively. Thus the
    composition takes values $-R_\epsilon$ and $R_\epsilon$ at $0$ and $1$, respectively.
\end{definition}

As will be shown in \cref{claim:YES-alpha-bar}, $\overline{\alpha}_f$ can be estimated at any point $y$ to $\poly(\epsilon)$
accuracy using $\poly(1/\epsilon)$ queries. To simplify the analysis, it will be helpful to consider
a single pointwise estimate of $\overline{\alpha}_f$, which is fixed at the beginning of the 
algorithm and then treated as a deterministic quantity. A simple but sufficient
approach\footnote{The proposed strategy also avoids inconvenient measurability concerns.} is to
sample once and for all a ``neighborhood tuple'' $\bm{N}$ of \iid Gaussian points playing the role
of $\bm{g}$ above, and then use the same $\bm{N}$ to estimate $\overline{\alpha}_f$ at every $y$.
(Looking ahead, we will ultimately take the number of points $p$ comprising $\bm{N}$ to be a large $\poly(1/\eps)$ value, independent of $n$.)

\begin{definition}
    \label{def:local-distance-estimate}
    \label{def:alpha-hat}
    Given a \emph{neighborhood tuple} $N = ({\ith{g}{1}, \dotsc, \ith{g}{p}})$ of points in
    $\bR^n$, we define for each point $y$ the \emph{local uniformity estimate} of $y$ with respect to
    $f$ induced by $N$ as
    \[
        \ell_{f,N}(y) \define \frac{1}{p} \sum_{i=1}^p \ind{f(y+\ith{g}{i}) = f(y)} \,,
    \]
    and the \emph{local distance estimate} of $y$ with respect to $f$ induced by $N$ as
    \[
        \wh{\alpha}_{f,N}(y) \define \trunc_{\pm R_\epsilon}\pbra{\Phi^{-1}(\ell_{f,N}(y))} \,.
    \]
\end{definition}

\subsection{Whittling Down the Distribution}
\label{sec:whittling}

To control the variance of our estimator for the distance disagreement measure,
and ultimately to achieve sublinear query complexity, we need the input
distribution $\mathcal D$ in the NO case to behave nicely, in the sense that it
places little (or even no) mass on regions where, in the YES case, the standard
Gaussian places negligible mass. Intuitively this is a mild requirement: if the
algorithm receives a sample from $\mathcal D$ that would be very unlikely under
the Gaussian, it can safely reject.

We formalize this by defining a “whittled distribution” $\mathcal D_*^{\hat{t}}$
from $\calD$ in \Cref{def:whittled-distribution}, and we give a test that, with
high probability, (1) accepts in the YES case and (2) rejects whenever
$\TV(\mathcal D,\mathcal D_*)$ is large. When the test passes we may therefore
analyze the rest of the algorithm in the NO case\footnote{For the YES case, we
always analyze the performance of the algorithm using the original standard
Gaussian.} as if it draws samples directly from $\mathcal D_*^{\hat{t}}$. We
continue the discussion on the right view one should take for the rest of the
algorithm after giving the test in
\Cref{lemma:testing-gaussian-versus-far-from-whittled}.

\begin{definition}[Whittled distribution]
    \label{def:whittled-distribution}
    Given a function $f$ and  a threshold parameter $\hat{t} \in \bR$, we define the \emph{band-distance} function $\smash{G_f^{\hat{t}}} :
    \bR^n \to \bR$ by\footnote{Note that the algorithm does not have explicit access to $G_f^t$, but
    rather will compute an estimate using samples.}
    \[
        G_f^{\hat{t}}(y) \define \abra{y, \Ex{\wt{\bm{v}}(\hat{t})}} \,,
    \]
    where $\wt{\bm{v}}(\hat{t})$ is the normal vector estimator\footnote{Recall that the random variable $\wt{\bm{v}}(\hat{t})$ uses $f$ and $\hat{t}$ but only draws from the standard Gaussian instead of $\calD$.} from \cref{def:normal-vector-estimator}.
We say a point $y \in \bR^n$ is \emph{$B$-typical} with respect to $(f,\hat{t})$ if it satisfies the following two conditions:
    \begin{enumerate}
        \item Ball condition: $\vabs{y}_2 \le 2\sqrt{n}$; and 
        \item Band condition: $\big|{G_f^{\hat{t}}(y)}\big| \le B$.
    \end{enumerate}
    Define the \emph{whittled distribution} $\cD_*^{\hat{t}}$ from $\calD$ with respect to $(f,\hat{t})$  as follows. To draw a
    sample from $\smash{\cD_*^{\hat{t}}}$, first sample $\bm{y} \sim \cD$ and then output $\bm{y} \cdot \ind{\bm{y}
    \text{ is $B$-typical}}$, that is, output $\bm{y}$ if it is $B$-typical and the origin
    otherwise.
    Therefore, $\smash{\calD_*^{\hat{t}}}$ is supported only on $B$-typical points with respect to $(f,\hat{t})$.
\end{definition}

Next we give the test:

\begin{lemma}[Testing YES case versus far-from-whittled]
    \label{lemma:testing-gaussian-versus-far-from-whittled}
    Let $K > 0$ be any constant. There exists an algorithm which, given query access to function
    $f$, sample access to distribution $\cD$, and parameters $\smash{\hat{t}} \in \bR$ and $m \in \bN$
    satisfying $|{\hat{t}}| \le T_\epsilon$ and $$\frac{\sqrt{n} \log n}{\epsilon^2} \le m
    \le \left(\frac{n}{\epsilon}\right)^{10},$$ uses $O(m)$ queries and samples and either \emph{accepts} or
    \emph{rejects} such that, with probability at least $99/100$, the following conditions hold:
\begin{flushleft}\begin{enumerate}
        \item If $f$ is a halfspace, $\cD$ is the standard Gaussian, and $\hat{t}$ is an $\epsilon^4$-good
            threshold estimate for $f$, the algorithm accepts.
        \item If $\TV(\cD, \cD_*^{\hat{t}}) > 1/{(Km)}$, the algorithm rejects.
    \end{enumerate}\end{flushleft}
\end{lemma}

We first record a definition motivated by 
\Cref{lemma:testing-gaussian-versus-far-from-whittled}:
\begin{definition}
    We say a distribution $\calD$ is $B$-\emph{typical} with respect to
    $(f,\hat{t})$ if it is supported only on $B$-typical points with respect to
    $(f,\hat{t})$. 
\end{definition}

Our main algorithm will start by computing a threshold estimate $\hat{t}\in \mathbb{R}$ (see \Cref{sec:threshold}) that is, most likely, an $\eps^4$-good threshold estimate for $f$ and satisfies $|\hat{t}|\le T_\eps$ in the YES case. It continues to run the test given in \Cref{lemma:testing-gaussian-versus-far-from-whittled} with parameters such that $Km$ is larger than the total number of samples drawn from $\calD$ in the rest of the algorithm.
With this in mind, one can view the goal of our algorithm as   distinguishing the following cases:
\begin{flushleft}
\begin{enumerate}
\item Accept on $(f,\calD,\hat{t})$, where $f$ is a halfspace, $\calD$ is the standard Gaussian and $\hat{t}$ is a good threshold estimate for $f$ with $|\hat{t}|\le T_\eps$; and 
\item Reject on $(f,\calD,\hat{t})$, where $|\hat{t}|\le T_\eps$, \begin{equation}\label{eq:haha3}\dist_\cD(f,h)\ge \eps-\frac{1}{Km}\end{equation} for every halfspace $h$ and $\calD$ is $B$-typical with respect to $(f,\hat{t})$.
\end{enumerate}
\end{flushleft}
We may assume the latter in the NO case, with $\calD_*^{\hat{t}}$ in  place of $\calD$, because either \begin{flushleft}\begin{enumerate}
    \item $\TV(\calD,\calD_*^{\hat{t}})> 1/(Km)$: in which case  the algorithm already rejected with high probability due to the test in \Cref{lemma:testing-gaussian-versus-far-from-whittled}; or
    \item $\TV(\calD,\calD_*^{\hat{t}})\le 1/(Km)$: in which case the algorithm can pretend to have  access to $\smash{\calD_*^{\hat{t}}}$ as $\calD$, given that it uses less than $Km$ samples in the rest of its execution,    
    and \Cref{eq:haha3} follows by the triangle inequality. 
\end{enumerate}\end{flushleft}

\begin{proof}[Proof of \Cref{lemma:testing-gaussian-versus-far-from-whittled}]
    Let $m' \define L m$, where $L > 0$ is a sufficiently large absolute constant. We use the
    shorthand $\tau \define \tau(\hat{t})$ for convenience. The algorithm works as follows:
    \begin{enumerate}
        \item Draw a tuple of \iid Gaussian points $\bm{X} \sim \cN(0, I_n)^{m'}$, and query
            $f(\bm{x})$ at each $\bm{x} \in \bm{X}$.
        \item Repeat for each $i \in [Km']$:
            \begin{enumerate}
                \item Draw a sample $\ith{\bm{y}}{i} \sim \cD$, and \textbf{reject} if
                    $\vabs{\ith{\bm{y}}{i}}_2 > 2\sqrt{n}$.
                    \label{item:test-norms}
                \item Compute
                    \[
                        \wh{\bm{G}}_{\ith{\bm{y}}{i}} \define \frac{1}{m'} \sum_{\bm{x} \in \bm{X}}
                            \abra{\ith{\bm{y}}{i}, \frac{\bm{x} f(\bm{x})}{\tau}} \,.
                    \]
                \item \textbf{Reject} if $\abs{\wh{\bm{G}}_{\ith{\bm{y}}{i}}} > B/2$.
                    \label{item:test-g}
            \end{enumerate}
        \item \textbf{Accept}.
    \end{enumerate}

    The algorithm above has the announced query and sample complexity. Before establishing its
    correctness, we make the main observation about its probabilistic structure. For each fixed $y
    \in \bR^n$, define random variables $\wh{\bm{G}}_y \define \wh{\bm{G}}_{\ith{\bm{y}}{i}} \mid
    \ith{\bm{y}}{i}=y$ (whose randomness comes from $\bm{X}$) and
    \[
        \bm{G}_y^\err \define \wh{\bm{G}}_y - \Ex{\wh{\bm{G}}_y}
        = \wh{\bm{G}}_y - G_f^{\hat{t}}(y) \,,
    \]
    the last equality by \cref{def:whittled-distribution}.

    \begin{claim}
        \label{claim:subgaussian}
        Let $y \in \bR^n$. Then $\bm{G}_y^\err$ is subgaussian with subgaussian norm at most
        $\frac{C \vabs{y}_2}{\tau \sqrt{m'}}$.
    \end{claim}
    \begin{proof}
        Define random variable $\wt{\bm{G}}_y \define \abra{y, \frac{\bm{z} f(\bm{z})}{\tau}}$,
        where $\bm{z} \sim \cN(0, I_n)$. Then $\wt{\bm{G}}_y$ is subgaussian with subgaussian norm
        $O(\vabs{y}_2/\tau)$, since for all $q \ge 1$,
        \[
            \Ex{\abs{\wt{\bm{G}}_y}^q}^{1/q}
            = \frac{\vabs{y}_2}{\tau} \cdot \Exu{\bm{g} \sim \cN(0,1)}{\abs{\bm{g}}^q}^{1/q}
            = O\pbra{\frac{\vabs{y}_2}{\tau} \cdot \sqrt{q}} \,,
        \]
        see \eg \cite[Proposition~2.6.1]{Ver26}. Hence the centered $\wt{\bm{G}}_y -
        \Ex{\wt{\bm{G}}_y}$ is subgaussian with the same asymptotic upper bound on its subgaussian
        norm \cite[Lemma~2.7.8]{Ver26}, and since $\bm{G}_y^\err$ is the empirical average over
        $m'$ \iid copies of that centered random variable, the conclusion follows by
        \cite[Proposition~2.7.1]{Ver26}.
    \end{proof}

    Let $\tau_{\min} \define \frac{1}{\sqrt{2\pi}} \cdot \epsilon^2$, so that $\tau \ge
    \tau_{\min}$ by \cref{lemma:tau-lower-bound}. Define the following event:
    \begin{quote}
        Event $E$: For all $i \in [Km']$, if $\vabs{\ith{\bm{y}}{i}}_2 \le 2\sqrt{n}$
        then $\abs{\bm{G}_{\ith{\bm{y}}{i}}^\err} \le B/4$.
    \end{quote}

    \begin{claim}
        \label{claim:subgaussian-event}
        Event $E$ occurs with probability $1 - o_n(1)$.
    \end{claim}
    \begin{proof}
        For any $y \in \bR^n$ satisfying $\vabs{y}_2 \le 2\sqrt{n}$, \cref{claim:subgaussian} and a
        subgaussian tail bound \cite[Proposition~2.6.1]{Ver26} yield
        \begin{align*}
            \Pr{\abs{\bm{G}_y^\err} \ge \frac{B}{4}}
            &\le 2\exp\pbra{-\frac{B^2 \tau_{\min}^2 m'}{C n}}
            \le 2\exp\pbra{
                -\frac{\sqrt{n}}{\epsilon^2} \cdot \epsilon^4
                \cdot \frac{L \sqrt{n} \log n}{\epsilon^2} \cdot \frac{1}{C n}
            } \\[0.8ex]
            &\le 2\exp\pbra{-\frac{L \log n}{C}}
            \le (1/n)^{35}
            \le \frac{(\epsilon/n)^{15}}{L K}
        \end{align*}
        as long as $L$ is chosen sufficiently large compared to $C$, where the last inequality
        holds for sufficiently large $n$ using the bound $\epsilon \ge 1/n$. By a union bound over
        all $i \in [Km']$, where $m' \le L (n/\epsilon)^{10}$, we conclude that event $E$ occurs
        with probability $1 - o_n(1)$.
    \end{proof}

    We now finish the proof of \Cref{lemma:testing-gaussian-versus-far-from-whittled}. We separately consider completeness and soundness, and
    then the lemma will follow by a union bound.

    \noindent\textbf{Completeness.} Suppose $f$ is a halfspace, $\cD = \cN(0, I_n)$, and $\hat{t}$ is an
    $\epsilon^4$-good threshold estimate for $f$. We first show that Step~\ref{item:test-norms}
    rejects with small probability. Indeed, for a single iteration with $\ith{\bm{y}}{i} \sim \cN(0,
    I_n)$, a standard concentration inequality gives $\Pr{\vabs{\ith{\bm{y}}{i}}_2^2 > 4n} \le
    e^{-n/2}$ and hence, recalling that $\epsilon \ge 1/n$ and by a union bound, with probability $1
    - o_n(1)$ none of the $Km' = O(n^{20})$ iterations of Step~\ref{item:test-norms} rejects.

    We now analyze a single iteration of Step~\ref{item:test-g}. Recall that
    $\vabs{\Ex{\wt{\bm{v}}(\hat{t})}}_2^2 \le 3$ by \cref{cor:yes-ex-v-small}. Hence, by a Gaussian
    tail bound and union bound, with probability $1 - o_n(1)$ it holds for every $i$ that
    \[
        \abs{G_f^{\hat{t}}(\ith{\bm{y}}{i})}
        = \abs{\abra{\ith{\bm{y}}{i}, \Ex{\wt{\bm{v}}(\hat{t})}}}
        \le C \sqrt{\log n} \ll B/4 \,.
    \]
    Suppose that this event occurs, as well as event $E$, and that for every $i$ we have
    $\vabs{\ith{\bm{y}}{i}}_2 \le 2\sqrt{n}$. This occurs with probability $1 - o_n(1)$ as shown
    above. In this case, for every $i$ we have that $\abs{\wh{\bm{G}}_{\ith{\bm{y}}{i}}} \le
    \abs{G_f^{\hat{t}}(\ith{\bm{y}}{i})} + \abs{\bm{G}_{\ith{\bm{y}}{i}}^\err} \le B/2$, and hence
    no iteration of Step~\ref{item:test-g} rejects. By a union bound, the algorithm accepts with
    probability $1 - o_n(1)$.

    \noindent\textbf{Soundness.} Suppose $\TV(\cD, \cD_*^{\hat{t}}) > \frac{1}{Km}$.
    Then
    \[
        \Pr{\forall i \in [Km'] \,. \,\, \ith{\bm{y}}{i} \text{ is $B$-typical}}
        \le \pbra{1 - \frac{1}{Km}}^{Km'}
        \le e^{-L}
        \le 1/200 \,.
    \]
    Suppose there exists index $i$ such that $\ith{\bm{y}}{i}$ is not $B$-typical. If
    $\ith{\bm{y}}{i}$ fails the ball condition, then Step~\ref{item:test-norms} rejects and we are
    done. Otherwise, $\ith{\bm{y}}{i}$ fails the band condition. In this case, if event $E$ occurs
    we obtain
    \[
        \abs{\wh{\bm{G}}_{\ith{\bm{y}}{i}}}
        \ge \abs{G_f^{\hat{t}}(\ith{\bm{y}}{i})} - \abs{\bm{G}_{\ith{\bm{y}}{i}}^\err}
        \ge B - B/4
        > B/2 \,,
    \]
    and thus Step~\ref{item:test-g} rejects. By \cref{claim:subgaussian-event} and a union bound,
    the algorithm rejects with probability $1 - 1/200 - o_n(1) > 99/100$.
\end{proof}

\subsection{Spectral Control}\label{sec:spectral}

In addition to the pointwise whittling above, we will also require spectral
conditions on the input, which are again testable. In this subsection, we give
two such tests, both of which   satisfy (informally) that (1) the YES case would
pass the test and (2) when $\calD$ is supported on the $2\sqrt{n}$-radius ball
centered at the origin (which is implied when $\calD$ is $B$-typical) and the
spectral condition being tested is violated, the test rejects, with high
probability.

The first one is a condition on the Frobenius norm of the uncentered
covariance matrix of $\calD$:

\begin{lemma}[Frobenius tester]
    \label{lemma:frobenius-tester}
    There is an algorithm which, 
    given sample access to a distribution $\cD$, draws
    $\smash{O(\sqrt{n} )}$  samples from $\calD$, and either
    \emph{accepts} or \emph{rejects} such that, with probability at least $99/100$, the following
    conditions hold:\begin{flushleft}\begin{enumerate}
        \item If $\cD$ is the standard Gaussian, the algorithm accepts.
        \item If  $\calD$ is 
        supported only on the $2\sqrt{n}$-radius ball centered at the origin\footnote{Note that $\calD$ satisfies this condition when it is $B$-typical with respect to $(f,\hat{t})$.} and $$\vabs{\Exu{\bm{y} \sim \cD}{\bm{y} \bm{y}^\top}}_F > 2\sqrt{n},$$ the
            algorithm rejects.
    \end{enumerate}\end{flushleft}
\end{lemma}
\begin{proof}
    The algorithm works as follows:
    \begin{enumerate}
\item[] For $s \define L\sqrt{n}$, where $L > 0$ is a sufficiently large absolute constant,
            draw \iid samples\\ $\ith{\bm{y}}{i}, \ith{\bm{z}}{i}$ from $\cD$ for $i \in [s]$. Let
            \[
                \wh{\bm{Q}} \define
                \frac{1}{s^2} \sum_{i=1}^s \sum_{j=1}^s \abra{\ith{\bm{y}}{i}, \ith{\bm{z}}{j}}^2
                \,.
            \]
            \textbf{Accept} if $\wh{\bm{Q}} \le 2n$, otherwise \textbf{reject}.
    \end{enumerate}
    The sample complexity is clear
    so it remains to show correctness.
To this end we first   observe that
    $$\Ex{\wh{\bm{Q}}} = \Exu{\bm{y}, \bm{z} \sim \cD}{\abra{\bm{y}, \bm{z}}^2} = \vabs{Y}_F^2,
    \quad\text{where}\ Y \define \Exu{\bm{y} \sim \cD}{\bm{y} \bm{y}^\top}.$$

    \noindent\textbf{Completeness.} Suppose that $\calD$ is the standard Gaussian. 
The test statistic $\wh{\bm{Q}}$ has variance
    \begin{align*}
        \Var{\wh{\bm{Q}}}
        &= \frac{1}{s^2} \Var{\abra{\ith{\bm{y}}{1}, \ith{\bm{z}}{1}}^2}
            + \frac{2}{s^4} \cdot s \cdot s(s-1) \cdot \Cov{
                \abra{\ith{\bm{y}}{1}, \ith{\bm{z}}{1}}^2,
                \abra{\ith{\bm{y}}{2}, \ith{\bm{z}}{1}}^2
            } \\
        &\le C \pbra{\frac{n^2}{s^2} + \frac{n}{s}}
        \le \frac{C n}{L^2} + \frac{C \sqrt{n}}{L}
        \le n \,,
    \end{align*}
    where the equality is by counting summands and the first inequality is by standard Gaussian calculations. By
    Chebyshev's inequality, with probability at least $99/100$ we have
    \[
        \wh{\bm{Q}} \le \Ex{\wh{\bm{Q}}} + 10\sqrt{n} = \vabs{I_n}_F^2 + 10\sqrt{n} < 2n \,,
    \]
    in which case the tester accepts.

    \noindent\textbf{Soundness.} Here $\mathbb E[\widehat{\bm{Q}}]=\|Y\|_F^2>4n$, whereas the algorithm rejects when $\widehat{\bm{Q}}>2n.$ So it suffices to show that $\smash{\widehat{\bm{Q}}}$ is unlikely to fall below half its expectation. We will do this by bounding its variance by a small constant times $\|Y\|_F^4.$ Similar to the completeness case,  we have
    \[
        \Var{\wh{\bm{Q}}}
        = \frac{1}{s^2} \Var{\abra{\ith{\bm{y}}{1}, \ith{\bm{z}}{1}}^2}
            + \frac{2}{s^4} \cdot s \cdot s(s-1) \cdot \Cov{
                \abra{\ith{\bm{y}}{1}, \ith{\bm{z}}{1}}^2,
                \abra{\ith{\bm{y}}{2}, \ith{\bm{z}}{1}}^2
            } \,.
    \]
    Using the assumption that $\cD$ is supported on the radius-$2\sqrt{n}$ ball
    centered at the origin, we have 
    \[
        \Var{\abra{\ith{\bm{y}}{1}, \ith{\bm{z}}{1}}^2}
        \le \Ex{\abra{\ith{\bm{y}}{1}, \ith{\bm{z}}{1}}^4}
        \le Cn^2 \Ex{\abra{\ith{\bm{y}}{1}, \ith{\bm{z}}{1}}^2}
        = C n^2 \vabs{Y}_F^2 \,.
    \]
    Also,
    \begin{align*}
        \Cov{\abra{\ith{\bm{y}}{1}, \ith{\bm{z}}{1}}^2, \abra{\ith{\bm{y}}{2}, \ith{\bm{z}}{1}}^2}
        &\le \Ex{
            \abra{\ith{\bm{y}}{1}, \ith{\bm{z}}{1}}^2
            \cdot \abra{\ith{\bm{y}}{2}, \ith{\bm{z}}{1}}^2
        }
        = \Exuc{\bm{z}}{\Exu{\bm{y}}{\abra{\bm{y}, \bm{z}}^2}^2}{\bm{z}}
        = \Exu{\bm{z}}{\pbra{\bm{z}^\top Y \bm{z}}^2} \\
        &\le \Exu{\bm{z}}{\vabs{Y}_{\mathrm{op}} \vabs{\bm{z}}_2^2 \cdot \bm{z}^\top Y \bm{z}}
        \le Cn \vabs{Y}_F \Exu{\bm{y,z}}{\abra{\bm{y}, \bm{z}}^2}
        = Cn \vabs{Y}_F^3 \,.
    \end{align*}
    Therefore
    \[
        \Var{\wh{\bm{Q}}} \le \frac{C n \vabs{Y}_F^2}{L^2} + \frac{C \sqrt{n} \vabs{Y}_F^3}{L}
        \le \frac{C \sqrt{n} \vabs{Y}_F^3}{L}
        \le \frac{C \vabs{Y}_F^4}{L} \,.
    \]
    By Chebyshev's inequality, for a sufficiently large choice of $L > 0$, we have
    \[
        \wh{\bm{Q}} \ge \Ex{\wh{\bm{Q}}} - 10\sqrt{\frac{C \vabs{Y}_F^4}{L}}
        \ge \vabs{Y}_F^2 \pbra{1 - 10\sqrt{C/L}}
        > \frac{1}{2} \vabs{Y}_F^2
        > 2n
    \]
    with probability at least
    $99/100$,  
    and hence the tester rejects as desired.
\end{proof}

The second spectral condition that we require is a joint condition on the distribution $\calD$, the function $f$, and the local
distance estimate $\wh{\alpha}_{f,N}$.

\begin{lemma}[Weighted centroid tester]
    \label{lemma:2-norm-tester}
    There is an algorithm which, given any neighborhood~$p$-tuple $N$ (as in \cref{def:local-distance-estimate}),  query access to a function $f$ and sample access to a distribution~$\cD$, 
    uses $O(\sqrt n R_\eps^2)$ samples from $\cD$ and  
    $O(p\sqrt n R_\eps^2)$ queries to $f$,
    and either \emph{accepts} or \emph{rejects}~such
    that, with probability at least $99/100$, the following conditions hold:
\begin{flushleft}
\begin{enumerate}
        \item If $f$ is a halfspace and $\calD$ is the standard Gaussian,    the algorithm accepts.
        \item 
        If $\calD$ is supported 
        only on the $2\sqrt{n}$-radius ball centered at the origin and 
\[
                \vabs{\Exu{\bm{y} \sim \cD}{
                    f(\bm{y}) \wh{\alpha}_{f,N}(\bm{y}) \bm{y}}}_2^2
                > 4\pbra{\sqrt{n} + R_\epsilon^2} \,,
            \]
            the algorithm rejects.
    \end{enumerate}\end{flushleft}
\end{lemma}
\begin{proof}
    The algorithm works as follows:
    \begin{enumerate}
        \item[] Draw $s \define L \sqrt{n} R_\epsilon^2$ \iid samples $\ith{\bm{y}}{i} \sim \cD$,
            where $L > 0$ is a sufficiently large constant. Let
            \[
                \wh{\bm{\mu}} \define \frac{1}{s} \sum_{i=1}^s
                    f\pbra{\ith{\bm{y}}{i}} \wh{\alpha}_{f,N}\pbra{\ith{\bm{y}}{i}} \ith{\bm{y}}{i}
                    \,.
            \]
        \textbf{Accept} if $\vabs{\wh{\bm{\mu}}}_2^2 \le 2\pbra{\sqrt{n} + R_\epsilon^2}$, and
            \textbf{reject} otherwise.
    \end{enumerate}

The sample and query complexities are clear.
We define a random variable $\bm{X} = f\pbra{\bm{y}} \wh{\alpha}_{f,N}\pbra{\bm{y}} \bm{y}$ for
    $\bm{y} \sim \cD$, and write $\mu \define \Ex{\bm{X}} = \Ex{\wh{\bm{\mu}}}$. We first have
    the following variance bound:
    \[
        \Ex{\vabs{\wh{\bm{\mu}}-\mu}_2^2}
        = \frac{1}{s} \Ex{\vabs{\bm{X}-\mu}_2^2}
        \le \frac{1}{s} \Ex{\vabs{\bm{X}}_2^2} \,.
    \]
    We now show completeness and soundness. 

    \noindent\textbf{Completeness.} 
    By rotation invariance, we may assume that $f(y) = \sign(y_1 - t)$ for some threshold $t$. For
    any point $y \in \bR^n$ and coordinate $i \in [n]$, write $y^{(i)}$ for the point obtained by
    flipping the sign of the $i^{\text{th}}$ coordinate of $y$. The key observation is that, for any
    $y \in \bR^n$ and $i \ne 1$, we have
    \[
        f(y) \wh{\alpha}_{f,N}(y) = f(y^{(i)}) \wh{\alpha}_{f,N}(y^{(i)}) \,,
    \]
    since $y_1 = y^{(i)}_1$ and $\pbra{y + \ith{g}{j}}_1 = \pbra{y^{(i)} + \ith{g}{j}}_1$ for each
    neighborhood tuple point $\ith{g}{j} \in N$. Therefore every coordinate but the first vanishes,
    and we obtain
    \[
        \vabs{\mu}_2^2
        = \Ex{f(\bm{y}) \wh{\alpha}_{f,N}(\bm{y}) \bm{y}_1}^2
        \le \Ex{R_\epsilon^2 \bm{y}_1^2}
        = R_\epsilon^2 \,.
    \]
    Also,
    \[
        \Ex{\vabs{\bm{X}}_2^2} \le R_\epsilon^2 \Ex{\vabs{\bm{y}}_2^2} = n R_\epsilon^2 \,,
    \]
    so that $\Ex{\vabs{\wh{\bm{\mu}}-\mu}_2^2} \le \frac{n R_\epsilon^2}{s} = \frac{\sqrt{n}}{L}$.
    By Markov's inequality, we have
    \[
        \vabs{\wh{\bm{\mu}}-\mu}_2^2 \le \frac{100 \sqrt{n}}{L}
    \]
    with probability at least $99/100$. 
As a result,    \[
        \vabs{\wh{\bm{\mu}}}_2
        \le \vabs{\mu}_2 + \frac{10 n^{1/4}}{\sqrt{L}}
        \le R_\epsilon + n^{1/4}
    \]
    for sufficiently large choice of $L$. In this case $\vabs{\wh{\bm{\mu}}}_2^2 \le
    2(R_\epsilon^2+\sqrt{n})$, so the algorithm accepts.

    \noindent\textbf{Soundness.} 
    We  have from the assumption that $\vabs{\mu}_2^2 >
    4(\sqrt{n}+R_\epsilon^2)$. Moreover, by the assumption that $\calD$ is supported on the $2\sqrt{n}$-radius ball centered at the origin,
    we have
    \[
        \Ex{\vabs{\bm{X}}_2^2} \le R_\epsilon^2 \cdot 4n \,,
    \]
    so that $\Ex{\vabs{\wh{\bm{\mu}}-\mu}_2^2} \le \frac{4n R_\epsilon^2}{s} = \frac{4\sqrt{n}}{L}$.
    By Markov's inequality,  we have
    \[
        \vabs{\wh{\bm{\mu}}-\mu}_2^2
        \le \frac{400 \sqrt{n}}{L}
        \le \frac{\sqrt{n} + R_\epsilon^2}{100}
    \]
    with probability at least $99/100$,
    for a sufficiently large choice of $L$. In this case, we have
    \[
        \vabs{\wh{\bm{\mu}}}_2
        \ge \vabs{\mu}_2 - \frac{\sqrt{\sqrt{n} + R_\epsilon^2}}{10}
        > \pbra{2 - \frac{1}{10}} \sqrt{\sqrt{n} + R_\epsilon^2}
        > \sqrt{2} \cdot \sqrt{\sqrt{n} + R_\epsilon^2} \,,
    \]
    so that $\vabs{\wh{\bm{\mu}}}_2^2 > 2(\sqrt{n} + R_\epsilon^2)$ and the algorithm rejects as desired.
\end{proof}

\subsection{Estimating the Threshold}\label{sec:threshold}

We give a simple lemma establishing the guarantees we can ensure on an estimate of the threshold
parameter ${t}$. Informally, the estimate has its absolute value  always bounded by $T_\eps$ and is accurate under the YES case --- with an escape hatch $\bot$ allowed for cases
when the function $f$ is extremely biased (in which case one can safely accept).

\begin{lemma}[Threshold estimator]
    \label{lemma:threshold-estimator}
    There is an algorithm which, given query access to a function $f$ and sample access to a
    distribution $\cD$, uses $O(1/\epsilon^{12})$ queries and samples, and outputs either $\bot$ or
    a real number $\hat{\bm{t}}$ with $|\hat{\bm{t}}|\le T_\eps$ such that, with probability at least $99/100$, the following
    hold:
\begin{flushleft}\begin{enumerate}
        \item If $\smash{\Big|
        {\Exu{\bm{y} \sim \cD}{f(\bm{y})}}\Big| \le 1 - \epsilon}$, then the algorithm
            outputs a real number $\hat{\bm{t}}$.\vspace{0.15cm}
        \item If $f$ is a halfspace and $\cD$ is the standard Gaussian, then  the algorithm outputs either
            $\bot$, or a real number $\hat{\bm{t}}$ that is an $\epsilon^4$-good threshold estimate for $f$.
    \end{enumerate}
    \end{flushleft}
\end{lemma}
\begin{proof}
    The algorithm works as follows. Let $L > 0$ be a sufficiently large constant.
\begin{flushleft}\begin{enumerate}
        \item Draw $s \define {L}/{\epsilon^{12}}$ independent samples from $\cD$, query $f$
            on these points, and let $\wh{\bm{p}}$ be the fraction of points labeled $-1$.
        \item If $\wh{\bm{p}} \in \sbra{ {\epsilon}/{4}, 1-({\epsilon}/{4})}$, output
            $\Phi^{-1}(\wh{\bm{p}})$. Otherwise, output $\bot$.
    \end{enumerate}\end{flushleft}

    The sample and query complexity claims are immediate, and  the algorithm outputs either $\bot$ or
    $$
        |{\hat{\bm{t}}}|
        \le \Phi^{-1}\pbra{1-\frac{\epsilon}{4}}
        \le \sqrt{2\log(4/\epsilon)}
        \le T_\epsilon \,,
    $$
    We now prove correctness. Write $p$ for the true probability $p
    \define \Pr{f(\bm{y}) = -1}$. By a Hoeffding bound, as long as $L$ is
    sufficiently large, with probability at least $99/100$ we have $\abs{p-\wh{\bm{p}}} \le
    \epsilon^6$. We show correctness assuming this event.

    For the first part, we have  $p \in
[ {\epsilon}/{2}, 1-({\epsilon}/{2})]$. 
    Then $\wh{\bm{p}} \in
    \sbra{{\epsilon}/{4}, 1-({\epsilon}/{4})}$, and hence the algorithm outputs a real number $\hat{\bm{t}}$.

    For the second part, suppose now that $f$ is a halfspace and $\cD = \cN(0, I_n)$. Without loss
    of generality, say $f$ is the halfspace $f(x) = \sign(x_1 - t)$. Then $p =  \Phi(t)$, so that $t = \Phi^{-1}(p)$.
    Now suppose the estimate $\wh{\bm{p}}$ satisfies $\wh{\bm{p}} \in \sbra{ {\epsilon}/{4},
    1-({\epsilon}/{4})}$, since otherwise the algorithm outputs $\bot$ and we are done. Then
    $p, \wh{\bm{p}} \in \sbra{{\epsilon}/{8}, 1-{(\epsilon}/{8})}$. Therefore, for all $q$
    between $p$ and $\wh{\bm{p}}$, the inverse function rule gives
    \[
        \abs{\pbra{\Phi^{-1}}'(q)}
        = \frac{1}{\phi\pbra{\Phi^{-1}(q)}}
        \le \frac{1}{\phi\pbra{\sqrt{2\log(8/\epsilon)}}}
        \le O\left(\frac{1}{\epsilon}\right) \,.
    \]
    Hence the mean value theorem gives
    \[
        \abs{\hat{\bm{t}} - t}
        = \abs{\Phi^{-1}(\wh{\bm{p}}) - \Phi^{-1}(p)}
        \le O\left(\frac{1}{\eps}\right) \abs{\wh{\bm{p}} - p}
        \le O(\eps^5) \le \epsilon^4
    \]
    using $\eps\le \eps_0$ and the assumption that $\eps_0$ is a sufficiently small constant.
\end{proof}

\section{Upper Bound: Tester and Proof of Correctness}\label{sec:maintester}

The tester is given in \cref{alg:tester}. Note that it is convenient for the sake of notation to
write the input distance parameter as $\epsilon^*$, and then work with $\epsilon \define
\epsilon^*/2$ through most of the analysis.

\def\bZ{\boldsymbol{Z}}

We begin by giving a high-level sketch of how the algorithm works. It first estimates the threshold $\hat{\bm{t}}$ (using \Cref{lemma:threshold-estimator}) and accepts early if the function that is being tested appears to be sufficiently close to a constant.
It then runs the ``YES-versus-far-from-whittled'' tester of \Cref{lemma:testing-gaussian-versus-far-from-whittled}, the Frobenius tester of \Cref{lemma:frobenius-tester}, and the weighted centroid tester of \Cref{lemma:2-norm-tester}, rejecting if any of these reject.
As shown in these lemmas, the YES case passes all these testers with high probability. 
For the NO case, on the other hand, these testers ensure that we will be able to control the variance of the final statistic (from line 7 to 10) as required.
Next, it checks that the local distance estimate $\smash{\wh{\alpha}_{f,\bm{N}}}$ is not too small too often, since this is required by the NO-case distance disagreement lemma (\Cref{lemma:distance-disagreement-no}); the YES case will again pass this test with high probability.
Finally, the~algorithm ends by estimating the distance disagreement measure, accepting if the estimate $\bZ$ is small and rejecting otherwise. With high probability, the YES case passes all checks and is accepted, while the NO case is rejected either by a preliminary check or by the final estimate.

We discuss the final statistic $\bZ$ in more detail. It is obtained by expanding the squared difference 
$$\big(f(\bm{y})(G_f^{\widehat{\bm{t}}}(\bm{y})-\widehat{\bm{t}}) - \widehat{\alpha}_{f,N}(\bm{y})\big)^2
=\big(G_f^{\widehat{\bm{t}}}(\bm{y})-\widehat{\bm{t}}\big)^2 
-2f(\bm{y}) \big(G_f^{\widehat{\bm{t}}}(\bm{y})-\widehat{\bm{t}}\big)
\widehat{\alpha}_{f,N}(\bm{y})+
\big(\widehat{\alpha}_{f,N}(\bm{y})\big)^2
$$ 
and uses $\bZ_1,\bZ_2$ and $\bZ_3$ to estimate the expectation of each of the three terms:
$\boldsymbol Z_1$ and $\boldsymbol Z_3$ estimate their mean squares over $\bm{y} \sim \cD$, and $\boldsymbol Z_2$ estimates their mean product.
This is why the final statistic is $\boldsymbol Z=\boldsymbol Z_1-2\boldsymbol Z_2+\boldsymbol Z_3.$

The main technical lemma for the analysis of the final statistic $\bZ$ is \Cref{lemma:test-statistic}. We delay its proof to \Cref{sec:test-statistic}. There are three assumptions that need to be met in order to apply this lemma so we briefly discuss the two scenarios for which we apply the lemma and how these conditions are met.
The two scenarios are:
\begin{flushleft}\begin{enumerate}
\item YES case: $f$ is a halfspace, $\calD$ is the standard Gaussian and $\hat{t}$ is a good threshold estimate for $f$. In this case, condition 1(a) of \Cref{lemma:test-statistic} holds trivially, while conditions 2 and 3 can be verified using the assumption that $\calD$ is the Gaussian;
\item NO case: While we cannot assume $\calD$ to satisfy any of the three conditions of  \Cref{lemma:test-statistic}, its whittled version $\calD_*^{\hat{t}}$ is $B$-typical with respect to $(f,\hat{t})$ by definition, and hence satisfies 1(b). 
Moreover, we may assume that $\smash{\TV(\calD,\calD_*^{\hat{t}})\le 1/Km}$ since otherwise the algorithm would reject with high probability due to the ``YES-versus-far-from-whittled'' tester.
In this case $f$ remains $\varepsilon$-far from every halfspace under $\mathcal D_*^{\hat t}$, since $\varepsilon^*-1/(Km)\ge\varepsilon=\varepsilon^*/2$.
As it can be verified that \Cref{alg:tester} draws less than $Km$ samples from $\calD$ in the rest of its execution, it is safe to assume that the rest of \Cref{alg:tester} is run on $\calD_*^{\hat{t}}$ instead of $\calD$; this includes the Frobenius tester, the weighted centroid tester, as well as the final estimate $\bZ$.
Passing the former two ensures that the whittled distribution satisfies both conditions 2 and 3 of  \Cref{lemma:test-statistic}.
This allows us to apply \Cref{lemma:test-statistic} on the whittled distribution as $\calD$ in the lemma to show that the final estimate must reject with high probability.
\end{enumerate}\end{flushleft}

    \begin{restatable}{lemma}{lemmateststatistic}
        \label{lemma:test-statistic}
        Let $f : \bR^n \to \pmset$ be a function and $\cD$ be a distribution over $\bR^n$. Let
        $\hat{t} \in \bR$ satisfy $|{\hat{t}}| \le T_\epsilon$, and $N$ be a neighborhood
        tuple. Recall the definition of the band-distance function $\smash{G_f^{\hat{t}}}$~from
        \cref{def:whittled-distribution}. Suppose the following conditions are satisfied:
        \begin{enumerate}
            \item \label{item:radius-bound} \label{item:band-distance-bound}
                One of the following two conditions holds:
                \begin{enumerate}
                    \item (YES case) $\cD = \cN(0, I_n)$, $f$ is a halfspace, and $\hat{t}$ is an
                        $\epsilon^4$-good threshold estimate for $f$.
                    \item (Bounded case) 
            $\calD$ is a $B$-typical distribution with respect to $(f,\hat{t})$.
\end{enumerate}
            \item \label{item:covariance-bound}
                Frobenius bound: $\vabs{
                    \Exu{\bm{y} \sim \cD}{\bm{y} \bm{y}^\top}
                }_F \le 2\sqrt{n}$.
            \item \label{item:weighted-centroid-bound}
                Weighted centroid bound: $\vabs{\Exu{\bm{y} \sim \cD}{
                f(\bm{y}) \wh{\alpha}_{f,N}(\bm{y}) \bm{y}
            }}_2^2 \le 4(\sqrt{n} + R_\epsilon^2)$.
        \end{enumerate}
        Then with probability at least $99/100$, the $\bm{Z}$ from \cref{alg:tester}
        satisfies the following two properties.
        \begin{enumerate}
            \item If $\distdis_{\wh{\alpha}_{f,N}}^{\hat{t}}(f, \cD) \le {\epsilon^3}\big/{128}$,
                then $\bm{Z} \le {\epsilon^3}\big/{64}$.
            \item If $\distdis_{\wh{\alpha}_{f,N}}^{\hat{t}}(f, \cD) > {\epsilon^3}\big/{32}$, then
                $\bm{Z} > {\epsilon^3}\big/{64}$.
        \end{enumerate}
    \end{restatable}

\begin{algorithm}[t!]
    \caption{Distribution-Robust Halfspace Tester}
    \label{alg:tester}
    \begin{algorithmic}[1]
        \Statex \textbf{Input}: Dimension $n$, parameter $\epsilon^* \in (0,1)$, sample access to a
        distribution $\cD$ over $\bR^n$, 
        \Statex and query access to a function $f : \bR^n \to \pmset$.

        \Statex \textbf{Output}: \textbf{Accept} or \textbf{reject}\vspace{-0.2cm}

        \Statex\hspace*{-\algorithmicindent}\rule{\dimexpr\linewidth+\algorithmicindent\relax}{0.4pt}\vspace{0.05cm}
\State Set $\epsilon \define \epsilon^*/2$ and $m \define
            \Theta(n^{2/3}/\epsilon^{10})$.

            \AlgSection{Estimate and test the threshold:}

            \State Compute a threshold estimate $\hat{\bm{t}}$ with $
        |\hat{\bm{t}}|\le T_\eps$ using \cref{lemma:threshold-estimator},
            or \textbf{accept} if it outputs $\bot$.
            \label{line:threshold}

            \AlgSection{Test that the distribution is well-behaved:}

            \State Set $p \define \Theta(1/\epsilon^{28})$ and sample  a neighborhood
            tuple $\bm{N} \define (\ith{\bm{g}}{1}, \dotsc, \ith{\bm{g}}{p}) \sim \cN(0, I_n)^p$.

            \State Run the algorithms from
            \cref{lemma:testing-gaussian-versus-far-from-whittled,lemma:frobenius-tester,lemma:2-norm-tester}
            (the first with parameter $K = 1000$ and $m$ defined above),
            and \textbf{reject} if any of them rejects.
            \label{line:distribution-tests}

            \AlgSection{Test the mass near the boundary:}

            \State Sample a tuple $\bm{S}$ of $\Theta(1/\epsilon)$ \iid points from $\cD$ and compute
            $\wh{\alpha}_{f,\bm{N}}(y)$ at each $y \in \bm{S}$.
            \label{line:sample-S}

            \State \textbf{Reject} if
            \[
                \frac{\abs{\cbra{y \in \bm{S}: \wh{\alpha}_{f,\bm{N}}(y)
                < \epsilon/4}}}{\abs{\bm{S}}} > \frac{3\epsilon}{8} \,.
            \]
            \label{line:alpha-hat}

            \AlgSection{Compute the main test statistic:}

            \State Draw \iid samples
            \[
                \ith{\bm{w}}{1}, \dotsc, \ith{\bm{w}}{m} \sim \cN(0, I_n) \,, \quad
                \ith{\bm{z}}{1}, \dotsc, \ith{\bm{z}}{m} \sim \cN(0, I_n)  \quad \text{and} \quad
                \ith{\bm{y}}{1}, \dotsc, \ith{\bm{y}}{m} \sim \cD \,.
            \]

            \State Set
            \[
                \wh{\bm{w}} \define \frac{1}{\tau(\hat{\bm{t}})} \cdot
                    \frac{1}{m} \sum_{i=1}^m \ith{\bm{w}}{i} f(\ith{\bm{w}}{i})
                \quad \text{and} \quad
                \wh{\bm{z}} \define \frac{1}{\tau(\hat{\bm{t}})} \cdot
                    \frac{1}{m} \sum_{j=1}^m \ith{\bm{z}}{j} f(\ith{\bm{z}}{j}) \,.
            \]

            \State
            \label{item:line-alpha}
            Compute $\wh{\alpha}_{f,\bm{N}}(\ith{\bm{y}}{k})$ using $p+1$ queries to $f$ for each $k
            \in [m]$, and set
            \begin{align}
                \label{eq:Z_1}
                \bm{Z}_1 &\define \frac{1}{m} \sum_{k=1}^m
                    \pbra{\abra{\ith{\bm{y}}{k}, \wh{\bm{w}}} - \hat{\bm{t}}}
                    \pbra{\abra{\ith{\bm{y}}{k}, \wh{\bm{z}}} - \hat{\bm{t}}}
                    \,, \\[0.3ex]
                \label{eq:Z_2}
                \bm{Z}_2 &\define \frac{1}{m} \sum_{k=1}^{m}
                    f(\ith{\bm{y}}{k}) \pbra{\abra{\ith{\bm{y}}{k}, \wh{\bm{w}}} - \hat{\bm{t}}}
                    \wh{\alpha}_{f,\bm{N}}(\ith{\bm{y}}{k}) \,, \text{ and} \\[0.3ex]
                \label{eq:Z_3}
                \bm{Z}_3 &\define \frac{1}{m} \sum_{k=1}^{m}
                    \wh{\alpha}_{f,\bm{N}}(\ith{\bm{y}}{k})^2 \,.
            \end{align}

            \State Set $\bm{Z} \define \bm{Z}_1 - 2\bm{Z}_2 + \bm{Z}_3$. \textbf{Accept} if $\bm{Z}
            \le \epsilon^3/64$, and \textbf{reject} otherwise.
            \label{line:test-statistic}
    \end{algorithmic}
\end{algorithm}

Assuming \Cref{lemma:test-statistic}, we now prove the correctness of the algorithm. 
We show that the algorithm accepts with high probability in the YES case in \Cref{sec:yescase},
  and rejects with high probability in the NO case in \Cref{sec:nocase}.
We combine them to prove \Cref{mainthm:upper} in \Cref{sec:together}.

\subsection{YES Case}\label{sec:yescase}
We first show that \cref{alg:tester} accepts with high constant probability in the YES case:

\begin{lemma}
    \label{lemma:yes-case}
    Suppose $f$ is a halfspace and $\cD$ is the standard Gaussian. Then \cref{alg:tester} accepts with
    probability at least $3/4$.
\end{lemma}
\begin{proof}
    Let $E$ be the event that the output of the threshold estimator algorithm from \cref{lemma:threshold-estimator} in
    \cref{line:threshold} satisfies the two correctness conditions stated therein. By
    \cref{lemma:threshold-estimator}, we have $\Pr{E} \ge 99/100$. Going forward, suppose that $E$
    occurs. 
If the output is $\bot$ then the algorithm accepts in \cref{line:threshold}. So assume in the rest of the proof that the output is not $\bot$ but a real number  $\hat{\bm{t}}=\hat{t}$. Given $E$ we know that $\hat{t}$ is an $\eps^4$-good threshold estimate for $f$ and satisfies $|\hat{t}|\le T_\eps$.

    We consider each step of \cref{alg:tester} that may reject, upper bound the probability that
    each such step causes the algorithm to reject, and then conclude by a union bound.

    \noindent\textbf{Distribution test (\cref{line:distribution-tests}).} 
    It follows from \cref{lemma:testing-gaussian-versus-far-from-whittled,lemma:frobenius-tester,lemma:2-norm-tester} and a union bound that this line rejects with probability at most $3/100$.

    \noindent\textbf{Mass near the boundary (\cref{line:alpha-hat}).} We defer for now the proofs of the
    following two claims.

    \begin{restatable}{claim}{claimSalphabar}
        Suppose $f$ is a halfspace and $\cD = \cN(0, I_n)$. When $\bm{S}$ is sampled in
        \cref{line:sample-S}, we have
        \[
            \Pru{\bm{S}}{
                \frac{\abs{\cbra{y \in \bm{S}: \overline{\alpha}_f(y)
                < \frac{\epsilon}{4} + \frac{\epsilon}{100}}}}{\abs{\bm{S}}} > \frac{3\epsilon}{8}
            } \le \frac{1}{100} \,.
        \]
    \end{restatable}

    \begin{restatable}{claim}{claimYESalphabar}
        \label{claim:YES-alpha-bar}
        Suppose $f$ is a halfspace and $\cD = \cN(0, I_n)$. Let $S$ be any tuple of
        $\Theta(1/\epsilon)$ points in $\bR^n$. Then the following holds with probability at least
        $99/100$ over the choice of $\bm{N}$:
        \[
            \big|{\overline{\alpha}_f(x) - \wh{\alpha}_{f,\bm{N}}(x)}\big| \le \frac{\epsilon}{100},\quad\text{for every $x\in S$.}
        \]
    \end{restatable}

    Assuming these two claims and applying a union bound, we conclude that, with probability at
    least $98/100$ over the choice of $\bm{S}$ and $\bm{N}$, independently of event $E$, we have
    \[
        \frac{\abs{\cbra{y \in \bm{S}: \wh{\alpha}_{f,\bm{N}}(y) < \epsilon/4}}}{\abs{\bm{S}}}
        \le \frac{\abs{\cbra{y \in \bm{S}: \overline{\alpha}_f(y)
        < \frac{\epsilon}{4} + \frac{\epsilon}{100}}}}{\abs{\bm{S}}}
        \le \frac{3\epsilon}{8}
    \]
    and hence \cref{line:alpha-hat} does not reject.

    \noindent\textbf{Main test statistic (\cref{line:test-statistic}).} We rely on three results. The
    first result, which we prove below, says that the distance disagreement measure is small with
    high constant probability.

    \begin{restatable}{claim}{claimYESdistdis}\label{claim:hehe1}
        Suppose $f$ is a halfspace and $\cD = \cN(0, I_n)$. Let $\hat{t}$ be an $\epsilon^4$-good
        threshold estimate for $f$ satisfying $|{\hat{t}} |\le T_\epsilon$. Then, with probability
        at least $99/100$ over the choice of tuple $\bm{N}$, we have
        \begin{equation}
            \label{eq:distdis-YES-statement}
            \distdis_{\wh{\alpha}_{f,\bm{N}}}^{\hat{t}}(f, \cN(0,I_n)) \le \frac{\epsilon^3}{128}
            \,.
        \end{equation}
    \end{restatable}

    The second result, which we prove below, shows that the conditions 2 and 3 of \cref{lemma:test-statistic}
    are satisfied with high constant probability in the YES case.

    \begin{restatable}{claim}{claimYESteststatisticconditions}
        \label{claim:YES-test-statistic-conditions}
        Suppose $f$ is a halfspace and $\cD = \cN(0, I_n)$. Let $\hat{t}$ be an $\epsilon^4$-good
        threshold estimate for $f$ with $|{\hat{t}}| \le T_\epsilon$ and $N$ be any
        neighborhood tuple. Then all conditions of \cref{lemma:test-statistic} are satisfied.
    \end{restatable}

    Putting the results above together,
      we have that \cref{line:test-statistic} accepts with probability at least $98/100$ by a union bound using \Cref{claim:hehe1} and \Cref{lemma:test-statistic}.
    The proof of    \Cref{lemma:yes-case} follows by a union bound over all lines where the algorithm may reject.  
\end{proof}

We now prove the claims we used above, restated for convenience (except for
\cref{lemma:test-statistic} which is deferred to \cref{sec:test-statistic}).

\claimSalphabar*
\begin{proof}
    Write $\epsilon' \define \frac{\epsilon}{4} + \frac{\epsilon}{100}$. Note that, by
    \cref{def:alpha-bar} and choice of $R_\epsilon \gg 1$, for each $x \in \bR^n$ we have
    $\overline{\alpha}_f(x) < \epsilon' \iff \alpha_f(x) < \epsilon'$, so we can replace
    $\overline{\alpha}_f$ with $\alpha_f$ in the statement.

    Let $v$ be a unit vector and $t \in \bR$ a threshold such that $f(\cdot) = \sign(\abra{\cdot, v}
    - t)$. Then $\alpha_f$ satisfies
    \begin{align*}
        \alpha_f(x)
        &= \Phi^{-1}(\ell_f(x))
        = \Phi^{-1}\pbra{\Pru{\bm{g} \sim \cN(0, I_n)}{f(x+\bm{g}) = f(x)}} \\
        &= \Phi^{-1}\pbra{\Pru{\bm{g} \sim \cN(0, I_n)}{
            \sign(\abra{x+\bm{g},v}-t) = \sign(\abra{x,v}-t)}
        } \\
        &= \Phi^{-1}\pbra{\Pru{\bm{g} \sim \cN(0, I_n)}{
            \sign(\abra{\bm{g},v} + \abra{x,v}-t) = \sign(\abra{x,v}-t)}
        } \\
        &= \Phi^{-1}\pbra{\Pru{\bm{g} \sim \cN(0, 1)}{
            \sign(\bm{g} + \abra{x,v}-t) = \sign(\abra{x,v}-t)}
        } \\
        &= \Phi^{-1}\pbra{\Phi(\abs{\abra{x,v} - t})}
            \tag{By inspecting the sign of $\abra{x,v} - t$} \\
        &= \abs{\abra{x,v} - t} \,,
    \end{align*}
    which is the distance from $x$ to the boundary hyperplane as discussed earlier. Therefore
    \[
        \Pru{\bm{x} \sim \cN(0,I_n)}{\alpha_f(\bm{x}) < \epsilon'}
        = \Pru{\bm{x} \sim \cN(0,I_n)}{\abs{\abra{\bm{x}, v} - t} < \epsilon'}
        = \Pru{\bm{g} \sim \cN(0,1)}{\abs{\bm{g} - t} < \epsilon'}
        \le \frac{2\epsilon'}{\sqrt{2\pi}}
        < \epsilon' \,.
    \]
    It follows that
    \[
        \Exu{\bm{S}}{\frac{\abs{\cbra{y \in \bm{S}: \alpha_f(y) < \epsilon'}}}{\abs{\bm{S}}}}
        = \Pru{\bm{x} \sim \cN(0,I_n)}{\alpha_f(\bm{x}) < \epsilon'}
        < \epsilon' \,,
    \]
    and hence the claim holds by a Chernoff bound as long as we set $\abs{\bm{S}} =
    \Theta(1/\epsilon)$ with sufficiently large constant.
\end{proof}

Before proving the next claims, we need a technical lemma on the truncated inverse Gaussian CDF.
Define function $F_\epsilon : [0,1] \to \bR$ by $F_\epsilon \define \trunc_{\pm R_\epsilon} \circ
\Phi^{-1}$, where we again implicitly take the limits $\Phi^{-1}(t) \to -\infty$ as $t \to 0^+$ and
$\Phi^{-1}(t) \to +\infty$ as $t \to 1^-$ so that $F_\epsilon(0) = -R_\epsilon$ and $F_\epsilon(1) =
R_\epsilon$. Then, recalling \cref{def:alpha-bar,def:alpha-hat}, we see that
\[
    \overline \alpha_f(x) = F_\epsilon(\ell_f(x))
    \quad \text{and} \quad
    \wh{\alpha}_{f,N}(x) = F_\epsilon(\ell_{f,N}(x))
\]
for each point $x$ and neighborhood tuple $N$. Then $F_\epsilon$ has bounded first derivative, as
follows.

\begin{lemma}
    \label{lemma:F-derivative}
    The function $F_\epsilon : [0,1] \to \bR$ is absolutely continuous and satisfies
    $\vabs{F_\epsilon'}_\infty \le \frac{C}{\epsilon^{25 / 2}}$.
\end{lemma}
\begin{proof}
    Since $\Phi^{-1} : (0,1) \to \bR$ is continuously differentiable and diverges to $+\infty$
    (resp.\ $-\infty$) at 1 (resp.\ 0), there exists $\delta \in (0,1)$ such that $\Phi^{-1}$ is
    Lipschitz (and hence absolutely continuous) on $[\delta, 1-\delta]$ and moreover we have
    $\Phi^{-1}(\delta) < -R_\epsilon$ and $\Phi^{-1}(1-\delta) > R_\epsilon$. Hence the truncated
    $F_\epsilon$ is absolutely continuous on $[\delta, 1-\delta]$, with $F_\epsilon(\delta) =
    -R_\epsilon$ and $F_\epsilon(1-\delta) = R_\epsilon$. It follows that the extension to all of
    $[0,1]$ (by keeping $F_\epsilon$ constant outside $[\delta, 1-\delta]$) is also absolutely
    continuous.

    By the same reasoning, it suffices to upper bound $\abs{\pbra{\Phi^{-1}}'(t)}$ for $t \in (0,1)$
    at which $\abs{\Phi^{-1}(t)} \le R_\epsilon$. At any such $t$, the inverse function rule gives
    \[
        \abs{\pbra{\Phi^{-1}}'(t)}
        = \abs{\frac{1}{\phi\pbra{\Phi^{-1}(t)}}}
        \le \frac{1}{\phi(R_\epsilon)}
        = \sqrt{2\pi} \exp\pbra{\frac{1}{2} \cdot 5^2 \log(1/\epsilon)}
        = \frac{\sqrt{2\pi}}{\epsilon^{25 / 2}} \,. \qedhere
    \]
\end{proof}

\claimYESalphabar*
\begin{proof}
    Using \cref{lemma:F-derivative}, for any $x \in \bR^n$ we have
    \begin{align*}
        \Exu{\bm{N}}{(\overline \alpha_f(x) - \wh{\alpha}_{f,\bm{N}}(x))^2}
        &= \Exu{\bm{N}}{\pbra{F_\epsilon(\ell_f(x)) - F_\epsilon(\ell_{f,\bm{N}}(x))}^2}
        \le \Exu{\bm{N}}{\vabs{F'_\epsilon}_\infty^2 (\ell_f(x) - \ell_{f,\bm{N}}(x))^2} \\
        &\le \frac{C}{\epsilon^{25}} \cdot \Varu{\bm{N}}{\ell_{f,\bm{N}}(x)}
        \le \frac{C}{p \epsilon^{25}}
        \le \frac{\epsilon^3}{100^3 K} \,,
    \end{align*}
    where we used the fact that random variable $\ell_{f,\bm{N}}(x)$ is an average over
    $\abs{\bm{N}} = p$ independent binary trials with expectation $\ell_f(x)$, and $K > 0$ can be
    made an arbitrarily large absolute constant by appropriately choosing the constant in the
    definition of $p$. We conclude by Markov's inequality that, with probability at least $1 -
    \frac{\epsilon}{100 K}$,
    \begin{equation}
        \label{eq:alpha-error-on-x}
        \abs{\overline \alpha_f(x) - \wh{\alpha}_{f,\bm{N}}(x)} \le \frac{\epsilon}{100} \,.
    \end{equation}
    Therefore, by a union bound over the $\abs{S} \le K/\epsilon$ elements of $S$ (here we choose
    the constant $K$ sufficiently large to match the asymptotic bound on $S$), with probability at
    least $99/100$, \eqref{eq:alpha-error-on-x} holds for every $x \in S$.
\end{proof}

\claimYESdistdis*
\begin{proof}
    For each neighborhood tuple $N$, write
    \begin{align*}
        \Delta(N) &\define \Exu{\bm{x} \sim \cN(0,I_n)}{
            (\alpha_f(\bm{x}) - \wh{\alpha}_{f,N}(\bm{x}))^2
        } \quad \text{and} \\
        \bar{\Delta}(N) &\define \Exu{\bm{x} \sim \cN(0,I_n)}{
            (\overline \alpha_f(\bm{x}) - \wh{\alpha}_{f,N}(\bm{x}))^2
        } \,.
    \end{align*}
    By \cref{lemma:distdis-YES},
    \begin{equation}
        \label{eq:distdis-YES}
        \distdis_{\wh{\alpha}_{f,\bm{N}}}^{\hat{t}}(f, \cN(0,I_n))
        \le C \Delta(\bm{N})
            + C \epsilon^8 \pbra{1 + \frac{\abs{\hat{t}} + \epsilon^4}{\tau(\hat{t})}}^2
        \,.
    \end{equation}
    Plugging \cref{lemma:tau-lower-bound} into the second term in the RHS of \eqref{eq:distdis-YES},
    we obtain
    \[
        C \epsilon^8 \pbra{1 + \frac{\abs{\hat{t}} + \epsilon^4}{\tau(\hat{t})}}^2
        \le C \epsilon^4 T_\epsilon^2
        \le C \epsilon^4 \log(1/\epsilon)
        \le \frac{\epsilon^3}{1000}
    \]
    for sufficiently small $\epsilon$.

    We now return to the first term in the RHS of \eqref{eq:distdis-YES}. As in the proof of
    \cref{claim:YES-alpha-bar}, for any $x \in \bR^n$ we have
    \[
        \Exu{\bm{N}}{(\overline \alpha_f(x) - \wh{\alpha}_{f,\bm{N}}(x))^2}
        \le \frac{\epsilon^3}{K} \,,
    \]
    where $K > 0$ can be made an arbitrarily large absolute constant by appropriately choosing the
    constant in the definition of $p$. Hence
    \[
        \Exu{\bm{N}}{\bar{\Delta}(\bm{N})}
        = \Exu{\bm{x} \sim \cN(0, I_n)}{
            \Exu{\bm{N}}{(\overline \alpha_f(\bm{x}) - \wh{\alpha}_{f,\bm{N}}(\bm{x}))^2}
        }
        \le \frac{\epsilon^3}{K} \,.
    \]
    However, we need a bound for $\Delta$ rather than just $\bar{\Delta}$. Let $f$ be the halfspace
    $f(x) = \sign(x_1 - t)$ without loss of generality. Then we also have
    \begin{align*}
        \Exu{\bm{x} \sim \cN(0, I_n)}{(\alpha_f(\bm{x}) - \overline{\alpha}_f(\bm{x}))^2}
        &\le \Exu{\bm{x} \sim \cN(0, I_n)}{
            \alpha_f(\bm{x})^2
            \cdot \ind{\alpha_f(\bm{x}) \ge R_\epsilon}
        } \\
        &\le \Exu{\bm{x} \sim \cN(0, I_n)}{
            (\abs{\bm{x}_1} + \abs{t})^2
            \cdot \ind{\abs{\bm{x}_1} \ge R_\epsilon - \abs{t}}
        } \\
        &= \Exu{\bm{g} \sim \cN(0,1)}{
            (\abs{\bm{g}} + \abs{t})^2
            \cdot \ind{\abs{\bm{g}} \ge R_\epsilon - \abs{t}}
        } \,.
    \end{align*}
    Now, we have
    \[
        \abs{t} \le \abs{\hat{t}} + \epsilon^4
        \le T_\epsilon + \epsilon^4
        = 2\sqrt{\log(1/\epsilon)} + \epsilon^4
        \le \frac{5}{2} \sqrt{\log(1/\epsilon)} \,,
    \]
    so that $R_\epsilon - \abs{t} = 5\sqrt{\log(1/\epsilon)} - \abs{t} \ge \frac{5}{2}
    \sqrt{\log(1/\epsilon)}$ and hence, by a Gaussian tail bound,
    \[
        \Exu{\bm{x} \sim \cN(0, I_n)}{(\alpha_f(\bm{x}) - \overline{\alpha}_f(\bm{x}))^2}
        \le \Exu{\bm{g} \sim \cN(0,1)}{
            (\abs{\bm{g}} + \abs{t})^2
            \cdot \ind{\abs{\bm{g}} \ge R_\epsilon - \abs{t}}
        }
        \le C \log(1/\epsilon) \epsilon^{25/8}
        \le \frac{\epsilon^3}{K} \,.
    \]
    Therefore
    \begin{align*}
        \Exu{\bm{N}}{\Delta(\bm{N})}
        &= \Exu{\bm{x} \sim \cN(0, I_n)}{
            \Exu{\bm{N}}{(\alpha_f(\bm{x}) - \wh{\alpha}_{f,\bm{N}}(\bm{x}))^2}
        } \\
        &\le 2\Exu{\bm{x} \sim \cN(0, I_n)}{(\alpha_f(\bm{x}) - \overline{\alpha}_f(\bm{x}))^2}
        + 2\Exu{\bm{x} \sim \cN(0, I_n)}{
            \Exu{\bm{N}}{(\overline{\alpha}_f(\bm{x}) - \wh{\alpha}_{f,\bm{N}}(\bm{x}))^2}
        } 
        \le \frac{4\epsilon^3}{K} \,.
    \end{align*}
    It follows by Markov's inequality that, with probability at least $99/100$ over $\bm{N}$, the
    first term in the RHS of \eqref{eq:distdis-YES} satisfies $C \Delta(\bm{N}) \le
    \frac{\epsilon^3}{1000}$, and in this case \eqref{eq:distdis-YES-statement} holds.
\end{proof}

\claimYESteststatisticconditions*
\begin{proof}
    The conditions in \cref{item:radius-bound}(a) and \Cref{item:covariance-bound} are trivially satisfied by the
    standard multivariate Gaussian distribution.
It remains to analyze \cref{item:weighted-centroid-bound}. By rotation invariance, we may assume
    that $f(x) = \sign(x_1 - t)$ for some $t \in \bR$. As observed in the proof of
    \cref{lemma:2-norm-tester}, by symmetry we have
    \begin{align*}
        \vabs{\Exu{\bm{y} \sim \cN(0, I_n)}{f(\bm{y}) \wh{\alpha}_{f,N}(\bm{y}) \bm{y}}}_2^2
        &= \Exu{\bm{y} \sim \cN(0, I_n)}{\sign(\bm{y}_1-t) \bm{y}_1 \wh{\alpha}_{f,N}(\bm{y})}^2 \\[1ex]
        &\le \Exu{\bm{y} \sim \cN(0, I_n)}{\bm{y}_1^2 \wh{\alpha}_{f,N}(\bm{y})^2}
        \le R_\epsilon^2 \Exu{\bm{y} \sim \cN(0, I_n)}{\bm{y}_1^2}
        = R_\epsilon^2 \,,
    \end{align*}
    thus satisfying \cref{item:weighted-centroid-bound}.
\end{proof}

\subsection{NO Case}\label{sec:nocase}
We now show that the algorithm rejects with high constant probability in the NO case:

\begin{lemma}
    \label{lemma:no-case}
    Suppose that function $f : \bR^n \to \pmset$ and distribution $\cD$ are such that
    $\dist_\cD(f, h) > \epsilon^*$ for every halfspace $h$. Then \cref{alg:tester} rejects with
    probability at least $3/4$.
\end{lemma}
\begin{proof}
Given that $\dist_\cD(f,h)> \eps^*$ for every halfspace $h$, we have that
$$\Pru{\bm{y} \sim \cD}{f(\bm{y}) = 1} \in [\epsilon^*, 1-\epsilon^*],$$ since
    otherwise there exists a constant function $g$ such that $\dist_\cD(f,g)<  \eps^*$,
    and hence~a halfspace $h$ with $\dist_\cD(f, h) < \epsilon^*$ (obtained by using a
    threshold sufficiently large), a contradiction. Hence 
    \[
        \abs{\Exu{\bm{y} \sim \cD}{f(\bm{y})}}
        = \abs{\Pru{\bm{y} \sim \cD}{f(\bm{y}) = 1} - \Pru{\bm{y} \sim \cD}{f(\bm{y}) = -1}}
        \le 1 - 2\epsilon^* \,.
    \]
    Hence, by \cref{lemma:threshold-estimator}, with probability at least $99/100$,
    \cref{line:threshold} returns a real number instead of $\bot$. 
We assume this is the case and fix $\hat{t}$ to denote the real number returned, satisfying $|\hat{t}|\le T_\eps$. 

    Recall $\cD_*^{\hat{t}}$ as the whittled distribution for threshold $\hat{t}$
    (\cref{def:whittled-distribution}). If $\TV(\cD, \cD_*^{\hat{t}}) > {1}/({Km})$, then
    \cref{line:distribution-tests} rejects with probability at least $99/100$ by
    \cref{lemma:testing-gaussian-versus-far-from-whittled}, and we are done. Therefore, we may assume
 $\smash{\TV(\cD, \cD_*^{\hat{t}}) \le {1}/({Km})}$. This in particular implies that, for every
    halfspace $h$,
    \begin{equation}
        \label{eq:no-dist}
        \dist_{\cD_*^{\hat{t}}}(f, h)
        \ge \dist_{\cD}(f, h) - \TV(\cD, \cD_*^{\hat{t}})
        > \eps^*-1\big/(Km)\ge \epsilon^* - \epsilon
        = \epsilon \,.
    \end{equation}
It is also easy to verify that the total number of samples drawn from $\calD$ in the rest of the algorithm is at most $10m$.
Using $K=1000$ and $\TV(\calD,\calD_*^{\hat{t}})\le 1/(Km)$, we have that, except for probability $1/100$, the rest of the algorithm can be viewed as running on the whittled distribution $\smash{\calD_*^{\hat{t}}}$.

We~write $\calD'$ for $\smash{\calD_*^{\hat{t}}}$, fix $\boldsymbol{N}=N$ to be any neighborhood tuple and consider the following cases:
\begin{enumerate}
\item $\calD'$  violates condition 2 of \cref{lemma:test-statistic} (with $\calD'$ in the role of $\calD$);
\item $\calD'$, with $f$ and $N$, violates the third condition of 
\cref{lemma:test-statistic} (with $\calD'$ in the role of $\calD$);
\item $\calD'$ and   $\wh{\alpha}_{f,\bm{N}}$ violate the condition required by \Cref{lemma:distance-disagreement-no} (with $\calD'$ and $\wh{\alpha}_{f,\bm{N}}$ as $\calD$ and $\beta$);
\item None of these three conditions is violated.
\end{enumerate}
It is clear that case  1 and case 2 would lead the testers of \Cref{lemma:frobenius-tester} and \Cref{lemma:2-norm-tester} to reject with probability at least $99/100$, respectively (because we are taking the view of running them on $\calD'$, which satisfies the $2\sqrt{n}$-radius ball assumption in both of these two lemmas). 

For case 3, we show that  \cref{line:alpha-hat} rejects with probability at least $99/100$.
To see this, given that
    \[
        \Pru{\bm{y} \sim \cD'}{\wh{\alpha}_{f,N}(\bm{y}) < \frac{\epsilon}{4}}
        > \frac{\epsilon}{2} \,,
    \]
    by a Chernoff bound, with probability at least $99/100$ over the choice of $\bm{S}$, we
    have
    \[
        \frac{\abs{\cbra{y \in \bm{S}: \wh{\alpha}_{f,{N}}(y)
        < \epsilon/4}}}{\abs{\bm{S}}} > \frac{3\epsilon}{8} \,,
    \]
    in which case \cref{line:alpha-hat} rejects.

Finally for case 4, 
\cref{lemma:distance-disagreement-no} gives
    (via \eqref{eq:no-dist}) that
\[
        \distdis^{\hat{\bm{t}}}_{\wh{\alpha}_{f,\bm{N}}}(f, \cD') > \frac{\epsilon^3}{32} \,,
    \]
    and \cref{lemma:test-statistic} (whose conditions are all fulfilled by $\cD' = \cD_*^{\hat{t}}$) shows that the estimate $\bZ$ leads the algorithm to reject with probability at least $99/100$.
The lemma follows by a union bound. \end{proof}

\subsection{Putting Everything Together}\label{sec:together}

We conclude our claimed upper bound result.

\begin{proof}[Proof of \Cref{mainthm:upper}]
    The complexity claim is immediate from inspecting \cref{alg:tester}; the dominating operation is
    the computation of $\smash{\wh{\alpha}_{f,\bm{N}}}$ per sample point $\smash{\ith{\bm{y}}{k}}$ in
    \cref{item:line-alpha}, which queries $p = O(1/\epsilon^{28})$ neighbor points for each of the
    $m = O(n^{2/3}/\epsilon^{10})$ sample points.
    The correctness of~\cref{alg:tester} is immediate from \cref{lemma:yes-case,lemma:no-case}.
\end{proof}

\subsection{Estimating the Test Statistic}
\label{sec:test-statistic}

We now prove the main lemma on the concentration of the test statistic $\bm{Z}$, restated below:

\lemmateststatistic*
\begin{proof}
    Write $\bm{Z}(\hat{t}, N)$ for the test statistic conditional on $\hat{\bm{t}} = \hat{t}$ and
    $\bm{N} = N$, and similarly for its constituents $\bm{Z}_1$, $\bm{Z}_2$, and $\bm{Z}_3$. We
    analyze the expected value and variance of these terms.

    \paragraph*{Expected value.} First, recalling the definition of normal vector estimator from
    \cref{def:normal-vector-estimator} and using the independence of $\bm{y}_k, \wh{\bm{w}},
    \wh{\bm{z}}$, from \eqref{eq:Z_1} we have
    \[
        \Ex{\bm{Z}_1(\hat{t}, N)}
        = \Exu{\bm{y} \sim \cD}{\pbra{\abra{\bm{y}, \Ex{\wt{\bm{v}}(\hat{t})}} - \hat{t}}^2}
        = \Exu{\bm{y} \sim \cD}{
            \pbra{f(\bm{y}) \pbra{\abra{\bm{y}, \Ex{\wt{\bm{v}}(\hat{t})}} - \hat{t}}}^2
        } \,,
    \]
    where we used the fact that $f(\cdot)^2 = 1$. Similarly, from \eqref{eq:Z_2} we have
    \[
        \Ex{\bm{Z}_2(\hat{t}, N)}
        = \Exu{\bm{y} \sim \cD}{
            f(\bm{y}) \pbra{\abra{\bm{y}, \Ex{\wt{\bm{v}}(\hat{t})}} - \hat{t}}
            \cdot \wh{\alpha}_{f,N}(\bm{y})
        } \,.
    \]
    Finally, from \eqref{eq:Z_3} we have
    \[
        \Ex{\bm{Z}_3(\hat{t}, N)}
        = \Exu{\bm{y} \sim \cD}{\wh{\alpha}_{f,N}(\bm{y})^2} \,.
    \]
    It follows that
    \begin{align*}
        \Ex{\bm{Z}(\hat{t}, N)}
        &= \Ex{\bm{Z_1}(\hat{t}, N) - 2\bm{Z}_2(\hat{t}, N) + \bm{Z}_3(\hat{t}, N)} \\
        &= \Exu{\bm{y} \sim \cD}{\pbra{
            f(\bm{y}) \pbra{\abra{\bm{y}, \Ex{\wt{\bm{v}}(\hat{t})}} - \hat{t}}
            - \wh{\alpha}_{f,N}(\bm{y})
        }^2}
        = \distdis^{\hat{t}}_{\wh{\alpha}_{f,N}}(f, \cD) \,.
    \end{align*}

    \paragraph*{Variance.} The terms $\bm{Z}_1(\hat{t}, N)$ and $\bm{Z}_2(\hat{t}, N)$ require
    somewhat lengthier calculations, which we defer to standalone claims.

    \begin{restatable}{claim}{claimvarZone}
        \label{claim:var-Z_1}
        In the setting of \cref{lemma:test-statistic}, define $a \define \Exu{\bm{y} \sim
        \cD}{\pbra{G_f^{\hat{t}}(\bm{y}) - \hat{t}}^2}$. Then
        \[
            \Var{\bm{Z}_1(\hat{t}, N)} \le C\pbra{
                \frac{B^2 a}{m}
                + \frac{na}{m^2 \tau(\hat{t})^2}
                + \frac{\sqrt{n} a}{m \tau(\hat{t})^2}
                + \frac{n^2}{m^3 \tau(\hat{t})^4}
                + \frac{n}{m^2 \tau(\hat{t})^4}
            } \,.
        \]
    \end{restatable}

    \begin{restatable}{claim}{claimvarZtwo}
        \label{claim:var-Z_2}
        In the setting of \cref{lemma:test-statistic}, we have
        \[
            \Var{\bm{Z}_2(\hat{t}, N)}
            \le \frac{C R_\epsilon^2}{\epsilon^4} \pbra{\frac{n}{m^2} + \frac{\sqrt{n}}{m}} \,.
        \]
    \end{restatable}

    The variance of the third term is easier to bound. By the independence of the $\bm{y}_k$'s, we
    have
    \[
        \Var{\bm{Z}_3(\hat{t}, N)}
        = \frac{1}{m} \cdot \Varu{\bm{y} \sim \cD}{\wh{\alpha}_{f,N}(\bm{y})^2}
        \le \frac{1}{m} \cdot \Exu{\bm{y} \sim \cD}{\wh{\alpha}_{f,N}(\bm{y})^4}
        \le \frac{R_\epsilon^4}{m} \,.
    \]
    Plugging in $m = \Theta(n^{2/3} / \epsilon^{10})$ with sufficiently large constant, as well as
    the lower bound $\tau(\hat{t}) \ge \sqrt{2/\pi} \cdot \epsilon^2$ from
    \cref{lemma:tau-lower-bound} and the assumption that $n\ge n_0$ and $\eps\ge 1/n$, we obtain
    \begin{equation}
        \label{eq:vars}
        \Var{\bm{Z}_1(\hat{t}, N)}, \Var{\bm{Z}_2(\hat{t}, N)}, \Var{\bm{Z}_3(\hat{t}, N)}
        \le \frac{(a+1) \epsilon^6}{L R_\epsilon^2} \,.
    \end{equation}
    where the constant $L$ can be made as large as desired.

    \paragraph*{Concentration using the variance bounds.} Recall that $\bm{Z}(\hat{t}, N) = \bm{Z}_1(\hat{t}, N) -
    2\bm{Z}_2(\hat{t}, N) + \bm{Z}_3(\hat{t}, N)$. We consider two cases. First, suppose $a \le a_0
    \define 16 R_\epsilon^2$. Then the RHS of \eqref{eq:vars} is at most $17 \epsilon^6 / L$ and
    hence three applications of Chebyshev's inequality together with the union bound and triangle
    inequality yield that, with probability at least $99/100$, we have
    \[
        \abs{\bm{Z}(\hat{t}, N) - \Ex{\bm{Z}(\hat{t}, N)}} \le \frac{\epsilon^3}{128} \,,
    \]
    which gives the conclusion.

    Second, suppose $a > a_0$. Then we claim that $\bm{Z}(\hat{t}, N)$ has large expectation.
    Indeed, we have $\bm{Z}_3(\hat{t}, N) \ge 0$, while
    the inequality $2xy \le \frac{1}{4}x^2 + 4y^2$
gives
    \[
        2\Ex{\bm{Z}_2(\hat{t}, N)}
        \le \frac{1}{4} \Ex{\pbra{\abra{\bm{y}, \Ex{\wt{\bm{v}}(\hat{t})}} - \hat{t}}^2}
            + 4 \Ex{\wh{\alpha}_{f,N}(\bm{y})^2}
        \le \frac{a}{4} + 4 R_\epsilon^2 \,,
    \]
    so that
    \[
        \Ex{\bm{Z}(\hat{t}, N)}
        \ge \frac{3a}{4} - 4 R_\epsilon^2
        = \frac{3a}{4} - \frac{a_0}{4}
        > \frac{a}{2} \,.
    \]
    Moreover, in this case the RHS of \eqref{eq:vars} is at most $a^2/L$, so that another
    application of Chebyshev's and triangle inequalities via a union bound gives that, with
    probability at least $99/100$,
    \[
        \abs{\bm{Z}(\hat{t}, N) - \Ex{\bm{Z}(\hat{t}, N)}}
        \le \frac{a}{4}
        < \frac{\Ex{\bm{Z}(\hat{t}, N)}}{2} \,,
    \]
    which also gives the conclusion.
\end{proof}

We now prove the two variance claims. We start with the second one, which is simpler.

\claimvarZtwo*
\begin{proof}
    Write $\overline{\bm{y}} \define (\ith{\bm{y}}{1}, \dotsc, \ith{\bm{y}}{m})$, and similarly for
    $\overline{\bm{w}}$. Define
    \[
        \mu(f, N, \overline{y})
        \define \frac{1}{m} \sum_{k=1}^{m}
            f(\ith{y}{k}) \ith{y}{k} \wh{\alpha}_{f,N}(\ith{y}{k}) \,.
    \]
    By the law of total variance,
    \[
        \Var{\bm{Z}_2(\hat{t}, N)}
        = \Exu{\overline{\bm{y}}}{\Varuc{\wh{\bm{w}}}{\bm{Z}_2(\hat{t}, N)}{\overline{\bm{y}}}}
            + \Varu{\overline{\bm{y}}}{\Exuc{\wh{\bm{w}}}{\bm{Z}_2(\hat{t}, N)}{\overline{\bm{y}}}}
        \,.
    \]
    The first term is
    \begin{align*}
        &\Exu{\overline{\bm{y}}}{\Varuc{\wh{\bm{w}}}{\bm{Z}_2(\hat{t}, N)}{\overline{\bm{y}}}} \\
        &= \Exu{\overline{\bm{y}}}{\Varuc{\wh{\bm{w}}}{
            \frac{1}{m} \sum_{k=1}^{m}
            f(\ith{\bm{y}}{k}) \pbra{\abra{\ith{\bm{y}}{k}, \wh{\bm{w}}} - \hat{t}}
            \wh{\alpha}_{f,N}(\ith{\bm{y}}{k})
        }{\overline{\bm{y}}}} \\
        &= \Exu{\overline{\bm{y}}}{\Varuc{\wh{\bm{w}}}{
            \frac{1}{m} \sum_{k=1}^{m}
            f(\ith{\bm{y}}{k}) \abra{\ith{\bm{y}}{k}, \wh{\bm{w}}}
            \wh{\alpha}_{f,N}(\ith{\bm{y}}{k})
        }{\overline{\bm{y}}}}
        \tag{The subtracted term is constant conditional on $\overline{\bm{y}}$}
        \\
        &= \Exu{\overline{\bm{y}}}{\Varuc{\wh{\bm{w}}}{
            \abra{
                \mu(f, N, \overline{\bm{y}}),
                \wh{\bm{w}}
            }
        }{\overline{\bm{y}}}} \\
        &= \Exu{\overline{\bm{y}}}{\Varuc{\overline{\bm{w}}}{
            \abra{
                \mu(f, N, \overline{\bm{y}}),
                \frac{1}{\tau(\hat{t})} \cdot
                    \frac{1}{m} \sum_{i=1}^m \ith{\bm{w}}{i} f(\ith{\bm{w}}{i})
            }
        }{\overline{\bm{y}}}} \\
        &= \frac{1}{m \tau(\hat{t})^2} \cdot
            \Exu{\overline{\bm{y}}}{\Varuc{\bm{w} \sim \cN(0, I_n)}{
                \abra{\mu(f, N, \overline{\bm{y}}), \bm{w} f(\bm{w})}
            }{\overline{\bm{y}}}}
            \tag{$\ith{\bm{w}}{i}$'s are \iid}
            \\
        &\le \frac{1}{m \tau(\hat{t})^2} \cdot
            \Exu{\overline{\bm{y}}}{\Exuc{\bm{w} \sim \cN(0, I_n)}{
                \abra{\mu(f, N, \overline{\bm{y}}), \bm{w} f(\bm{w})}^2
            }{\overline{\bm{y}}}} \\
        &= \frac{1}{m \tau(\hat{t})^2} \cdot
            \Exu{\overline{\bm{y}}}{\Exuc{\bm{w} \sim \cN(0, I_n)}{
                \abra{\mu(f, N, \overline{\bm{y}}), \bm{w}}^2
            }{\overline{\bm{y}}}}
            \tag{Since $f(\cdot)^2$ = 1}
            \\
        &= \frac{1}{m \tau(\hat{t})^2} \cdot
            \Exu{\overline{\bm{y}}}{\vabs{\mu(f, N, \overline{\bm{y}})}_2^2}
            \tag{Expectation of scaled $\chi^2_1$} \\
        &= \frac{1}{m \tau(\hat{t})^2} \cdot \Exu{\overline{\bm{y}}}{\abra{
            \frac{1}{m} \sum_{k=1}^{m}
                f(\ith{\bm{y}}{k}) \ith{\bm{y}}{k} \wh{\alpha}_{f,N}(\ith{\bm{y}}{k}),
            \frac{1}{m} \sum_{k'=1}^{m}
                f(\ith{\bm{y}}{k'}) \ith{\bm{y}}{k'} \wh{\alpha}_{f,N}(\ith{\bm{y}}{k'})
        }} \\
        &= \frac{1}{m^3 \tau(\hat{t})^2} \pbra{
            m \Exu{\bm{y} \sim \cD}{\vabs{f(\bm{y}) \bm{y} \wh{\alpha}_{f,N}(\bm{y})}_2^2}
            + m(m-1) \Exu{\bm{y}, \bm{y}' \sim \cD}{\abra{
                f(\bm{y}) \bm{y} \wh{\alpha}_{f,N}(\bm{y}),
                f(\bm{y}') \bm{y}' \wh{\alpha}_{f,N}(\bm{y}')
            }}
        } \\
        &= \frac{1}{m^3 \tau(\hat{t})^2} \pbra{
            m \Exu{\bm{y} \sim \cD}{\vabs{\bm{y} \wh{\alpha}_{f,N}(\bm{y})}_2^2}
            + m(m-1) \vabs{
                \Exu{\bm{y} \sim \cD}{f(\bm{y}) \bm{y} \wh{\alpha}_{f,N}(\bm{y})}
            }_2^2
        } \\
        &\le \frac{C}{m^3 \tau(\hat{t})^2} \pbra{
            m \cdot R_\epsilon^2 n
            + m(m-1) (\sqrt{n} + R_\epsilon^2)
        }
        \tag{\cref{item:radius-bound,item:weighted-centroid-bound} of \cref{lemma:test-statistic},
        bound on $\wh{\alpha}_{f,N}$}
        \\
        &\le \frac{C R_\epsilon^2}{\epsilon^4} \pbra{
            \frac{n}{m^2} + \frac{\sqrt{n}}{m}
        } \,. \tag{\cref{lemma:tau-lower-bound} and $\eps\ge 1/n$}
    \end{align*}
    For the second term, we first note the bound $\Exu{\bm{y} \sim \cD}{G_f^{\hat{t}}(\bm{y})^2} \le
    B^2$, which holds by hypothesis in the bounded case of \cref{item:band-distance-bound}, and
    by a Gaussian moment bound via \cref{cor:yes-ex-v-small} in the YES case. Now, the second term
    is
    \begin{align*}
        &\Varu{\overline{\bm{y}}}{\Exuc{\wh{\bm{w}}}{\bm{Z}_2(\hat{t}, N)}{\overline{\bm{y}}}} \\
        &= \Varu{\overline{\bm{y}}}{\Exuc{\wh{\bm{w}}}{
            \frac{1}{m} \sum_{k=1}^{m}
            f(\ith{\bm{y}}{k}) \pbra{\abra{\ith{\bm{y}}{k}, \wh{\bm{w}}} - \hat{t}}
            \wh{\alpha}_{f,N}(\ith{\bm{y}}{k})
        }{\overline{\bm{y}}}} \\
        &= \Varu{\overline{\bm{y}}}{
            \frac{1}{m} \sum_{k=1}^{m}
            f(\ith{\bm{y}}{k}) \pbra{\abra{\ith{\bm{y}}{k}, \Ex{\wh{\bm{w}}}} - \hat{t}}
            \wh{\alpha}_{f,N}(\ith{\bm{y}}{k})
        } \\
        &= \Varu{\overline{\bm{y}}}{
            \frac{1}{m} \sum_{k=1}^{m}
            f(\ith{\bm{y}}{k}) \pbra{G_f^{\hat{t}}(\ith{\bm{y}}{k}) - \hat{t}}
            \wh{\alpha}_{f,N}(\ith{\bm{y}}{k})
        }
        \tag{\cref{def:whittled-distribution}} \\
        &= \frac{1}{m} \cdot \Varu{\bm{y} \sim \cD}{
            f(\bm{y}) \pbra{G_f^{\hat{t}}(\bm{y}) - \hat{t}} \wh{\alpha}_{f,N}(\bm{y})
        }
        \tag{$\ith{\bm{y}}{k}$'s are \iid} \\
        &\le \frac{1}{m} \cdot \Exu{\bm{y} \sim \cD}{
            \pbra{G_f^{\hat{t}}(\bm{y}) - \hat{t}}^2 \wh{\alpha}_{f,N}(\bm{y})^2
        } \\
        &\le \frac{2 R_\epsilon^2}{m} \pbra{
            \Exu{\bm{y} \sim \cD}{G_f^{\hat{t}}(\bm{y})^2}
            + T_\epsilon^2
        }
        \tag{Bound on $\wh{\alpha}_{f,N}$} \\
        &\le \frac{2 R_\epsilon^2}{m} \pbra{B^2 + T_\epsilon^2} \,.
        \tag{Bound above} \\
        &\le \frac{C R_\epsilon^2}{m} \cdot \frac{\sqrt{n}}{\epsilon^2} \,.
    \end{align*}
    The conclusion follows.
\end{proof}

\claimvarZone*
\begin{proof}
    For convenience of notation, let
    \[
        v \define \Ex{\wt{\bm{v}}(\hat{t})} \,, \quad
        h(y) \define \abra{y,v} - \hat{t} = G_f^{\hat{t}}(y) - \hat{t} \,.
    \]
    Define random variables
    \[
        \bm{w}_{\err} \define \wh{\bm{w}} - v \,, \quad
        \bm{z}_{\err} \define \wh{\bm{z}} - v \,,
    \]
    which are centered and independent, identically distributed, and also independent of the
    $\ith{\bm{y}}{k}$'s. Write $\Sigma \define \Cov{\bm{w}_{\err}} = \Cov{\bm{z}_{\err}}$. Then
    \begin{equation}
        \label{eq:err-cov}
        \Sigma
        = \frac{1}{m \tau(\hat{t})^2} \Covu{\bm{x} \sim \cN(0, I_n)}{\bm{x} f(\bm{x})}
        \preceq \frac{1}{m \tau(\hat{t})^2} \Ex{\bm{x} \bm{x}^\top f(\bm{x})^2}
        = \frac{I_n}{m \tau(\hat{t})^2} \,,
    \end{equation}
    where the first inequality is in the PSD order.

    We first establish the following three moment bounds:
    \begin{equation}
        \label{eq:moment-bounds}
        \Exu{\bm{y} \sim \cD}{h(\bm{y})^4} \le C B^2 a \,, \quad
        \Exu{\bm{y} \sim \cD}{h(\bm{y})^2 \vabs{\bm{y}}_2^2} \le Cna \,, \quad \text{and} \quad
        \Exu{\bm{y} \sim \cD}{\vabs{\bm{y}}_2^4} \le Cn^2 \,.
    \end{equation}
    In the bounded case of \cref{item:radius-bound}, the third bound is immediate from the uniform
    bound on $\vabs{\bm{y}}_2$, the first bound follows via
    \[
        h(\bm{y})^2
        \le \pbra{\abs{G_f^{\hat{t}}(\bm{y})} + \abs{\hat{t}}}^2
        \le 2B^2 + 2T_\epsilon^2
        \le CB^2
    \]
    along with the definition of $a = \Ex{h(\bm{y})^2}$, and the second bound follows from combining
    the bound on $\vabs{\bm{y}}_2$ and the definition of $a$.

    In the YES case, using that $\bm{y}$ is standard Gaussian and \cref{cor:yes-ex-v-small}, we have
    \[
        a = \Exu{\bm{y} \sim \cD}{\pbra{\abra{\bm{y},v} - \hat{t}}^2}
        = \vabs{v}_2^2 + \hat{t}^2
        \le 3 + T_\epsilon^2
        \le C B^2 \,.
    \]
    Now, the first bound follows by
    \[
        \Exu{\bm{y} \sim \cD}{h(\bm{y})^4}
        \le C\pbra{\Exu{\bm{y} \sim \cD}{\abra{\bm{y},v}^4} + \hat{t}^4}
        \le C\pbra{\vabs{v}_2^4 + \hat{t}^4}
        \le C\pbra{\vabs{v}_2^2 + \hat{t}^2}^2
        = Ca^2
        \le C B^2 a \,,
    \]
    the third bound is a standard multivariate Gaussian moment, and the second bound follows from
    the third and the penultimate step above via the Cauchy-Schwarz inequality.

    We now proceed with the variance calculation. Let $\overline{\bm{y}} \define (\ith{\bm{y}}{1},
    \dotsc, \ith{\bm{y}}{m})$, and define
    \begin{align*}
        R(\overline{y}) &\define \frac{1}{m} \sum_{k=1}^m h(\ith{y}{k})^2 \,, \\
        q(\overline{y}) &\define \frac{1}{m} \sum_{k=1}^m h(\ith{y}{k}) \ith{y}{k} \,, \text{ and}
        \\
        M(\overline{y}) &\define \frac{1}{m} \sum_{k=1}^m \ith{y}{k} (\ith{y}{k})^\top \,.
    \end{align*}
    It is helpful to use the decompositions $\wh{\bm{w}} = v + \bm{w}_{\err}$ and $\wh{\bm{z}} = v +
    \bm{z}_{\err}$ to write
    \[
        \bm{Z}_1(\hat{t}, N)
        = \frac{1}{m} \sum_{k=1}^m
            \pbra{\abra{\ith{\bm{y}}{k}, \wh{\bm{w}}} - \hat{t}}
            \pbra{\abra{\ith{\bm{y}}{k}, \wh{\bm{z}}} - \hat{t}}
        = R(\overline{\bm{y}}) + \abra{q(\overline{\bm{y}}), \bm{w}_{\err} + \bm{z}_{\err}}
            + \bm{w}_{\err}^\top M(\overline{\bm{y}}) \bm{z}_{\err} \,.
    \]
    Now, by the law of total variance,
    \begin{equation}
        \label{eq:z1-total-variance}
        \Var{\bm{Z}_1(\hat{t}, N)}
        = \Exu{\overline{\bm{y}}}{
            \Varuc{\wh{\bm{w}},\wh{\bm{z}}}{\bm{Z}_1(\hat{t}, N)}{\overline{\bm{y}}}
        }
        + \Varu{\overline{\bm{y}}}{
            \Exuc{\wh{\bm{w}},\wh{\bm{z}}}{\bm{Z}_1(\hat{t}, N)}{\overline{\bm{y}}}
        }
        \,.
    \end{equation}
    The second term is easier to handle. Since $\bm{w}_{\err}$ and $\bm{z}_{\err}$ are centered,
    while the $\ith{\bm{y}}{k}$'s are independent, we have
    \begin{equation}
        \label{eq:z1-tot-second}
        \Varu{\overline{\bm{y}}}{
            \Exuc{\wh{\bm{w}},\wh{\bm{z}}}{\bm{Z}_1(\hat{t}, N)}{\overline{\bm{y}}}
        }
        = \Varu{\overline{\bm{y}}}{R(\overline{\bm{y}})}
        = \frac{1}{m} \Varu{\bm{y} \sim \cD}{h(\bm{y})^2}
        \le \frac{1}{m} \Exu{\bm{y} \sim \cD}{h(\bm{y})^4}
        \le \frac{C B^2 a}{m} \,,
    \end{equation}
    where the last inequality is by \eqref{eq:moment-bounds}. We now turn to the first term. The
    $R(\overline{\bm{y}})$ term vanishes in the conditional variance, and we have
    \[
        \Exu{\overline{\bm{y}}}{
            \Varuc{\wh{\bm{w}},\wh{\bm{z}}}{\bm{Z}_1(\hat{t}, N)}{\overline{\bm{y}}}
        }
        = \Exu{\overline{\bm{y}}}{
            \Varuc{\wh{\bm{w}},\wh{\bm{z}}}{
                \abra{q(\overline{\bm{y}}), \bm{w}_{\err} + \bm{z}_{\err}}
                + \bm{w}_{\err}^\top M(\overline{\bm{y}}) \bm{z}_{\err}
            }{\overline{\bm{y}}}
        }
    \]
    The two summands above are uncorrelated, because each term in the expansion of their product
    contains one of the centered random variables $\bm{w}_{\err}$ or $\bm{z}_{\err}$ exactly once.
    Thus we have
    \begin{equation}
        \label{eq:z1-tot-first}
        \Exu{\overline{\bm{y}}}{
            \Varuc{\wh{\bm{w}},\wh{\bm{z}}}{\bm{Z}_1(\hat{t}, N)}{\overline{\bm{y}}}
        }
        = \Exu{\overline{\bm{y}}}{
            \Varuc{\wh{\bm{w}},\wh{\bm{z}}}{
                \abra{q(\overline{\bm{y}}), \bm{w}_{\err} + \bm{z}_{\err}}
            }{\overline{\bm{y}}}
        } + \Exu{\overline{\bm{y}}}{
            \Varuc{\wh{\bm{w}},\wh{\bm{z}}}{
                \bm{w}_{\err}^\top M(\overline{\bm{y}}) \bm{z}_{\err}
            }{\overline{\bm{y}}}
        } \,.
    \end{equation}
    We control each term separately. First, by the independence of $\wh{\bm{w}}$ and $\wh{\bm{z}}$
    along with \eqref{eq:err-cov}, we have
    \[
        \Varuc{\wh{\bm{w}},\wh{\bm{z}}}{
            \abra{q(\overline{\bm{y}}), \bm{w}_{\err} + \bm{z}_{\err}}
        }{\overline{\bm{y}}}
        = q(\overline{\bm{y}})^\top (\Cov{\bm{w}_{\err}}+\Cov{\bm{z}_{\err}}) q(\overline{\bm{y}})
        \le \frac{2\vabs{q(\overline{\bm{y}})}_2^2}{m \tau(\hat{t})^2} \,,
    \]
    and moreover
    \[
        \Ex{\vabs{q(\overline{\bm{y}})}_2^2}
        = \frac{1}{m} \Exu{\bm{y} \sim \cD}{h(\bm{y})^2 \vabs{\bm{y}}_2^2}
            + \frac{m-1}{m} \vabs{\Exu{\bm{y} \sim \cD}{h(\bm{y}) \bm{y}}}_2^2 \,.
    \]
    Using \eqref{eq:moment-bounds}, the first term above is
    \[
        \frac{1}{m} \Exu{\bm{y} \sim \cD}{h(\bm{y})^2 \vabs{\bm{y}}_2^2}
        \le \frac{Cna}{m} \,.
    \]
    For the second term, use the variational characterization of $\ell_2$ norm to write
    \begin{align*}
        \vabs{\Exu{\bm{y} \sim \cD}{h(\bm{y}) \bm{y}}}_2^2
        &= \sup_{\vabs{u}_2=1} \abra{u, \Ex{h(\bm{y}) \bm{y}}}^2 \\
        &= \sup_{\vabs{u}_2=1} \Ex{h(\bm{y}) \abra{u, \bm{y}}}^2 \\
        &\le \sup_{\vabs{u}_2=1} \Ex{h(\bm{y})^2} \Ex{\abra{u, \bm{y}}^2}
            \tag{Cauchy-Schwarz} \\
        &\le a \sup_{\vabs{u}_2=1} u^\top \Ex{\bm{y} \bm{y}^\top} u
            \tag{Definition of $a$} \\
        &= a \vabs{\Ex{\bm{y} \bm{y}^\top}}_{\mathrm{op}} \\
        &\le a \vabs{\Ex{\bm{y} \bm{y}^\top}}_F \\
        &\le 2a\sqrt{n}
            \tag{Frobenius bound of \cref{lemma:test-statistic}} \,.
    \end{align*}
    Therefore
    \[
        \Ex{\vabs{q(\overline{\bm{y}})}_2^2} \le C\pbra{\frac{na}{m} + a\sqrt{n}} \,,
    \]
    so that
    \begin{equation}
        \label{eq:linear-term}
        \Exu{\overline{\bm{y}}}{
            \Varuc{\wh{\bm{w}},\wh{\bm{z}}}{
                \abra{q(\overline{\bm{y}}), \bm{w}_{\err} + \bm{z}_{\err}}
            }{\overline{\bm{y}}}
        }
        \le C\pbra{\frac{na}{m^2 \tau(\hat{t})^2} + \frac{a\sqrt{n}}{m \tau(\hat{t})^2}} \,.
    \end{equation}
    It remains to control the term
    $\Exu{\overline{\bm{y}}}{\Varuc{\wh{\bm{w}},\wh{\bm{z}}}{\bm{w}_{\err}^\top M(\overline{\bm{y}})
    \bm{z}_{\err} }{\overline{\bm{y}}}}$. For fixed $\overline{y}$, we have
    \begin{align*}
        \Varu{\wh{\bm{w}},\wh{\bm{z}}}{
            \bm{w}_{\err}^\top M(\overline{y}) \bm{z}_{\err}
        }
        &= \Exu{\wh{\bm{w}},\wh{\bm{z}}}{
            \pbra{\bm{w}_{\err}^\top M(\overline{y}) \bm{z}_{\err}}^2
        } \\
        &= \Exu{\wh{\bm{w}},\wh{\bm{z}}}{
            \Tr\pbra{
                \bm{w}_{\err}^\top M(\overline{y}) \bm{z}_{\err}
                \bm{z}_{\err}^\top M(\overline{y})^\top \bm{w}_{\err}
            }
        } \\
        &= \Tr\pbra{
            M(\overline{y}) \Sigma M(\overline{y})^\top \Sigma
        }
        = \vabs{\Sigma^{1/2} M(\overline{y}) \Sigma^{1/2}}_F^2
        \le \vabs{\Sigma}_{\mathrm{op}}^2 \vabs{M(\overline{y})}_F^2
        \le \frac{\vabs{M(\overline{y})}_F^2}{m^2 \tau(\hat{t})^4} \,,
    \end{align*}
    the last step by \eqref{eq:err-cov}. Finally, we have
    \[
        \Ex{\vabs{M(\overline{\bm{y}})}_F^2}
        = \Ex{\Tr\pbra{M(\overline{\bm{y}})^\top M(\overline{\bm{y}})}}
        = \frac{1}{m} \Exu{\bm{y} \sim \cD}{\vabs{\bm{y}}_2^4}
            + \frac{m-1}{m} \vabs{\Exu{\bm{y} \sim \cD}{\bm{y} \bm{y}^\top}}_F^2
        \le C\pbra{\frac{n^2}{m} + n} \,,
    \]
    the last step by the Frobenius bound of \cref{lemma:test-statistic} and
    \eqref{eq:moment-bounds}. Therefore
    \begin{equation}
        \label{eq:bilinear-term}
        \Exu{\overline{\bm{y}}}{
            \Varuc{\wh{\bm{w}},\wh{\bm{z}}}{
                \bm{w}_{\err}^\top M(\overline{\bm{y}}) \bm{z}_{\err}
            }{\overline{\bm{y}}}
        }
        \le C\pbra{\frac{n^2}{m^3 \tau(\hat{t})^4} + \frac{n}{m^2 \tau(\hat{t})^4}} \,.
    \end{equation}
    Plugging \cref{eq:z1-tot-second,eq:z1-tot-first,eq:linear-term,eq:bilinear-term} into
    \eqref{eq:z1-total-variance} concludes the proof.
\end{proof}

\section{Lower Bound}
\label{sec:lb}

Our goal is to prove the following lower bound on distribution-picky testing (restated from the
introduction):

\thmintrolbmain*

\subsection{Construction}

\newcommand{\samp}{\mathsf{samp}}
\begin{definition}[Meta-Distributions]
For any domain $\cX$, a \emph{meta-distribution} $\mathfrak D$ over $\cX$ is a
probability distribution where random variable $\cD \sim \mathfrak D$ is a probability distribution
over $\cX \times \pmset$. For any distribution $\cD$ over $\cX \times \pmset$,
we write $\cD_\cX$ for its marginal over $\cX$.

For a given meta-distribution $\mathfrak D$ and a sample-size parameter $m \in
\bN$, we define $\samp^m(\mathfrak D)$ as the distribution over $m$ labelled
$(\bm z^{(1)}, \dotsc, \bm z^{(m)}) \sim (\cX \times \pmset)^m$ obtained by
choosing $\cD \sim \mathfrak D$ and then letting each $\bm z^{(i)} \sim \cD$ be
independently drawn from $\cD$.
\end{definition}

We will construct two meta-distributions $\DYES$ and $\DNO$ over $\bR^n$ as follows. 
$\DYES$ is straightforward:
\begin{definition}[YES Distribution]
To obtain $\cD \sim \DYES$,
\begin{enumerate}
\item Choose $\bm v \sim \cN(0, \tfrac{1}{n} I_n)$.
\item Let $\cD$ be the distribution over $\bR^n \times \pmset$ where a sample $(\bm x, \bm b) \sim
\cD$ is distributed with $\bm x \sim \cN(0, I_n)$ and $\bm b = \sign\inn{\bm x, \bm v}$.
\end{enumerate}
\end{definition}
A $\cN(0,I)$-picky tester should $\Accept$ when given $m$ samples from $\cD \sim \DYES$, since the
marginal over $\bR^n$ is the standard Gaussian and the labels are given by a halfspace.  Next, we
define $\DNO$.  Now the underlying probability distribution over $\bR^n$ will be the \emph{spiked
Gaussian}.

\newcommand{\Spike}{\mathsf{Spike}}
\begin{definition}[Spiked Gaussian]
Fix any dimension $n$ and parameter $\theta \geq 0$. Define
$\Spike_\theta$ as a distribution over probability distributions over $\bR^n$ as follows. To
obtain $\cS \sim \Spike_\theta$:
\begin{enumerate}
\item Choose vector $\bm v \sim \cN(0, \tfrac{1}{n} I_n)$.
\item Let $\cS$ be the distribution $\cN(0, I_n + \theta \bm v \bm v^\top )$.
\end{enumerate}
\end{definition}

Then $\DNO$ is defined as:
\begin{definition}[NO Distribution]
To obtain $\cD \sim \DNO$:
\begin{enumerate}
\item Choose $\bm v \sim \cN(0, \tfrac{1}{n} I_n)$.
\item Let $\cD$ be the distribution over $\bR^n \times \pmset$ where a sample $(\bm x, \bm b) \sim
\cD$ is distributed with:
\begin{enumerate}
    \item  $\bm x \sim \cN(0, I_n + \theta \bm v \bm v^\top)$;
    \item  $\bm b$ is obtained by choosing $\bm z \sim \cN(0, 1)$ and setting
        $$\bm b = \sign\left( \frac{\rho}{\sqrt{1+\theta}} \bm v^\top x + \sqrt{1-\rho^2} \bm z
            \right), \quad\text{where}\ 
            \rho \define \frac{1}{\sqrt{1+\theta}} .$$
\end{enumerate}
\end{enumerate}
\end{definition}

The distribution defined above is not sufficient because it randomly labels each domain element
instead of labeling according to a measurable function $f$. Since we are only concerned with sample-based testers, this is fixed by a basic
discretization. We will handle this along with showing that the NO distribution is
far from all halfspaces in the appendix, since the arguments are standard:

\begin{restatable}{lemma}{lemmalbdistance}
\label{lem:lb-distance}
For all constants $\theta \in (0,1)$, there exists a proximity parameter
$\epsilon > 0$ such that, for all $m \in \bN$, there exists a meta-distribution
$\DNO^\square$  such that:
\begin{enumerate}
    \item For every $\cD$ in the support of $\DNO^\square$, there exists a measurable function $f \colon \bR^n \to \pmset$ such that, for $(\bm x, \bm b) \sim \cD$,
    we have $\bm b = f(\bm x)$.
    \item With probability at least $99/100$ over $\cD \sim \DNO^\square$, the
    function $f$ obtained above satisfies
    \[
        \dist_{\cD_{\bR^n}}(f, h) > \epsilon
    \]
    for all halfspaces $h$.
    \item $\TV(\samp^m(\DNO), \samp^m(\DNO^\square)) = o(1)$.
\end{enumerate}
\end{restatable}
\begin{proof}
    See \cref{section:lb-discretization}.
\end{proof}

Also note that distributions sampled from the spiked Gaussian may
have largest eigenvalue larger than $1+\theta$, since $\bm{v}$ is not a unit
vector in general. However, $\bm{v}$ has bounded norm with high constant
probability, so the distribution $\cD \sim \DNO$ has covariance matrix
$\bm{\Sigma}$ satisfying, say, $\lambda_{\max}(\bm{\Sigma}) \le
1 + \theta\vabs{v}^2 \le 1 + 100 \theta$
with probability at least $99/100$. Moreover, the discretization operates on the
labels without changing the marginal of $\cD$. Therefore, since we can choose
$\theta$ as small as desired, this is sufficient.

Now, the following lemma suffices to prove \cref{thm:intro-lb-main}.

\begin{lemma}
\label{lemma:lb-main}
For all sufficiently small constant $0 < \theta < 1$,
for sufficiently large $n$ and for $m = o(n^{2/3})$,
\[
    \TV\big( \samp^m(\DYES), \samp^m(\DNO)\big ) = o(1) .
\]
\end{lemma}

The remainder of the section proves this lemma.

\subsection{Indistinguishability of $\DYES$ and $\DNO$}

To prove \cref{lemma:lb-main} we define the following random variables, which we will use for the
remainder of the section.
\begin{itemize}
    \item $\vyes \sim \cN(0, \tfrac{1}{n} I_n)$ is a Gaussian normalized so that $\Ex{\|\vyes\|^2}=1$;
    \item $\Xyes \in \bR^{m \times n}$ is the random matrix where each row $i \in [m]$ is an
            independent $\bm x_i \sim \cN(0,I_n)$;
    \item $\yyes \define \Xyes \vyes$.
    \item $\vno \sim \cN(0, \tfrac{1}{n} I_n)$;
    \item $\Xno \in \bR^{m \times n}$ is the random matrix where each row $i \in [m]$ is an
            independent $\bm x_i \sim \cN(0, I_n + \theta \vno \vno^\top)$;
    \item $\yno \define \Xno \vno$;
    \item $\yno^\rho \define \rho \frac{1}{\sqrt{1+\theta}} \yno + \sqrt{1-\rho^2} \bm z$
        for independent $\bm z \sim \cN(0,I_m)$.
\end{itemize}

With these variables, the proof of \cref{lemma:lb-main} is as follows, with the necessary claims
being proved in the section below:

\begin{proof}[Proof of \cref{lemma:lb-main}]
\sloppy
The distribution $\samp^m(\DYES)$ is equivalent to the joint distribution 
$(\Xyes, \sign(\yyes))$ where $\sign(\yyes) \in \pmset^m$ is taken entrywise. The distribution
$\samp^m(\DNO)$ is equivalent to the joint distribution
$(\Xno, \sign(\yno^\rho))$.  Our goal is to prove
\begin{equation}
\label{eq:lb-goal}
    \TV\big( (\Xyes, \sign(\yyes)), (\Xno, \sign(\yno^\rho)) \big) = o(1) .
\end{equation}
We first show (in \cref{prop:Xyes-vs-Xno} in \Cref{sec:delayed1}) that
\begin{equation}
\label{eq:lb-X-TV}
\TV( \Xyes, \Xno ) = o(1) .
\end{equation}
Next we will condition on $\Xyes = \Xno = X$ where $X$ has a ``good'' structure.  Given an $X \in \bR^{m
\times n}$, we let $X = U\Sigma V^\top$ be its singular-value decomposition  with singular values $\Sigma =
\diag(\sigma_1, \dotsc, \sigma_m)$.  We say $X \in \bR^{m \times n}$ is \emph{good} if for every $i\in [m]$:
\begin{equation}
\label{eq:lb-good-X}
\sigma_i^2 = n(1+\delta_i), \quad\text{for some $\delta_i$ satisfying}\  |\delta_i| \leq
K\sqrt{m/n} ,
\end{equation}
where $K > 0$ is a constant that is fixed in \cref{prop:lb-good-X}.  In \cref{prop:lb-good-X} in \Cref{sec:delayed2}, we establish that
$\Xyes$ is good with probability $1-o(1)$, and the same holds for $\Xno$ by \cref{eq:lb-X-TV}.

We now consider the distributions of $\yyes, \yno$ and $\yno^\rho$ conditional on $\Xyes = \Xno = X$ for
good $X$. Let $\yyes\mid_X, \yno\mid_X, \yno^\rho\mid_X$ denote these variables. It suffices to
prove that
\begin{equation}
    \TV( \yyes\mid_X, \yno^\rho \mid_X ) = o(1) .
\end{equation}
Since $\vyes \sim \cN(0, \tfrac{1}{n} I_n)$ independently of $\Xyes$, the variable $(\yyes\mid_X) =
X\vyes$ is itself Gaussian. Let $X = U\Sigma V^\top$ be the singular value decomposition. Then
\[
    (\yyes \mid_X) \sim \cN(0, \tfrac{1}{n} U\Sigma^2 U^\top),
    \quad\text{where}\quad 
    \tfrac{1}{n} U\Sigma^2 U^\top = U\diag(1+\delta_i)_{i \in [m]} U^\top.
\]
We must compare $\yno\mid_X$ and $\yno^\rho\mid_X$ to this Gaussian. Consider
the distribution of $\vno \mid_X$ conditional on $\Xno = X$. If $\vno \mid_X$ was Gaussian, then the
random variable $(\yno \mid_X) = X (\vno\mid_X)$ would also be Gaussian, and it would be easy to compare to
$\yyes\mid_X$, but this is not the case: let $\qno(v \mid X)$ be the PDF of $v \in \bR^n$ according
to $\yno\mid_X$. We calculate in \cref{prop:lb-vno-density} in \Cref{sec:delayed3} that
\[
    \qno(v \mid X) \propto \frac{1}{(1+\theta \|v\|^2)^{m/2}}
        \exp\left(\frac{\theta}{2(1+\theta\|v\|^2)} \|Xv\|^2\right) \gamma(v) \,,
\]
where $\gamma(v)$ is the PDF of $\cN(0, \tfrac{1}{n} I_n)$ and the notation $\propto$ hides a
normalizing factor depending on $X$ but not on $v$. This is not a Gaussian density, but
using the distribution $\vno \sim \cN(0, \tfrac{1}{n}I_n)$
to justify the approximation $\smash{\vabs{\vno}^2 \approx 1}$ in the bulk of the support of $\vno$,
we can
define a related random variable $\wt{\vno}$ with PDF
\[
    \widetilde \qno(v) \propto \exp\left(\frac{\theta}{2(1+\theta)} \|Xv\|^2\right) \gamma(v) .
\]
We may then define random variables
\begin{equation}
\label{eq:lb-approx-def}
    \widetilde{\yno} \define X \widetilde{\vno}
    \quad\text{ and }\quad
    \widetilde{\yno^\rho}  \define \rho \frac{1}{\sqrt{1+\theta}}\widetilde{\yno} + \sqrt{1-\rho^2}
\bm z .
\end{equation}
We calculate in \cref{prop:lb-gaussian-approx-calc} in \Cref{sec:delayed3} that these variables have Gaussian densities
\[
    \yyes \sim \cN(0, A), \quad \widetilde{\yno} \sim \cN(0, B), \quad
    \widetilde{\yno^\rho}  \sim \cN(0, C) ,
\]
where $A,B$ and $C$ are given as follows:
\begin{align*}
A &\define U\diag(1+\delta_i)_{i \in [m]} U^\top ,\quad
B \define \left(A^{-1} - \frac{\theta}{1 + \theta} I_m\right)^{-1} ,\quad
C \define \frac{\rho^2}{1+\theta}B + (1-\rho^2) I_m .
\end{align*}
Using known formulas for bounding the total variation distance between Gaussian distributions, we show
(in \cref{prop:lb-approx-tv} in \Cref{sec:delayed4}) that
\begin{equation}
\label{eq:apple}
\TV\big( (\yyes\mid_X), \widetilde{\yno^\rho}\big) = O_\theta\left(\frac{m^{3/2}}{n}\right) .
\end{equation} 
All that remains is to show that the analysis for the Gaussian approximations $\widetilde{\yno}$ and
$\widetilde{\yno^\rho}$ holds also for $\yno$ and $\yno^\rho$. We show in
\cref{prop:lb-pno-relative} that
\begin{equation}
    \label{eq:banana}
    \TV\big((\yno\mid_X), \widetilde{\yno}\big) = O_\theta\left(\frac{m^{3/2}}{n}\right) ,
\end{equation}
which implies that $\TV((\yno^\rho\mid_X), \widetilde{\yno^\rho})$ satisfies the same bound.
This shows that \Cref{eq:lb-goal} holds when $m=o(n^{2/3})$, which completes the proof.
\end{proof}

\subsubsection{Indistinguishability of $\Xyes$ and $\Xno$}\label{sec:delayed1}

\begin{proposition}
\label{prop:Xyes-vs-Xno}
Suppose $m = o(n)$. Then $\TV(\Xyes, \Xno) = o(1)$.
\end{proposition}

This follows from the following known theorem (see \cite[Proposition 2]{CMW15};
similar statements appear in \cite{OMM13,PWBM18}):

\begin{theorem}[Indistinguishability of Spiked Gaussian]
\label{thm:spiked-gaussian}
For all $\theta \in (0,1)$ and $\epsilon > 0$, there~exists $\delta > 0$ such that, for all $m,n \in
\bN$ satisfying $m < \delta n/\theta^2$,
\[
    \TV\big( \cN(0, I_n)^{\otimes m}, \samp^m(\Spike_\theta) \big) < \epsilon . 
\]
\end{theorem}

\begin{remark}
There is a small discrepancy between the theorem as stated in \cite{CMW15} and
the one stated here, namely that they have $\bm v \sim \bS^{n-1}$ in
their spike model instead of our $\bm v \sim \cN(0, \tfrac{1}{n} I_n)$. Since
$\|\bm v\| < 1 + o(1)$ with probability $1 - o(1)$ for $\bm v \sim \cN(0,
\tfrac{1}{n} I_n)$, this does not affect the theorem statement.
\end{remark}

\subsubsection{Likelihood of Good $\Xyes$}\label{sec:delayed2}

Recall that $X \in \bR^{m \times n}$ is \emph{good} if its singular values $\sigma_i$ satisfy
\begin{equation}\label{eq:haha8}
\sigma_i^2 = n(1+\delta_i), \quad\text{for some $\delta_i$ satisfying}\  |\delta_i| \leq
K\sqrt{m/n} ,
\end{equation}
where $K > 0$ is the constant  provided by the proposition below. We  show that $\Xyes$ is good with
high probability:

\begin{proposition}
\label{prop:lb-good-X}
There exists a constant $K>0$ such that the following holds.
Let $\bm X \in \bR^{m \times n}$ be a random matrix with each entry drawn independently from $\cN(0,
1)$, with $m = o(n)$. Then $\bm X$ is good with probability at least $1-o(1)$.\end{proposition}
\begin{proof}
We use a standard bound on the spectrum of random matrices:

\begin{theorem}[\cite{Ver26}, Theorem 4.6.1 combined with Lemma 4.1.17]
\label{thm:gaussian-singular-values}
Let $\bm X \in \bR^{m \times n}$ be a random matrix where each entry is drawn independently from
$\cN(0,1)$. Then for any $t \geq 0$, with probability at least $1 - 2e^{-t^2}$ the following holds:
\[
    \left\| \frac{1}{n} \bm X \bm X^\top - I_m \right\|_{op} \leq O\big(\max\{\delta, \delta^2\}\big),\quad
    \text{where\ } \delta = O\left(\sqrt{\frac{m}{n}} + \frac{t}{\sqrt{n}}\right). 
\]
Consequently, for $t =
\sqrt{m}$, under such an
event, the singular values $\sigma_i(\bm X)$ satisfy
\[
    \sigma_i^2(\bm X) = n(1 + \delta_i), 
\quad\text{where}\ |\delta_i| = O\left(\sqrt{m/n}\right).
\]
\end{theorem}

Therefore, $\bm X \sim \cN(0,1)^{m \times n}$ satisfies \cref{eq:haha8} with probability at
least $1-o(1)$.
\end{proof}

\subsubsection{Distributions of $\yyes,\yno$ and $\yno^{\rho}$ Conditional on $X$}\label{sec:delayed3}

We use the notation $f(x) \propto g(x)$ for $x \in \cX$ to mean that there exists a constant $L
> 0$ independent of $x$ such that $f(x) = L \cdot g(x)$.
We need the probability density function (PDF) of $x \in \bR^d$ under $\cN(0, \Xi)$. When $\Xi \in
\bR^{d \times d}$ is invertible, this is
\begin{equation}
\label{eq:gaussian-density}
x \mapsto \frac{1}{(2\pi)^{d/2} \det(\Xi)^{1/2}}
    \exp\left( - \tfrac{1}{2} x^\top \Xi^{-1} x \right) .
\end{equation}
Let $\qyes(v \mid X)$ and $\qno(v \mid X)$ be the PDF of $\vyes$ and $\vno$ conditional on $\Xyes =
X$ and $\Xno = X$, respectively.

\begin{proposition}
\label{prop:lb-vno-density}
Let $\gamma(v)$ be the PDF of $\cN(0, \tfrac{1}{n} I_n)$. Let $X \in \bR^{m \times n}$, $v \in
\bR^n$, and $\theta \in (0,1)$. Then
\[
    \qyes(v \mid X) = \gamma(v) \quad \text{and} \quad
    \qno(v \mid X) \propto \frac{1}{(1+\theta \|v\|^2)^{m/2}}
        \exp\left(\frac{\theta}{2(1+\theta\|v\|^2)} \|Xv\|^2\right) \gamma(v) .
\]
\end{proposition}
\begin{proof}
That $\qyes(v \mid X) = \gamma(v)$ is because $\Xyes$ and $\vyes$ are independent. 
Let $\mu_{no}(X \mid v)$ be the PDF of $\Xno$ conditional on $\vno = v$.
By Bayes' theorem,
\begin{equation}
\label{eq:lb-bayes}
    \qno(v \mid X) = \frac{\mu_{no}(X \mid v) \gamma(v)}{\Exu{\vno}{\mu_{no}(X \mid \vno)}}
    \propto \mu_{no}(X \mid v) \gamma(v) .
\end{equation}
For fixed $\vno = v$, $(\Xno\mid \vno = v) \in \bR^{m \times n}$ has the distribution where each row
is drawn independently from $\cN(0, I_n + \theta vv^\top)$. By \cref{eq:gaussian-density}, since
$I_n + \theta vv^\top$ is invertible and its determinant is $1 + \theta \|v\|^2$, the PDF
$\gamma_\theta$ of $\cN(0, I_n + \theta vv^\top)$ is
\[
    \gamma_\theta(x) \propto \frac{1}{\sqrt{1+\theta\|v\|^2}} \exp\left(-\tfrac{1}{2} x^\top (I_n + \theta vv^\top)^{-1} x\right) .
\]
Observe that
\[
    (I_n + \theta vv^\top)^{-1} = I_n - \frac{\theta}{1+\theta \|v\|^2} vv^\top ,
\]
so $x^\top (I_n + \theta vv^\top)^{-1} x = \|x\|^2 - \frac{\theta}{1+\theta\|v\|^2} (v^\top x)^2$.
Let $x_i$ be the $i^{th}$ row of $X$. Then
\begin{align*}
    \mu_{no}(X \mid v) = \prod_{i=1}^m \gamma_\theta(x_i)
    &\propto \frac{1}{(1+\theta\|v\|^2)^{m/2}}
        \prod_{i=1}^m \exp\left(\frac{\theta}{2(1+\theta\|v\|^2)} (v^\top x_i)^2 \right) \\
    &= \frac{1}{(1+\theta\|v\|^2)^{m/2}} \exp\left(\frac{\theta}{2(1+\theta\|v\|^2)} \|Xv\|^2\right) ,
\end{align*}
where the notation $\propto$ hides a normalizing factor that does not depend on $v$.
Substituting it into \cref{eq:lb-bayes} concludes the proof.
\end{proof}

\begin{proposition}
\label{prop:lb-gaussian-approx-calc}
Fix a good $X \in \bR^{m \times n}$ with singular value decomposition $X = U\diag(\sigma_i)_{i \in [m]}
V^\top$ where $\sigma_i^2 = n(1+\delta_i)$, and let $\widetilde{\yno}$, $\widetilde{\yno^\rho}$ be
defined as in \cref{eq:lb-approx-def}. Then $\widetilde{\yno}$ and $\widetilde{\yno^\rho}$ are
distributed as $\cN(0, B)$ and $\cN(0, C)$, respectively, where
\begin{align*}
B \define \left(A^{-1} - \frac{\theta}{1 + \theta} I_m\right)^{-1} \quad\text{and}\ \quad
C \define \frac{\rho^2}{1+\theta}B + (1-\rho^2) I_m .\end{align*}
\end{proposition}
\begin{proof}
Let
$\widetilde{\pno}(y)$ denote the PDF of $X\widetilde{\vno}$. Recall that the PDF of
$\widetilde{\vno}$ is
\[
    \widetilde{\qno}(v) \propto \exp\left(\frac{\theta}{2(1+\theta)} \|Xv\|^2 \right) \gamma(v).
\]
where $\gamma$ is the PDF of $\cN(0, \tfrac{1}{n} I_n)$,

Therefore, by \cref{eq:gaussian-density},
\begin{align*}
    \wt{\qno}(v)
    &\propto \exp\pbra{
        \frac{\theta}{2(1+\theta)} \|Xv\|^2
        - \frac{1}{2} v^\top (\tfrac{1}{n} I_n)^{-1} v
    }\\
    &= \exp\pbra{
        - \frac{1}{2} v^\top V \pbra{
            n I_n - \frac{\theta}{1+\theta} \diag(\sigma_i^2)_{i \in [n]}
        } V^\top v
    } \\
    &= \exp\pbra{
        - \frac{1}{2} v^\top V \pbra{
            n \cdot \diag\pbra{1 - \frac{\theta}{1+\theta} (1+\delta_i)}_{i \in [n]}
        } V^\top v
    } \\
    &= \exp\pbra{
        - \frac{1}{2} v^\top V \pbra{
            n \cdot \diag\pbra{\frac{1 - \theta \delta_i}{1+\theta}}_{i \in [n]}
        } V^\top v
    } \\
    &= \exp\pbra{
        - \frac{1}{2} v^\top V \pbra{
            \tfrac{1}{n} \diag\pbra{\frac{1+\theta}{1 - \theta \delta_i}}_{i \in [n]}
        }^{-1} V^\top v
    } \,,
\end{align*}
where we implicitly take $\sigma_i = 0$ (and $\delta_i = -1$) for $i > m$.
Since $X$ is good, so that $\abs{\delta_i} = o(1)$ for $i \in [m]$, we conclude that
$1 - \theta \delta_i > 0$ for each $i \in  [m]$ and hence
\[
    \wt{\vno} \sim \cN\pbra{
        0,
        V \pbra{
            \tfrac{1}{n} \diag\pbra{\frac{1+\theta}{1 - \theta \delta_i}}_{i \in [n]}
        } V^\top
    } \,.
\]
Therefore $\wt{\yno} = X \wt{\vno}$ satisfies
\begin{align*}
    \wt{\yno} 
    &\sim \cN\pbra{
        0,
        U \diag(\sigma_i) \pbra{
            \tfrac{1}{n} \diag\pbra{\frac{1+\theta}{1 - \theta \delta_i}}_{i \in [m]}
        } \diag(\sigma_i) U^\top
    } \\
    &= \cN\pbra{
        0,
        U \pbra{
            \diag\pbra{\frac{(1+\delta_i)(1+\theta)}{1 - \theta \delta_i}}_{i \in [m]}
        } U^\top
    } \,.
\end{align*}
Then, we can also verify that
\begin{align*}
    B &= \pbra{A^{-1} - \frac{\theta}{1 + \theta} I_m}^{-1}\\
    &= \pbra{U \pbra{\diag\pbra{\frac{1}{1+\delta_i}}_{i \in [m]} - \frac{\theta}{1+\theta} I_m} U^\top}^{-1} \\
    &= U \diag\pbra{\frac{1 + \theta - \theta(1+\delta_i)}{(1+\delta_i)(1+\theta)}}_{i \in [m]}^{-1} U^\top
    = U \diag\pbra{\frac{1 - \theta \delta_i}{(1+\delta_i)(1+\theta)}}_{i \in [m]}^{-1} U^\top \\
    &= U \diag\pbra{\frac{(1+\delta_i)(1+\theta)}{1 - \theta \delta_i}}_{i \in [m]} U^\top \,,
\end{align*}
so indeed $\wt{\yno} \sim \cN(0, B)$ as claimed.
Now then, $\widetilde{\yno^\rho}$ is a sum of Gaussians, distributed as
\[
    \rho \frac{1}{\sqrt{1+\theta}} \cN(0, B) + \sqrt{1-\rho^2} \cN(0, I_m) ,
\]
so its distribution is $\cN(0, C)$.
\end{proof}

\subsubsection{Indistinguishability of Gaussians}\label{sec:delayed4}

\begin{proposition}
\label{prop:lb-approx-tv}
Given a good $X \in \bR^{m \times n}$,  we have
\[
    \TV\big( (\yyes|_X), \widetilde{\yno^\rho}|_X\big) = O_\theta\left(\frac{m^{3/2}}{n}\right) .
\]
\end{proposition}
\begin{proof}
Recall from \cref{prop:lb-gaussian-approx-calc} that
\[
    (\yyes|_X) \sim \cN(0, A) , \quad
    (\widetilde{\yno}|_X) \sim \cN(0, B) , \quad
    (\widetilde{\yno^\rho}|_X) \sim \cN(0, C) ,
\]
where $A,B$ and $C$ are as follows:
\begin{align*}
A &\define U\diag(1+\delta_i)_{i \in [m]} U^\top ,\quad
B \define \left(A^{-1} - \frac{\theta}{1 + \theta} I_m\right)^{-1} ,\quad
C \define \frac{\rho^2}{1+\theta}B + (1-\rho^2) I_m .
\end{align*}
We will use the following upper bound on the total variation distance between two Gaussians: 

\begin{theorem}[\cite{DMR18}]
\label{thm:gaussian-tv-distance}
Let $\Sigma_1, \Sigma_2 \in \bR^{m \times m}$ be two positive definite matrices, and $\lambda_1, \dotsc,
\lambda_m$ be the eigenvalues of $\Sigma_1^{-1}\Sigma_2 - I_m$. Then
\[
    \TV\big(\cN(0, \Sigma_1), \cN(0, \Sigma_2)\big) = O\left(\left(\sum_{i=1}^m \lambda_i^2
\right)^{1/2}\right) .
\]
\end{theorem}

To apply the theorem, we must compute the eigenvalues of $A^{-1} C - I_m$. These matrices share a
basis of eigenvectors (they are all matrices of the form $UDU^\top$ for diagonal $D$ of
eigenvalues). The eigenvalues $\lambda_i(B)$ of $B$ are:
\[
    \lambda_i(B) = \left( \lambda_i(A^{-1}) - \frac{\theta}{1+\theta} \right)^{-1}
    = \left(\frac{1}{1 + \delta_i} - \frac{\theta}{1+\theta}\right)^{-1}
    = \frac{(1 + \delta_i)(1+\theta)}{(1+\theta) - \theta(1+\delta_i)}
    = \frac{(1+\delta_i)(1+\theta)}{1-\theta\delta_i} .
\]
Now the eigenvalues $\lambda_i(C)$ of $C$:
\begin{align*}
\lambda_i(C)
&= 1 + \rho^2 \frac{\lambda_i(B) - (1+\theta)}{1+\theta} \\
&= 1 + \frac{1}{1+\theta} \cdot \frac{1+\delta_i - (1-\theta\delta_i)}{1-\theta\delta_i} \tag{Using
$\rho^2 = \tfrac{1}{1+\theta}$}\\
&=  1 + \frac{\delta_i}{1 - \theta\delta_i}
= \frac{1-\theta\delta_i + \delta_i}{1-\theta\delta_i} .
\end{align*}
Now $A^{-1} C - I_m$:
\[
\lambda_i
= \frac{1}{1+\delta_i}\left(\frac{1-\theta\delta_i + \delta_i}{1-\theta\delta_i}\right) - 1 
= \frac{1-\theta\delta_i + \delta_i - (1 + \delta_i -
\theta\delta_i - \theta\delta_i^2)}{(1+\delta_i)(1-\theta\delta_i)} 
= \frac{\theta\delta_i^2}{(1+\delta_i)(1-\theta\delta_i)} .
\]
Since $|\delta_i| = O(\sqrt{m/n})$, this is at most $2\theta\delta_i^2$. Then by 
\cref{thm:gaussian-tv-distance},
\begin{align*}
    \TV\big(\yyes|_X, \widetilde{\yno}^\rho|_X\big)
    &= O\left( \left(\sum_{i=1}^m \lambda_i^2 \right)^{1/2} \right)
    = O\left( \theta \left(\sum_{i=1}^m \delta_i^4 \right)^{1/2} \right) \\
    &= O\left( \theta \big(m (m^2/n^2)\big)^{1/2} \right) = O\left( \frac{m^{3/2}}{n} \right) . \qedhere
\end{align*}
\end{proof}

\subsubsection{Closeness of Gaussian Approximation}
\label{subsection:lb-approximation}

We need to prove the following:

\begin{proposition}
\label{prop:lb-pno-relative}
Let $\theta \in (0,1)$ be any constant.
Let $X \in \bR^{m \times n}$ be good. 
Let $\bm v$ and $\widetilde{\bm v}$ be random variables in $\bR^n$
with probability density functions proportional to
\[
    q(v) \propto \frac{1}{(1+\theta\|v\|^2)^{m/2}}
            \exp\left(\frac{\theta}{2(1+\theta \|v\|^2)} \|Xv\|^2\right) \gamma(v)
        \quad\text{ and }\quad
    \widetilde q(\widetilde v)
        \propto \exp\left(\frac{\theta}{2(1+\theta)} \|X\widetilde v\|^2\right) \gamma(\widetilde v) ,
\]
respectively, where $\gamma(\cdot)$ is the density of $\cN(0, \tfrac{1}{n} I_n)$. Then
\[
    \TV( X\bm v, X\widetilde{\bm v} )
        = O_\theta\left( \frac{m^{3/2}}{n} \right) .
\]
\end{proposition}
\begin{proof}

Our main tool is the following lemma, which is a consequence of the Gaussian Poincar\'e inequality.
We provide a proof in \cref{section:lb-poincare}.
\begin{restatable}{lemma}{lemmalbpoincare}
\label{lemma:lb-poincare}
Let $q$ be the PDF of $\cN(0, \Sigma)$ and let $p$ be the PDF of any
distribution that is absolutely continuous with respect to $q$. Then
\[
    \TV(p,q)^2 \leq \frac{1}{4} \|\Sigma\|_{op} \Exu{\bm y \sim p}{ \left\| \nabla \log\frac{p(\bm
y)}{q(\bm y)} \right\|^2 }
\]
\end{restatable}

Let $X = U\diag(\sigma_i)V^\top$ be the singular value decomposition of $X$. By
rotation invariance, we may assume without loss of generality that
\[
    X = \diag(\sigma_i)_{i \in [m]} [ I_m \mid 0 ] .
\]
Let $p$ be the density of $X\bm v$ and $\tilde p$ be the density of $X \widetilde{\bm v}$. Recall
from \cref{prop:lb-gaussian-approx-calc} that $\tilde p$ is the density of $\cN(0, B)$ where

\begin{align*}
    A \define \diag(1+\delta_i)_{i \in [m]} ,
    \qquad
    B &\define \left(A^{-1} - \frac{\theta}{1+\theta} I_m\right)^{-1} ,
\end{align*}
To prove the proposition, it suffices to apply \cref{lemma:lb-poincare} with the next two claims. We defer the proofs of these claims to the appendix
because they are tedious.

\begin{restatable}{claim}{claimlbderivativebound}
\label{claim:lb-derivative-bound}
For all $y \in \bR^m$,
\[
    \left\| \nabla \log \frac{p(y)}{\tilde p(y)} \right\|^2
        \leq O\left(\frac{\|y\|^2(\|y\|^2+m+1)^2}{n^2} \right) .
\]
\end{restatable}
\begin{proof}[Proof of claim]
See \cref{section:lb-derivative-bound}
\end{proof}

\begin{restatable}{claim}{claimlbnormbound}
\label{claim:lb-norm-bound}
Let $\bm y = X\bm v$. Then for $j \in \{1,2,3\}$,
\[
    \Ex{ \|\bm y\|^{2j} } = O(m^j) .
\]
\end{restatable}
\begin{proof}[Proof of claim]
See \cref{section:lb-norm-bound}.
\end{proof}

With these two claims, we have
\[
\TV(p,q)^2 = O\left( \|B\|_{op} 
    \Exu{\bm y \sim p}{\left\|\nabla \log \frac{p(\bm y)}{\tilde p(\bm y)} \right\|^2 }
    \right)
    = O_\theta\left( \frac{m^3}{n^2} \right) ,
\]
which suffices to complete the proof.
\end{proof}

\bibliographystyle{alpha}
\bibliography{references,allrefs}

\appendix

\section{Reduction to Distribution-Picky Testing}
\label{sec:picky}

The following is a special case of the distribution-robust to distribution-picky reduction proved in
\cite{CFHS26general}, but we provide the proof here for completeness and convenience of the reader.

\begin{restatable}{proposition}{proppickystrongness}
\label{prop:picky-strongness}
Suppose $\cA$ is a $(\cD^*, \mathfrak C)$-picky halfspace tester, and let
$m(n,\epsilon)$ and $q(n,\epsilon)$ the sample and query cost of $\cA$. Then
there exists a distribution-robust tester for halfspaces under $\{\cD^*\}$ with
sample and query cost $O( m(n,\epsilon) + q(n,\epsilon) )$.
\end{restatable}

\begin{algorithm}[ht]
    \caption{Reduction to Picky Testing}    
    \label{alg:DRT-via-DPT}
    
    \begin{algorithmic}
        \State \textbf{Requirements}:   
        \State \qquad $\bullet$ $\cA$ is a $(\cD^*, \mathfrak C)$-picky halfspace tester.
        \State \textbf{Input}:
        \State \qquad $\bullet$ Query access to $f : \cX \to \zo$, 
\State \qquad $\bullet$ Access to i.i.d.~samples $(\bm x,f(\bm x))$ where $\bm x \sim \cD$,
        \State \qquad $\bullet$ Proximity and failure parameters $\epsilon,\delta \in (0,1)$.
        \Procedure{DRT-via-DPT}{}
            \State 1. Set $r = O(\log(1/\delta))$ with an appropriate constant.
                \State \phantom{1.} Run $r$ iterations of $\cA$ on $f$ and $\cD^*$, with distance parameter $\epsilon$.
                \State \phantom{1.} Let $\bm{t}_1$ be the fraction of times it outputs $\Accept$.
            \State 2. Run $r$ iterations of $\cA$ on $f$ and $\cD$, with distance parameter $\epsilon$.
                \State \phantom{2.} Let $\bm{t}_2$ be the fraction of times it outputs $\Accept$.
            \State 3. If $|\bm{t}_1 - \bm{t}_2| > 0.01$ then output $\Abstain$.
            \State \phantom{4.} Otherwise output $\Accept$ if $\bm{t}_1 > 1/2$ and $\Reject$ if $\bm{t}_1 \leq 1/2$. 
        \EndProcedure
    \end{algorithmic}
\end{algorithm}

\begin{proof}
The algorithm is  \cref{alg:DRT-via-DPT}.
The sample and query complexity are immediate from inspection of the algorithm.
Let us now establish correctness.

Let $p_1$ be the probability that $\cA$ outputs $\Accept$ on $f$ and $\cD^*$, and let $p_2$ be the
probability that $\cA$ outputs $\Accept$ on $f$ and $\cD$. Let $\bm t_1, \bm t_2$ be the random
variables which are the observed fraction of times these subroutines output $\Accept$ in Steps~1 and~2, respectively. By choosing $r = O(\log(1/\delta))$ appropriately, the Chernoff bound ensures that
$|p_1 - \bm t_1| < 0.005$ and $|p_2 - \bm t_2| < 0.005$ with probability at least $1-\delta$. Let us
assume this event occurs.

\textbf{Completeness.} Suppose $f$ is a halfspace. Then $\cA$ running on $\cD^*$ will output
$\Accept$ with probability $p_1 \geq 3/4$, so $\bm t_1 > 3/4 - 0.005$.  If $\bm t_2 \leq 1/2$ then $|\bm t_1
- \bm t_2| > 0.2$ and the algorithm will output $\Abstain$, so it is not possible for the algorithm
to output $\Reject$.

\textbf{Soundness.} Suppose $\dist_\cD(f,h) > \epsilon$ for all halfspace $h$. Then $\cA$
running on $\cD$ will output $\Accept$ with probability $p_2 \leq 1/4$, so $\bm t_2 < 1/4
+ 0.005$, so it is not possible for the algorithm to output $\Accept$.

\textbf{Assertiveness.} Suppose $\cD = \cD^*$. 
Then $p_1 = p_2$ since the two runs of the tester $\cA$ are running on identical inputs. Then
$|\bm{t}_1 - \bm{t}_2| < 0.01$, so the algorithm will not output $\Abstain$.
\end{proof}

\section{Deferred Proofs from the Lower Bound}

\subsection{Discretization and Distance From Halfspaces}
\label{section:lb-discretization}

We prove the following lemma from \cref{sec:lb}
\lemmalbdistance*

Our discretized distributions are defined as follows:

\begin{definition}[Discretized NO Distribution]
For any parameter $\tau > 0$, we a distribution $\cD \sim \DNO^{\square \tau}$ is obtained by
\begin{enumerate}
    \item Choose $\bm v \sim \cN(0, \tfrac{1}{n} I_n)$.
    \item Choose any orthonormal basis $\cU$ for $\bR^n$ which includes $\bm v / \|\bm v\|$.
    \item Partition $\bR^n$ into cubes of length $\tau$, with sides aligned with the axes of $\cU$.
    \item For each cube $B$, assign a label as follows:
        \begin{enumerate}
            \item Choose $\bm x^* \sim \left(\cN(0, I_n + \theta \bm v \bm v^\top )
            \mid B\right)$ from the spiked Gaussian conditional on $\bm x^* \in B$.
            \item Choose $\bm z^* \sim \cN(0, 1)$.
            \item Assign the value $\sign\left(\frac{\rho}{\sqrt{1+\theta}} \bm v^\top \bm x^*
                    + \sqrt{1-\rho^2} \bm z^*\right)$ to all points in $B$.
        \end{enumerate}
\end{enumerate}
\end{definition}

The next two propositions establish the lemma.

\begin{proposition}
\label{prop:lb-discretization}
For every $m \in \bN$ and $\delta > 0$, there exists $\tau > 0$ such that
\[
    \TV(\samp^m(\DNO), \samp^m(\DNO^{\square \tau}))  < \delta .
\]
\end{proposition}
\begin{proof}[Proof sketch]
The proof is standard, so we give only a sketch. We may couple the distributions as follows. First
choose $\bm v \sim \cN(0, \tfrac{1}{n} I_n)$ and define the orthonormal basis $\cU$ together with
a partition into cubes. For each of the $m$ samples, we draw $\bm x' \sim \cN(0, I_n + \theta \bm v
\bm v^\top)$ and then select the unique box $\bm B$ in the partition which contains $\bm x'$. Draw
$\bm x \sim \cN(0, I_n + \theta \bm v \bm v^\top)$ conditional on $\bm x \in \bm B$ and $\bm z \sim
\cN(0, 1)$. Now, to produce the samples from the original and discretized distributions:
\begin{enumerate}
\item To obtain the sample from the original distribution, output $(\bm x, \bm b)$
      where $\bm b = \sign(\frac{\rho}{\sqrt{1+\theta}} \bm v^\top \bm x + \sqrt{1-\rho^2} \bm z)$.
\item To obtain the sample form the discretized distribution:
\begin{itemize}
    \item If a sample from box $\bm B$ has already been chosen earlier, then box $\bm B$ has already
        been assigned an $\bm x^*$ and $\bm z^*$ value and a corresponding label $\bm b^*$. In
        that case, output $(\bm x, \bm b^*)$.
    \item Otherwise, output the same as the original distribution, and set $\bm x^* \gets \bm x$,
        $\bm z^* \gets \bm z$, and label $\bm b^* \gets \bm b$ to this box $\bm B$.
    \end{itemize}
\end{enumerate}
The distributions will produce unequal samples only when more than one sample $\bm x$ is drawn from
a single box $B$. By choosing small enough $\tau > 0$, the maximum probability mass of any box can
be arbitrarily small, so the probability of this bad event is arbitrarily small as $\tau \to 0$.
\end{proof}

\begin{proposition}
\label{prop:lb-distance}
There exists a constant $\epsilon > 0$ such that, for all sufficiently large $n$, there is $\tau >
0$ such that for $\cD \sim \DNO^{\square \tau}$, the labels $\bm b$ of $(\bm x, \bm b) \sim \cD$ are
assigned by a measurable function $\bm b = f(\bm x)$ where $f$ is $\epsilon$-far from being a
halfspace under $\cD$.
\end{proposition}
\begin{proof}
Choose $\tau = \delta/\sqrt n$ so that the diameter of any cube in the partition is at most
$\delta$, where we choose $\delta > 0$ below.
We wish to show that $\cD \sim \DNO^{\square \tau}$ is $\epsilon$-far from every halfspace with high
probability. Let $\bm v \sim \cN(0, \tfrac{1}{n} I_n)$ be the normal vector defining $\cD$.  By
rotation invariance, it suffices to consider distributions $\cD$ where $\bm v = \bm s e_1$ where
$\bm s \sim \cN(0, 1)$. Let $S > 0$ be the constant such that
\[
    \Pr{|\bm s| > S } = 1/200
\]
Fix any $0 < s < S$ and consider $\cD$ conditional on $\bm
v = se_1$ (the same analysis holds for $-S < s \leq 0$). For each cube $B$ in the discretization, let
\[
    \bm \ell(B) \define 
        \sign\left(\frac{\rho}{\sqrt{1+\theta}} s\bm x_1^*(B) + \sqrt{1-\rho^2} \bm z^*(B)\right)
\]
be its randomly assigned label, where $\bm x^*(B)$ is drawn from $\cN(0, I + \theta s^2 e_1e_1^\top)$
conditional on $\bm x^*(B) \in B$ and $\bm z^*(B) \sim N(0,1)$. Let
\[
    \bm f(x) \define \bm \ell(B) \text{ where $B$ is the unique cube containing } x .
\]
We want to show that with probability at least $199/200$, $\bm f$ is $\epsilon$-far from every
halfspace.

Let $h(x) = \sign(w^\top x - t)$ be any halfspace and consider the function $\bm D_h =
\sum_B \bm D_h(B)$ where the sum is over all cubes $B$ in the discretization and
\[
    \bm D_h(B) = \Pr{ \bm x \in B } \ind{ h \text{ constant on } B  \wedge h(B) \neq \bm \ell(B) } ,
\]
where for convenience we assign $h(B) \in \pmset$ arbitrarily when $h$ is not constant on $B$,
and otherwise assign $h(B)$ to be the constant value $h$ takes on $B$. Observe that
\[
    \dist(h, \bm f) \geq \bm D_h .
\]
Now
\begin{align*}
\Ex{ \bm D_h }
&= \sum_{B : h(B) \text{ constant}} \Pr{ \bm x \in B } \Pr{ h(B) \neq \bm \ell(B) } \\
&\geq \left(\sum_{B} \Pr{ \bm x \in B } \Pr{ h(B) \neq \bm \ell(B) } \right)
    - \Pr{ |w^\top \bm x - t| < \delta } \\
&\geq \left(\sum_{B} \Pr{ \bm x \in B } \Pr{ h(B) \neq \bm \ell(B) } \right)
    - 2\delta ,
\end{align*}
since the probability that $\bm x \in B$ where $h$ is not constant on $B$ is at most the probability
that $|w^\top \bm x - t| < \delta$ (because each $B$ has diameter $\delta$), and this probability is
at most $\delta$ times the maximum density of $w^\top x$ (which is at most 1).

Let $B$ be any cube in the discretization. If $\bm x_1^* \geq 0$ then $\Pr{\bm \ell(B) = 1} \geq
1/2$ and
\begin{align*}
\Pr{ \bm \ell(B) = -1 }
&= \Pr{ \bm z^* > \frac{\rho}{\sqrt{(1+\theta)(1-\rho^2)}} s \bm x_1^* } 
\geq \frac{1}{2} - \frac{1}{\sqrt{2\pi}} \frac{\rho s \bm x_1^*}{\sqrt{(1+\theta)(1-\rho^2)}} \\
&\geq \frac{1}{2} - \frac{1}{\sqrt{2\pi}} \frac{\rho s B_1}{\sqrt{(1+\theta)(1-\rho^2)}} ,
\end{align*}
where we define $B_1 = \sup_{x \in B} |x_1|$. Set $T \define \frac{\sqrt{2\pi}}{4S}
\frac{\sqrt{(1+\theta)(1-\rho^2)}}{\rho}$. If $B_1 \leq T$ then
\[
    \Pr{ h(B) \neq \bm \ell(B) }
        \geq \min\left( \Pr{ \bm \ell(B) = 1 }, \Pr{ \bm \ell(B) = -1 } \right)
        \geq \frac{1}{2}\left(1 - \frac{s}{2S}\right) .
\]
A similar analysis holds when $\bm x_1^* < 0$.  Therefore $\Pr{ h(B) \neq \bm \ell(B) } \geq
\frac{1}{2}(1 - \tfrac{s}{8})$ when $B_1 \leq T$, so
\begin{align*}
    \Ex{\bm D_h}
    &\geq \frac{1}{2}\left(1 - \frac{s}{2S}\right) \cdot
            \left(\sum_{B : B_1 \leq T} \Pr{ \bm x \in B } \right) - 2\delta \\
    &\geq \frac{1}{4} \cdot \Pr{ \bm x_1 \leq T - \tau } - 2\delta
\end{align*}
where the last inequality is because the unique $B$ containing $\bm x$ will have $B_1 < \bm x_1 +
\tau$. Then
\[
    \Ex{\bm D_h} \geq \frac{1}{4} \frac{T/2}{\sqrt{2\pi(1+S\theta)}} - 2\delta > 4\epsilon 
\]
where $\epsilon$ is a sufficiently small constant that depends only on $\theta$ as long as we choose
$\delta$ (and therefore $\tau$) sufficiently small.

The difference in $\bm D_h$ by changing the value of any $\bm D_h(B)$ variable is at most $\Pr{ \bm
x \in B }$.  The maximum density of $\cN(0, I + \theta s^2 e_1e_1^\top)$ is at most
$\frac{1}{(2\pi)^{n/2} \sqrt{1+\theta s^2}} < 1$, so
\[
    \Pr{ \bm x \in B } \leq \tau^n .
\]
Therefore, McDiarmid's bounded differences concentration inequality gives
\begin{align*}
\Pr{ \dist(\bm f, h) < 2\epsilon }
\leq \Pr{ \bm D_h < 2\epsilon }
&\leq \Pr{ \bm D_h < \frac{1}{2} \Ex{\bm D_h} }
\leq \exp\left( - \frac{2\epsilon^2}{\sum_B \Pr{ \bm x \in B }^2 }\right) \\
&\leq \exp\left( - \frac{2\epsilon^2}{\tau^n \sum_B \Pr{ \bm x \in B } }\right)
\leq \exp\left( - \frac{2\epsilon^2}{\tau^n }\right) .
\end{align*}
We cannot use a union bound over all halfspaces, so we use the following standard claim about
$\epsilon$-nets for halfspaces (see \eg~\cite{SSBD14}):
\begin{claim}
For every $\epsilon > 0$ and distribution $\cE$ over $\bR^n$, there exists a set $H$ of
$(1/\epsilon)^{O(n/\epsilon)}$ halfspaces such that, for all halfspaces $h$, there exists $h' \in H$
with $\dist_\cE(h, h') < \epsilon$. 
\end{claim}
Let $H$ be the set of halfspaces from this claim (applied with $\epsilon' = \epsilon/2$).
Therefore, by the union bound over $h \in H$:
\[
\Pr{ \exists h \in H : \dist(\bm f, h) < 2\epsilon }
    \leq (1/\epsilon)^{O(n/\epsilon)} \cdot
        \exp\left( - \frac{2\epsilon^2}{\tau^n}\right) .
\]
Then by the properties of $H$,
\[
\Pr{ \exists \text{ halfspace } h : \dist(\bm f, h) < \epsilon }
    \leq (1/\epsilon)^{O(n/\epsilon)} \cdot
        \exp\left( - \frac{2\epsilon^2}{ \tau^n}\right) .
\]
Since we are free to choose $\tau > 0$ arbitrarily small, this probability is arbitrarily small.
The above holds when $\bm s < S$ which occurs with probability at least $199/200$, which gives the
conclusion.
\end{proof}

\subsection{Gaussian Poincar\'e Inequality}
\label{section:lb-poincare}
We prove the following lemma used in the proof of \cref{prop:lb-pno-relative}:

\lemmalbpoincare*

\begin{proof}
The Gaussian Poincar\'e inequality (see \eg~\cite[Theorem 3.20]{BLM13}) states that for any differentiable $f \colon \bR^n \to \bR$,
\[
    \Var{f} \define \Ex{f(\bm z)^2} - \Ex{f(\bm z)}^2 \leq
        \Ex{ \left\| \nabla f(\bm z) \right\|^2 } ,
\]
where $\bm z \sim \cN(0, I)$. Now let $g \colon \bR^n \to \bR$ be differentiable. For $\bm w \sim
\cN(0, \Sigma)$, we have $\bm w = \Sigma^{1/2} \bm z$.
\[
    \Varu{\bm w}{ g(\bm w) } = \Varu{\bm z}{ g(\Sigma^{1/2} \bm z) } .
\]
Applying the Gaussian Poincar\'e inequality with $f(z) = g(\Sigma^{1/2} z)$
and $\nabla f(z) = \Sigma^{1/2} [\nabla g](\Sigma^{1/2} z)$:
\begin{equation}
\label{eq:poincare-sigma}
    \Varu{\bm w}{ g(\bm w) } = \Varu{\bm z}{ f(\bm z) } \leq \Ex{ \|\nabla f(\bm z)\|^2 }
    = \Ex{ (\nabla g(\bm w))^\top \Sigma (\nabla g(\bm w)) }
    \leq \|\Sigma\|_{op} \Ex{ \|\nabla g(\bm w)\|^2 } .
\end{equation}
We will apply this inequality with $h(w) \define \sqrt{p(w)/q(w)}$.
Using $\bm w \sim \cN(0, \Sigma)$,
\begin{align*}
\TV(p,q)^2
&= \frac{1}{4} \left(\int_w |p(w) - q(w)| dw\right)^2 
= \frac{1}{4} \left(\int_w q(w) |h(w)^2 - 1| dw\right)^2  \\
&= \frac{1}{4} \left(\int_w q(w) |h(w) - 1|(h(w)+1) dw\right)^2  \\
&\leq \frac{1}{4} \Exu{\bm w \sim q}{(h(\bm w)-1)^2} \Exu{\bm w \sim q}{(h(\bm w)+1)^2}
    \tag{Cauchy-Schwartz} \\
&= \frac{1}{4} \Exu{\bm w \sim q}{h(\bm w)^2 - 2h(\bm w) + 1)}
    \Exu{\bm w \sim q}{ h(\bm w)^2 + 2h(\bm w) + 1 } \\
&= (1 - \Ex{h(\bm w)})(1 + \Ex{h(\bm w)})
    \tag{Since $\Ex{h(\bm w)^2} = \int_w p(w) dw = 1$} \\
&= 1 - \Ex{h(\bm w)}^2 = \Var{h(\bm w)} .
\end{align*}
With \cref{eq:poincare-sigma}, this is at most $\|\Sigma\|_{op} \Ex{ \| \nabla h(\bm w) \|^2 }$.
Now, since $h(w) = e^{\tfrac{1}{2} \ln(p(w)/q(w))}$,
\[
    \nabla h(w) = e^{\tfrac{1}{2} \ln(p(w)/q(w))} \frac{1}{2} \nabla \ln(p(w)/q(w))
        = \frac{1}{2} \sqrt{\frac{p(w)}{q(w)}}  \nabla \ln(p(w)/q(w)) .
\]
Therefore
\[
\Exu{\bm w \sim q}{ \|\nabla h(\bm w)\|^2 }
= \frac{1}{4} \Exu{\bm w \sim q}{ \frac{p(\bm w)}{q(\bm w)} \|\nabla \log(p(\bm w)/q(\bm w)) \|^2}
= \frac{1}{4} \Exu{\bm y \sim p}{ \| \nabla \log(p(\bm w)/q(\bm w)) \|^2 } . \qedhere
\]
\end{proof}

\subsection{Gradient Calculations}
\label{section:lb-gradient-calculations}

Recall from the proof of \cref{prop:lb-pno-relative} in
\cref{subsection:lb-approximation} that
\[
    X = \diag(\sigma_i)_{i \in [m]} [ I_m \mid 0 ],
\]
$p$ is the density of $X\bm v$, and $\tilde p$ is the density of $\cN(0, B)$ where
\begin{align*}
    A \define \diag(1+\delta_i)_{i \in [m]} ,
    \qquad
    B &\define \left(A^{-1} - \frac{\theta}{1+\theta} I_m\right)^{-1} .
\end{align*}
We need to prove the following two claims.

\subsubsection{Gradient Bound}
\label{section:lb-derivative-bound}

\claimlbderivativebound*
\begin{proof}
First observe
\[
    \tilde p(y) \propto
        \gamma_A(y) \cdot \exp\left(\frac{\theta}{2(1+\theta)} \|y\|^2 \right) ,
\]
where $\gamma_A$ is the PDF of $\cN(0, A)$.
Let us now calculate $p(y)$. For $Xv = y$ we can write $v = (u,w)$ where $u \in \bR^m$, $w \in
\bR^k$ with $k=n-m$, where $u_i = y_i/\sqrt{n(1+\delta_i)}$ and $y$ does
not depend on $w$, so
\[
    \|v\|^2 = \frac{1}{n} (y^\top A^{-1} y) + \|w\|^2 .
\]
Then
\begin{align*}
p(y) &\propto \int_{w \in \bR^{n-m}} \frac{1}{(1+\theta(\tfrac{1}{n} (y^\top A^{-1} y)
        + \|w\|^2))^{m/2}} \;\cdot \\
        &\qquad\qquad \exp\left(\frac{\theta}{2(1+\theta \tfrac{1}{n} (y^\top A^{-1} y)
        + \theta\|w\|^2)} \|y\|^2\right) \gamma_{I_m/n}(u) \gamma_{I_k/n}(w) dw \\
     &= \int_{w \in \bR^{n-m}} \frac{1}{D_y(n\|w\|^2)^{m/2}}
        \exp\left(\frac{\theta}{2 D_y(n\|w\|^2)} \|y\|^2\right) \gamma_{I_m/n}(u) \gamma_{I_k/n}(w) dw ,
\end{align*}
where we have written
\[
    D_y(W) \define 1 + \frac{\theta}{n}\left(y^\top A^{-1} y + W \right)  .
\]
It will be convenient to write this quantity as a fraction of $1+\theta$, so
observe that
\[
    D_y(W) = (1 + \theta)\left(1 + \frac{\theta}{n(1+\theta)}(y^\top A^{-1} y + (W-n)) \right)
           = (1 + \theta)(1 + \Delta(W)) ,
\]
where $\Delta(W)$ tracks the difference between $n = m + k$ and $y^\top A^{-1} y + W$:
\[
    \Delta(W) \define \frac{\theta}{n(1+\theta)}( (y^\top A^{-1} y - m) + (W-k) ) .
\]
Note that for $W \geq 0$,
\[
    \Delta(W) \geq - \frac{\theta}{1 + \theta} , \qquad \text{ and } \qquad D_y(W) \geq 1 .
\]
Observe that for $\bm w \sim \cN(0, \tfrac{1}{n} I_k)$, $n \|\bm w\|^2$ is distributed according to
the $\chi^2_k$ distribution, so for $\bm W \sim \chi^2_k$ we can write
\begin{align*}
p(y)
&\propto \gamma_{I_m/n}(u) \cdot
        \Exu{\bm W}{ \frac{1}{D_y(\bm W)^{m/2}}
            \exp\left(\frac{\theta}{2 D_y(\bm W)} \|y\|^2 \right) } \\
&= \gamma_A(y) \cdot
        \Exu{\bm W}{ \frac{1}{D_y(\bm W)^{m/2}}
            \exp\left(\frac{\theta}{2 D_y(\bm W)} \|y\|^2 \right) } .
\end{align*}
Then
\begin{align*}
    \frac{p(y)}{\tilde p(y)}
&\propto
    \Exu{\bm W}{ \frac{1}{D_y(\bm W)^{m/2}} \exp\left(\frac{\theta}{2}\|y\|^2\left(\frac{1}{D_y(\bm
W)} - \frac{1}{1+\theta}\right)\right) } \\
&=
    \Exu{\bm W}{ \frac{1}{D_y(\bm W)^{m/2}} \exp\left(- \frac{\theta}{2}\|y\|^2
        \frac{\Delta(\bm W)}{D_y(\bm W)} \right) }  \\
    &= \Exu{\bm W}{ F_y(\bm W) } ,
\end{align*}
where we have defined $F_y(\cdot)$ as the term within the expectation.
We now compute the gradient:
\begin{align*}
\nabla \log \frac{p(y)}{\tilde p(y)}
&= \frac{\nabla \Exu{\bm W}{ F_y(\bm W) }}{\Exu{\bm W}{ F_y(\bm W) } } 
= \frac{\Exu{\bm W}{ \nabla F_y(\bm W) }}{\Exu{\bm W}{ F_y(\bm W) } } 
\end{align*}
For fixed $W$:
\begin{align*}
\nabla F_y(W)
&= \exp\left(-\frac{\theta}{2}\|y\|^2 \frac{\Delta(W)}{D_y(W)} \right)
            \left(\nabla D_y(W)^{-m/2} 
    - D_y(W)^{-m/2} \cdot \nabla \left(\frac{\theta}{2} \|y\|^2 \frac{\Delta(W)}{D_y(W)}\right)
\right) .
\end{align*}
We compute the inner gradients as follows:
\begin{align*}
    \nabla D_y(W) &= \frac{2\theta}{n} A^{-1} y , \\
    \nabla D_y(W)^{-m/2} &= - \frac{\theta m}{n} D_y(W)^{-(m+2)/2} A^{-1} y , \\
    \nabla \frac{\theta}{2} \frac{\Delta(W)}{D_y(W)} \|y\|^2
    &= \theta \left(
             \frac{\Delta(W)}{D_y(W)} I_m
            - \|y\|^2\frac{\theta }{n \cdot D_y(W)^2}  A^{-1}
        \right) y .
\end{align*}
Therefore:
\begin{align*}
    \nabla F_y(W)
    &= \exp\left( - \frac{\theta}{2} \|y\|^2 \frac{\delta(W)}{D_y(W)} \right) \cdot \\
    &\qquad \left(-\frac{\theta m}{n} \frac{1}{D_y(W)^{1 + m/2}} A^{-1} y
        + \frac{1}{D_y(W)^{m/2}} \theta \left( \|y\|^2 \frac{\theta}{n \cdot D_y(W)^2 } A^{-1}
            - \frac{\Delta(W)}{D_y(W)} I_m \right) y \right) \\
    &= F_y(W) \cdot \theta \left( 
        \left(\frac{\theta}{n \cdot D_y(W)^2} \|y\|^2 - \frac{m}{n \cdot D_y(W)} \right) A^{-1} y
         -\frac{\Delta(W)}{D_y(W)} y \right) \\
    &= F_y(W) \cdot \theta \frac{1}{D_y(W)} \left( 
        \left(\frac{\theta}{n \cdot D_y(W)} \|y\|^2 - \frac{m}{n} \right) A^{-1} y
         -\Delta(W) y \right) \\
    &= F_y(W) \cdot \theta \frac{1}{D_y(W)} \left( R(W) A^{-1} - \Delta(W) I_m \right) y ,
\end{align*}
where we define $R(W)$ appropriately. By the triangle inequality,
\begin{align}
    \left\| \Ex{ \nabla F_y(\bm W) } \right\|
    &= \left\| \theta \Ex{ \frac{F_y(\bm W) \Delta(\bm W)}{D_y(\bm W)} } y 
        - \Ex{ \frac{F_y(\bm W) R(\bm W)}{D_y(\bm W)} } A^{-1} y \right\| \nonumber \\
    &\leq \theta \left| \Ex{ \frac{F_y(\bm W) \Delta(\bm W)}{D_y(\bm W)} } \right| \|y\|
        + \left| \Ex{\frac{F_y(\bm W) R(\bm W)}{D_y(\bm W)}} \right| \|A^{-1} y\| .
        \label{eq:lb-triangle}
\end{align}
To bound the term on the right, we using $D_y(\bm W) \geq 1$ in each inequality below:
\begin{equation}
\label{eq:lb-claim-r}
    \left| \Ex{ \frac{F_y(\bm W) R(\bm W)}{D_y(\bm W)}} \right|
    \leq \left| \Ex{ F_y(\bm W) R(\bm W)} \right|
    \leq \Ex{ F_y(\bm W) } \cdot \frac{1}{n}( \theta \|y\|^2 + m ) .
\end{equation}
For the remaining term, recall
$\Delta(\bm W) = \frac{\theta}{n(1+\theta)}( (y^\top A^{-1} y - m) + (W - k))$, so that:
\begin{equation}
\label{eq:lb-claim-l}
\left| \Ex{ \frac{F_y(\bm W) \Delta(\bm W)}{D_y(\bm W)} } \right|
\leq \frac{\theta}{n(1+\theta)} \left(
        \Ex{ F_y(\bm W)} \left|y^\top A^{-1} y - m \right|
        + \left|\Ex{ F_y(\bm W)  \frac{(\bm W - k)}{D_y(\bm W)}}\right| \right) .
\end{equation}
To bound the term on the right, we use the following identity, which holds for $\bm W \sim \chi_k^2$
and all $h$ where $h$ and its derivative $h'$ are continuously differentiable
(see~\cite{Hud78}):
\begin{equation}
\label{eq:chi2-identity}
\Ex{ (\bm W - k) h(\bm W) } = 2 \Ex{ \bm W h'(\bm W) } .
\end{equation}
We will apply this identity with $h(W) = \frac{F_y(W)}{D_y(W)}$. Then
\begin{equation}
\label{eq:lb-chi2-identity}
h'(W)
= - \frac{\theta}{n \cdot D_y(W)^2} F_y(W) + \frac{1}{D_y(W)} F'_y(W) ,
\end{equation}
where 
\begin{align*}
    F'_y(W)
    &= \exp\left(-\frac{\theta}{2} \|y\|^2 \frac{\Delta(W)}{D_y(W)}\right)
        \cdot  \left(
            - \frac{\theta m}{2 n} \frac{1}{D_y(W)^{1+m/2}} - \right.\\
              &\left.\qquad\qquad\frac{1}{D_y(W)^{m/2}} \frac{\theta}{2} \|y\|^2
    \left(-\frac{\theta \Delta(W)}{n \cdot D_y(W)^2}
    + \frac{\theta }{n(1+\theta) \cdot D_y(W)} \right)
    \right) \\
    &= F_y(W) \cdot \frac{\theta}{2n \cdot D_y(W)} \left(
        - m - \theta \|y\|^2 \left( - \frac{\Delta(W)}{D_y(W)}
        + \frac{1}{1 + \theta} \right) \right) \\
    &= F_y(W) \cdot \frac{\theta}{2n \cdot D_y(W)} \left(
        - m - \frac{\theta}{1+\theta} \|y\|^2 \left( 1 - \frac{\Delta(W)}{1+\Delta(W)}
        \right) \right) \\
    &= - F_y(W) \cdot \frac{\theta}{2n \cdot D_y(W)} \left(
        m + \frac{\theta}{1+\theta} \|y\|^2  \frac{1}{1+\Delta(W)} \right) \\
    &= - F_y(W) \cdot \frac{\theta}{2n \cdot D_y(W)} \left(
        m + \theta \|y\|^2  \frac{1}{D_y(W)} \right)
\end{align*}
Observe that
\[
    \frac{W}{D_y(W)} = \frac{W}{1 + \frac{\theta}{n}(y^\top A^{-1} y + W)} \leq \frac{n}{\theta} .
\]
Using this inequality (along with $D_y(W) \geq 1$ and substituting our calculation for $F'_y(W)$):
\begin{align*}
\left| \Ex{ (\bm W - k) h(\bm W) } \right|
&= 2\left|\Ex{ \bm W h'(\bm W) }\right| \\
&= 2\left|\Ex{ \frac{\bm W}{D_y(\bm W)} F'(\bm W) - \frac{\theta \bm W}{n \cdot D_y(\bm W)^2}
F_y(\bm W) } \right | \\
&\leq 2 \Ex{ \frac{1}{2} \left( m + \frac{\theta\|y\|^2}{D_y(\bm W)} \right) F_y(\bm W) + F_y(\bm W) } \\
&\leq (m + \theta \|y\|^2 + 2)  \Ex{F_y(\bm W)} .
\end{align*}
Substituting into \cref{eq:lb-claim-l}:
\[
    \left| \frac{F_y(\bm W)\Delta(\bm W)}{D_y(\bm W)} \right|
        \leq \Ex{ F_y(\bm W)} \cdot \frac{\theta}{n(1+\theta)} \left(y^\top A^{-1} y + 2m + \theta \|y\|^2 + 2\right) .
\]
Note that $\|A^{-1} y\| \leq (1 + O(\sqrt{m/n})) \|y\| \leq 2\|y\|$ since each $|\delta_i| =
O(\sqrt{m/n})$.  Now substituting into \cref{eq:lb-triangle}, we get
\begin{align*}
\|\Ex{\nabla F_y(\bm W)}\|
    &\leq \theta \Ex{F_y(\bm W)} \left(\frac{1}{n}(\theta \|y\|^2 + m)\|A^{-1} y\|
        + \frac{\theta}{n(1+\theta)} \left(y^\top A^{-1} y + 2m + \theta \|y\|^2 + 2\right)  \|y\| \right) \\
    &= O_\theta\left( \Ex{F_y(\bm W)} \cdot \frac{1}{n} \|y\| ( \|y\|^2 + m ) \right) .
\end{align*}
Finally, we conclude
\[
    \frac{\|\Ex{\nabla F_y(\bm W)}\|^2}{\Ex{F_y(\bm W)}^2} = O\left( \frac{1}{n^2} \|y\|^2(\|y\|^2
        + m)^2 \right) . \qedhere
\]
\end{proof}

\subsubsection{Bound on the Norm}
\label{section:lb-norm-bound}

\claimlbnormbound*
\begin{proof}
Recall that $p$ is the PDF of $\bm y = X\bm v$ and $\gamma_{\alpha I_m}$ is the PDF of $\cN(0,
\alpha I_m)$. We will show that, for appropriate choice of $\alpha$,  we can write $p(y) =
\frac{p(y)}{\gamma_{\alpha I_m}(y)} \gamma_{\alpha I_m}(y)$ where the ratio
$\frac{p(y)}{\gamma_{\alpha I_m}(y)}$
is non-increasing with the norm $\|y\|$. Thus $p$ is concentrating its mass closer to the origin and
we will have $\Ex{\|\bm y\|^{2j}} \leq \Ex{\|\bm z\|^{2j}}$ where $\bm z \sim \cN(0, \alpha I_m)$.
It is known that $\Ex{\|\bm z\|^{2j}} = O(\alpha^{j} m^{j})$; this follows from the
formula for the moments of the $\chi^2_m$ distribution,
since $\|\bm z\|^2$ is distributed as $\alpha \cdot \chi^2_m$:
\[
    \Ex{ (\chi^2_m)^j }
    = m(m+2)\dotsm (m+2j-2) .
\]

Set $\alpha = \frac{2}{1-\theta}$.  Fix any $y \in \bR^m$ and let $s \geq 1$. Recall the notation
$\bm W \sim \chi_k^2$ where $k = n-m$, and
\[
    D_y(W) = 1 + \frac{\theta}{n}\left( y^\top A^{-1} y + W \right) .
\]
The density ratio is
\[
    \frac{p(sy)}{\gamma_{\alpha I_m}(sy)}
    \propto \Ex{ \frac{1}{D_{sy}(\bm W)^{m/2}} \exp\left(G_{sy}(\bm W) \right) }
\]
where the exponent is written
\begin{align*}
G_{sy}(\bm W)
    &\define \frac{\theta}{2D_{sy}(\bm W)} \|sy\|^2
        - \frac{1}{2} (sy)^\top A^{-1} (sy) + \frac{1}{4} \|sy\|^2 \\
    &= s^2\left( \frac{\theta}{2 D_{sy}(\bm W)} \|y\|^2 - \frac{1}{2} y^\top A^{-1} y
        + \frac{1-\theta}{4} \|y\|^2 \right) .
\end{align*}
Recall that $A = \diag(1+\delta_i)_{i \in [m]}$ with each $|\delta_i| = O(\sqrt{m/n})$, so
\[
    y^\top A^{-1} y = (1 \pm o(1)) \|y\|^2 .
\]
Then
\begin{align*}
    \partial_s G_{sy}(W)
    &= 2s\left(\frac{\theta}{2 D_{sy}(W)} \|y\|^2 - \frac{1}{2} y^\top A^{-1} y + \frac{1-\theta}{4}\|y\|^2 \right)
        - 2 s^3 \frac{\theta \|y\|^2}{2 \cdot D_{sy}(W)^2} \frac{\theta}{n} y^\top A^{-1} y \\
    &< 2s\left( \left(\frac{\theta}{2} + \frac{1-\theta}{4}\right) \|y\|^2 - \frac{1}{2} (\|y\|^2 - o(1)) \right) 
    = 2s\left(\frac{1}{4} \theta - \frac{1}{2} + o(1) \right) \|y\|^2 < 0 .
\end{align*}
Therefore the exponent $G_{sy}(\bm W)$ is decreasing with $s$; since the denominator $D_{sy}(\bm W)^{m/2}$
is increasing with $s$, this means the whole density ratio is decreasing, as desired.

Finally, consider the distributions of $\|\bm y\|$ and $\|\bm z\|$ for $\bm z \sim \cN(0, \alpha
I_m)$, and let $\hat p$ and $\hat q$ be their respective densities. The above suffices to show that
$\hat p(\ell) / \hat q(\ell)$ is decreasing with $\ell$, which establishes that $\|\bm z\|$
stochastically dominates $\|\bm y\|$. Thus $\Ex{ \|\bm y\|^{2j} } \leq \Ex{\| \bm z \|^{2j} }$.
\end{proof}
 
 \end{document}